\documentclass[a4paper,11pt]{article}

\usepackage[T1]{fontenc}

\usepackage{latexsym}
\usepackage{longtable}
\usepackage{epsfig}
\usepackage{graphicx,bbm,psfrag} 
\usepackage{amssymb,amsmath,amsthm,booktabs,mathtools}
\usepackage{bm}% bold math
\usepackage{color}
\usepackage{dutchcal}
\usepackage{hyperref}
\hypersetup{
    colorlinks,
    citecolor=red,
    linkcolor=blue,
}

\newtheorem{rema}{Remark}[section]

\newcommand{\bc}{\begin{center}}
\newcommand{\ec}{\end{center}}
\def\ba#1{\begin{array}{#1}\displaystyle}
\newcommand{\ea}{\end{array}}
\newcommand{\z}{\\[0.3cm] \displaystyle}
\newcommand{\beq}{\begin{equation}}
\newcommand{\eeq}{\end{equation}}
\newcommand{\beqa}{\begin{eqnarray}}
\newcommand{\eeqa}{\end{eqnarray}}
\newcommand{\no}{\nonumber}
\newcommand{\n}{\nonumber\\}
\newcommand{\bi}{\begin{itemize}}
\newcommand{\ei}{\end{itemize}}
\renewcommand{\v}[1]{\boldsymbol{#1}}

\def\micro#1{\mathrm{#1}}

\def\lt#1{\left#1}
\def\rt#1{\right#1}
\def\t#1{\tilde{#1}}
\def\h#1{\hat{#1}}

\def\frc#1#2{\frac{#1}{#2}}

\newcommand{\p}{\partial}

\newcommand{\bra}{\langle}
\newcommand{\ket}{\rangle}
\newcommand{\dbra}{\langle\!\langle}
\newcommand{\dket}{\rangle\!\rangle}

\newcommand{\Z}{{\mathbb{Z}}}

\newcommand{\R}{{\mathbb{R}}}

\newcommand{\rx}{{\rm x}}
\newcommand{\ry}{{\rm y}}
\newcommand{\rp}{{\rm p}}

\newcommand{\ep}{\epsilon}

\newcommand{\re}{{\rm e}}
\newcommand{\dd}{{\rm d}}

\DeclareMathOperator{\sgn}{sgn}

\newcommand{\halmos}{\rule{1ex}{1.4ex}}
\newcommand{\eproof}{\hspace*{\fill}\mbox{$\halmos$}}

\begin{document}

\begin{center}
{\Large {\sc Hydrodynamic noise in one dimension:\\ projected Kubo formula\\ and how it vanishes in integrable models}}

\vspace{1cm}

{\large Benjamin Doyon
}

\vspace{0.2cm}
Department of Mathematics, King’s College London, Strand, London WC2R 2LS, U.K.
\ec

Hydrodynamic noise is the Gaussian process that emerges at larges scales of space and time in many-body systems. It is justified by the central limit theorem, and represents degrees of freedom forgotten when projecting coarse-grained observables onto conserved quantities. It is the basis for fluctuating hydrodynamics, where it appears along with ``bare'' diffusion terms via the Einstein relation. In one dimension of space, nonlinearities may modify the corrections to ballistic behaviours by superdiffusive effects. But in systems where no shocks appear, such as linearly degenerate and integrable systems, it turns out that the diffusive scaling stays intact. Nevertheless, anomalies remain. We show that in such systems, the noise covariance is given by a modification of the Kubo formula, where effects of ballistic long-range correlations -- quadratic charges -- have been projected out. We further show that nonlinearities are tamed by a point-splitting regularisation. We then obtain a well-defined hydrodynamic fluctuation theory in the ballistic scaling of space-time, as a stochastic PDE. It describes the asymptotic expansion in the inverse variation scale of connected correlation functions, self-consistently organised via a cumulant expansion. The resulting anomalous hydrodynamic equation takes into account both long-range correlations and bare diffusion, generalising recent works. Despite these anomalies, two-point functions satisfy an ordinary diffusion equation, with a normal Kubo formula. In integrable systems, we show that hydrodynamic noise, hence bare diffusion, must vanish, as was conjectured recently, and argue that under an appropriate gauge, this is true at all orders. Thus initial-state fluctuations do not affect coarse-grained currents, and the Ballistic Macroscopic  Fluctuation Theory give the all-order hydrodynamic theory for integrable models.

\medskip

\tableofcontents

\section*{List of notations}

In this paper, there are a number of thermodynamic ensembles, fluctuation theory, and associated observables and expectations, related to each other in subtle ways. We use specific notations for each of these. They are explained in the text, but for convenience I list them here.
\bi
\item $\micro x,\micro t$ are space-time variables of the microscopic model under consideration (spin chain, gas of particles, quantum field theory, etc.).
\item $\ell$ is the hydrodynamic spatial / temporal scale parameter, taken large, $\ell\to\infty$.
\item $x,t$ are scaled space-time variables used for the hydrodynamic theory: in the hydrodynamic limit $\micro x,\micro t$ are taken large, proportional to the scale parameter $\ell$, and $x=\micro x/\ell,\,t =\micro t/\ell$ are fixed.
\item $q_i(\micro x,\micro t)$ are microscopic local conserved densities, with associated currents $j_i(\micro x,\micro t)$, Eq.~\eqref{cons}, and $o(\micro x,\micro t)$ are more general microscopic local observables. We use $\underline q = (q_i)_i$ for the vector of conserved densities, etc.
\item $\overline o(x,t)$ is the fluid-cell mean of the microscopic observable $o$ at the point $\micro x = \ell x,\,\micro t = \ell t$, Eq.~\eqref{cellmean}. We can also take $o=q_i, j_i$.
\item $\beta^i$, also $\underline \beta = (\beta^i)_i$, is the set of Lagrange parameters for a space-time stationary, maximal entropy state.
\item $\bra \cdots\ket_{\underline\beta}$ is the expectation in the infinite-volume macrocanonical ensemble for Lagrange parameters $\underline\beta$, while $\bra \cdots\ket_{\underline{\mathsf q}}$ is the expectation in the infinite-volume  microcanonical ensemble for average conserved densities $\underline{\mathsf q}$. They are equivalent, with $\bra q_i\ket_{\underline{\mathsf q}}=\mathsf q_i = \bra q_i\ket_{\underline \beta}$, and this gives an invertible relation $\underline\beta\mapsto \underline{\mathsf q}$.
\item $\mathsf o = \mathsf o(\underline{\mathsf q})$ is a function of the ensemble's conserved densities $\mathsf q_i$'s, giving the infinite-volume microcanonical average of the generic microscopic observable $o$, Eq.~\eqref{defo}.
\item $\bra\cdots\ket_\ell$ is the expectation value in the initial state of the microscopic model, which possess a large-scale parameter $\ell$ characterising its long-wavelength variations. It is often taken as a locally-entropy-maximised state, Eq.~\eqref{lw}, but doesn't have to.
\item $\dbra\cdots\dket_\ell$ is the corresponding expectation value in the emergent fluctuation theory for fluctuating conserved densities  that describes the large-scale expectation values and correlation functions. The connection is Eq.~\eqref{corrfctexp}.
\item We use the superscript $^{\rm c}$ to represent connected correlation functions in all ensembles.
\item $q_i(x,t)$ are the fluctuating conserved densities in the emergent fluctuation theory. Note how we use the scaled $x,t$ variables, like for fluid-cell means $\overline{q_i}(x,t)$, but unlike for the microscopic observable $q_i(\micro x,\micro t)$.
\item $o(x,t)$ are likewise the random variables of the emergent fluctuation theory corresponding to microscopic observables $o(\micro x,\micro t)$. They are given in terms fluctuating conserved densities and emergent noise fields, with the main relation, at the basis of fluctuating hydrodynamics, Eq.~\eqref{NLFBG}.
\item $\xi_o(x,t)$ is an emergent noise field in the fluctuation theory -- there is one for every local microscopic observable $o$.
\item $\mathsf A$, $\mathsf C$, $\mathsf D$ are the flux Jacobian Eq.~\eqref{Amatrix}, covariance matrix Eq.~\eqref{Cmatrix} and Drude weight matrix Eq.~\eqref{drudedef}, respectively, in space-time stationary maximal entropy state, seen as functions of $\underline{\mathsf q}$ or equivalently $\underline{\beta}$.
\item $Q_i$ are total conserved quantities (either in the original microscopic model, or the emergent fluctuation theory).
\item $\mathfrak D,\,\mathfrak L$ are the macroscopic diffusion and Onsager matrices, respectively; while $\h{\mathfrak D},\,\h{\mathfrak L}$ are the associated  microscopic quantities. Their indices can be $i,j,\ldots$ representing the conserved quantities, or more generally $o_1,o_2,\ldots$ representing local observables. $\mathfrak L$ has conventional definition Eq.~\eqref{Onsager} and $\mathfrak D$ is given by Einstein's relation. Correctly defining $\h{\mathfrak D},\,\h{\mathfrak L}$, via Eq.~\eqref{NLFBG}, and establishing their relations with $\mathfrak D,\,\mathfrak L$, Sec.~\ref{secconst}, is one of the main results of this paper.
\ei

\section{Introduction}

At large scales of space and time, degrees of freedom in many-body systems separate into ``fast'' and ``slow'' ones. Fast ones can be assumed to quickly relax throughout space-time, at each time taking values that are determined by the spatial profile of slow ones. This is the projection principle of Mori and Zwanzig \cite{mori1965transport,zwanzig1966statistical}. Slow degrees of freedom are associated with conserved quantities \cite{brox1984equilibrium,demasi2006mathematical,spohn2012large,kipnis2013scaling}: coarse-grained conserved densities vary slowly due to  interchanges at the surface of ``fluid cells'', while generic coarse-grained observables fluctuate a lot, thus converging to their {\em microcanonical} values in each fluid cell, as total conserved quantities in each cell appear to be fixed at this time scale. At the linearised level, this is called the Boltzmann-Gibbs principle \cite{kipnis2013scaling}, and keeping microcanonical values as nonlinear functions of densities, it is the basis for the Ballistic Macrocopic Fluctution Theory (BMFT) \cite{doyon2023emergence,doyon2023ballistic}, a large-deviation theory.

This is exact in the strict limit where spatial variations are on infinite length scales $\ell$, as fluid cells can be taken as large as we want. But looking at corrections in the {\em hydrodynamic expansion} in $1/\ell$, this projection is not perfect. The microcanonical averages on fluid cells receive many corrections. In particular, by the central limit theorem, there must be an emergent Gaussian white noise for each coarse-grained observable. This noise is usually assumed to be delta-correlated in space-time, but correlated amongst different observables, and by the fluctuation-dissipation theorem and Einstein relation, is argued to be associated with diffusion terms \cite{demasi2006mathematical,spohn2012large,kipnis2013scaling,de2022correlation}. This is the basis for the theory of fluctuating hydrodynamics, a stochastic PDE. It applies both to deterministic systems, where the noise is an emergent phenomenon carrying initial fluctuations, and stochastic systems, where the emergent hydrodynamic noise is in general not simply related to the microscopic noise. Its applicability to many-body quantum systems has also been investigated \cite{wienand2024emergence}.

In driven-diffusive systems, and under the diffusive scaling of space and time $\vec{\rx}-\vec{v}\micro t\propto \ell,\,\micro t\propto \ell^2$, the combination of fluctuating hydrodynamics and fluctuations of initial conditions gives rise to the Macroscopic Fluctuation Theory (MFT) \cite{bertini2015macroscopic}, a large-deviation theory.

But fluctuating hydrodynamics goes beyond, and holds also when ballistic fluxes -- microcanonical averages of currents -- are nonlinear functions. For instance, take a system with a single conservation law $\p_{\micro t} q + \p_{\micro x} j=0$, where $\micro x,\micro t$ are microscopic coordinates. One writes the coarse-grained current $\overline{j}$ as a function of the coarse-grained conserved density $\overline{q}$
\beq
	\overline{j} = \mathsf j(\overline {q}) - \frc12\t{\mathfrak D}(\overline q)\p_{\micro x} \overline{q} + \eta
\eeq
where $\mathsf j(\overline {q})$ is the microcanonical average of the current in a state characterised by $\overline q$, and $\eta(\rx,\micro t)$ is an emergent noise, correlated on microscopic space-time scales only, with strength $\t{\mathfrak L}(\overline{q}) = \t{\mathfrak D}(\overline{q})\chi(\overline{q})$ where $\chi(\overline{q})$ is the susceptibility. The fluctuating hydrodynamics, valid on large scales of $\micro x,\micro t$, is the stochastic PDE
\beq\label{onecomp}
	\p_{\micro t}\overline{q} + \p_{\micro x} \Big(\mathsf j(\overline {q}) -\frc12 \t{\mathfrak D}(\overline{q})\p_{\micro x} \overline{q} + \eta\Big)=0.
\eeq

Crucially, the noise strength $\t{\mathfrak L}$ and bare diffusion $\t{\mathfrak D}$ are generically different from their macroscopic counterparts \cite{zwanzig1961memory,mori1973nonlinear,kawasaki1973simple,fujisaka1976fluctuation,zubarev1983statistical}. The evaluation of bare transport coefficients is delicate \cite{mori1973nonlinear,donev2011diffusive,donev2011enhancement,nakano2025looking,saito2021microscopic}, and the understanding of the effect of hydrodynamic noise on large-scale physics is an important question, potentially related to spontaneous stochasticity \cite{bandak2024spontaneous} and the dissipation anomaly in turbulence \cite{cheskidov2023dissipation}. For instance, in low-dimensional systems, $\t{\mathfrak L}$ is not given by the Onsager coefficient -- the Green-Kubo formula from the current-current correlator. In fact, the Onsager coefficient often diverges on infinite volumes, indicating the presence of superdiffusion. Notably, if the second derivative $\mathsf j''$ is non-vanishing, Eq.~\eqref{onecomp}, under the KPZ scaling of space and time, is known to lie in the KPZ universality class. With more conservation laws, $\p_{\micro t} q_i + \p_{\micro x}j_i=0$, $i=1,2,\ldots$, the KPZ universality class also emerges  \cite{spohn2014nonlinear}, as controlled by the ``diagonal 3-point couplings'', equilibrium 3-point correlation functions $\mathsf A_k^{~II} = \bra Q_I,Q_I,j_k^-\ket^{\rm c}$ simply related to the ballistic flux derivatives \cite{doyon2022diffusion} (see  Appendix \ref{apphydro}). %Universal KPZ functions do not require the knowledge of bare transport coefficients, but many applications of nonlinear fluctuating hydrodynamics do, so it is important to evaluate them, 

Interestingly, one can show that {\em if the Euler-scale equation is such that shocks are not produced, then the diagonal 3-point couplings vanish, $\mathsf A_k^{~II}=0$, thus there is no superdiffusion}, see \cite{PhysRevLett.134.187101} and Appendix \ref{app3ptcoupling}. A class of such no-shock systems is that of {\em linearly degenerate systems}. Linear degeneracy was first introduce by Lax \cite{lax2005hyperbolic}, see also \cite{ferapontov1991integration,el2011kinetic,pavlov2012generalized} and \cite[Sec 3.2]{bressan2013hyperbolic}. It means that hydrodynamic modes do not self-interact: the hydrodynamic velocity of any normal mode, does not depend on this normal mode, which precludes the appearance of shocks \cite{rozlinedeg,liu1979development,bressan2013hyperbolic}. Integrable many-body systems have linearly degenerate hydrodynamic equations \cite{el2011kinetic,pavlov2012generalized,doyon2020lecture} and are proven not to develop shocks \cite{hubner2024new,hubner2024existence}, and there are many other examples \cite{gopalakrishnan2024non,yoshimura2025anomalous,mcculloch2025ballistic,hydrodynamic2025Yoshimura}.

But despite the absence of superdiffusion, there are still large-scale anomalies in no-shock systems, including linearly degenerate systems.

Consider the {\em the ballistic scaling}
\beq\label{ballisticscale}
	\rx = \ell x,\quad \micro t = \ell t,\quad \ell\to\infty.
\eeq
In no-shock systems, the fluctuating theory at the leading order is simply the deterministic transport of initial fluctuations via the Euler equation: this makes sense as the absence of shocks makes solutions unique, and entropic effects do not occur. This is the BMFT \cite{doyon2023emergence,doyon2023ballistic}. But beyond the leading order, noise and diffusion appears: under the scaling \eqref{ballisticscale} we may expect to have non-zero bare contributions $\frc1{ 2\ell} \t{\mathfrak D}(\overline{q})\p_{x} \overline{q} + \frc1\ell\xi$ in \eqref{onecomp}, where $\xi$ is delta-correlated in space-time. The question of the exact diffusive scale fluctuating hydrodynamics, bare transport coefficients, and their effects on large-scale correlation functions and the hydrodynamic equation for average densities, has not been addressed. In fact, it was recently conjectured that in integrable systems, there is no bare diffusion neither noise, $\xi=0$ \cite{gopalakrishnan2024non,yoshimura2025anomalous,PhysRevLett.134.187101}, even though {\em there are diffusive-scale effects} on the hydrodynamic equation \cite{PhysRevLett.134.187101}, and two-point correlation functions satisfy a diffusive equation \cite{de2018hydrodynamic,de2019diffusion,gopalakrishnan2018hydrodynamics}.

In this paper, we propose the exact diffusive scale fluctuating hydrodynamic theory for no-shock systems in the ballistic scaling \eqref{ballisticscale}. Along with a nonlinear fluctuating Boltzmann-Gibbs principle for generic observables,  Eq.~\eqref{NLFBG} below,  this gives a framework for evaluating the leading and first subleading order in the asymptotic $\ell\to\infty$ expansion of all $n$-point connected correlation functions of local observables, $n=1,2,3,\ldots$. The leading order is given by the BMFT and worked out in \cite{doyon2023ballistic,doyon2025nonlinear}. The first subleading order is a $1/\ell$ relative correction. The resulting stochastic PDE, Eq.~\eqref{hydroeq} below, is well defined for this asymptotic expansion. In particular, we evaluate exactly the noise strength and bare diffusion in terms of a modification of the Green-Kubo formula.
%We also provide physically motivated explanations for the various elements of the fluctuating hydrodynamic equation, pointing to the origin of hydrodynamic noise. 
%A field-theoretic reformulation of the stochastic PDE, such as that proposed in \cite[Sec 3.3]{doyon2023ballistic}, should provide large deviations and their first subleading corrections. These will be studied in a future work.
Our findings are as follows:

(1) We show that there is a correction to the fluid-cell microcanonical average, in inverse power of the cell's size. This correction is usually not discussed in works on fluctuating hydrodynamics, but is crucial. It means that fluxes are subject to a ``regularisation'', $\mathsf j(\overline q) \to \mathsf j^{\rm reg}(\overline q)$, which we show is the point-splitting regularisation first conjectured in \cite{PhysRevLett.134.187101}. This in turn guarantees that the stochastic PDE is well defined in the $1/\ell$ expansion, despite delta-correlations (in macroscopic coordinates) in initial conditions.

(2) We show that the noise covariance is given by a ``projected'' version of the Green-Kubo formula. In this version, quadratic charges \cite{doyon2022diffusion} are projected out. These represent the effects of wave interaction which are at the source of long-range correlations \cite{doyon2023emergence}, and come from the nonlinearity of the flux as worked out in \cite{PhysRevLett.134.187101} in the context of integrability. The fact that flux nonlinearities give rise to diffusive contributions was first proposed as ``diffusion from convection'' in \cite{medenjak2020diffusion} in the context of integrability, and our proof uses a calculation that is somewhat similar.

(3) We show that the bare diffusion matrix is connected to the projected Onsager matrix by the Einstein relation. This relation is natural on the grounds of the fluctuation-dissipation theorem, but it remains an assumption in the context of fluctuating hydrodynamics. We show that it is a consequence of the self-consistency of the PDE as a principle for the $1/\ell$ expansion, along with the conservation laws.

(4) We show that in many-body integrable models, the noise for currents vanishes. This follows from our result, along with earlier calculations of the full Onsager matrix in integrable models \cite{boldrighini1997one,spohn2012large,de2018hydrodynamic,de2019diffusion,gopalakrishnan2018hydrodynamics} (that used different methods). This establishes the conjecture of \cite{gopalakrishnan2024non,yoshimura2025anomalous,PhysRevLett.134.187101}. This implies that in integrable systems, initial fluctuations do not influence the diffusive corrections to coarse-grained nonlinear currents as functions of conserved densities. We also propose a more heuristic derivation of the vanishing of noise from first principles, using the infinite family of commuting flows that many-body integrable models admit. The argument shows that under an appropriate choice of ``gauge'' for the currents, their noise vanish at all orders in $1/\ell$. It also effectively gives a new, first-principle derivation of the exact formula for the full Onsager matrix in integrable models.  We emphasise, however, that in generic linearly degenerate systems, the noise does not vanish (contrary to the conjecture made in \cite{PhysRevLett.134.187101}) -- this is a property of integrability.

From our fluctuating hydrodynamic theory, we write down the hydrodynamic equation up to, including, the diffusive order $1/\ell$, see Eqs.~\eqref{hydro1}, \eqref{hydro2} below. It is a simple modification of that proposed in \cite{PhysRevLett.134.187101} for integrable models. Long-range corelations and bare diffusion are explicitly taken into account separately.

Interestingly, the projected Kubo formula for the noise is also worked out, but in the diffusive instead of ballistic scaling, in the simultaneous paper \cite{hydrodynamic2025Yoshimura} of which we have been made aware at the time of writing. Thus this formula holds no matter the choice of scaling. In particular, in \cite{hydrodynamic2025Yoshimura} it is evaluated in a specific model, where it is shown to agree with microscopic calculations and to be non-zero.

We emphasise that bare diffusion is different from hydrodynamic diffusion. Linearised hydrodynamics for two-point correlation functions in linearly degenerate systems is not anomalous, and simply diffusive. There, the hydrodynamic diffusion constant is connected by the Einstein relation to the {\em full Onsager matrix}, not the projected one  (see e.g.~\cite{de2022correlation}). We show that this follows from fluctuating hydrodynamics theory; it was also argued for in integrable models in \cite{PhysRevLett.134.187101} using different ideas, and assumed to be true for some time \cite{de2018hydrodynamic,de2019diffusion,gopalakrishnan2018hydrodynamics}. Thus, integrable models have normal diffusive corrections to ballistic two-point functions. Beyond the linearised level, however, these diffusive effects desintegrate into long-range correlations \cite{PhysRevLett.134.187101}; higher-point functions have anomalous diffusive-scale corrections.

We believe our general explanations and precise derivation give a better understanding of the microscopic origin of hydrodynamic noise and bare diffusion, and of the subtle principles behind fluctuating hydrodynamics more generally.

The paper is organised as follows. In Section \ref{secconsqties} we specify our general setup, describing the conserved quantities and states. In Section \ref{sectbeyond}, we express our main result for the fluctuating hydrodynamic theory, and provide arguments as to how coarse-graining gives rise to the special effects seen beyond the Euler scale, including the noise. In Section \ref{secconst}, we prove points (1)-(3) described above. In Section \ref{secabsence} we show the absence of noise, point (4) above, in integrable systems. In Section \ref{sechydroeq}, we obtain, from the fluctuating theory, the hydrodynamic equation for one-point averages out of equilibrium, and for two-point correlations in stationary states. Finally, in Section \ref{sectconclusion} we conclude.

{\em Note added:} Shortly after the first version of this paper was posted on arXiv, a paper by F. H\"ubner was also posted \cite{hubner2025hydrodynamics} which presents an analtical and numerical study of the hard rods model confirming our finding that hydrodymamic noise in integrable systems vanish at the diffusive level.

\section{Conservation laws and states}\label{secconsqties}

Consider a dynamical system in one spatial dimension, which may be deterministic or stochastic, with short-range, translation invariant interactions. Fluctuating hydrodynamics, where a classical noise is assumed to emerge, is naturally a classical theory, and quantum effects may affect this in quantum systems. Many of the results below may still be applicable to quantum models \cite{wienand2024emergence}, however we will not discuss this.

In order to have a universal description, one powerful method is to concentrate on {\em local observables} -- such as in the algebraic formulation of statistical mechanics \cite{bratteli1987operator,bratteli1997operator}. Roughly, for our purposes, a local observable, denoted $o(\rx,\micro t)$, is a function on phase space, or a random variable of the stochastic process, etc, which is supported (in a natural sense) on some finite region of space in microscopic units, around position $\rx\in\R$ at time $\micro t\in\R$. Observables can be multiplied with each other (so they form an algebra), and states (see below) are positive linear maps on this algebra.

We will write $\rx = \ell x,\,\micro t = \ell t$, defining the space-time point coordinates $(x,t)$ in macroscopic unit, where $\ell$ is the large macroscopic scale\footnote{We consistently use the different fonts $\rx, \micro t$ and $x,t$ for microscopic and macroscopic coordinates, respectively.}.

The system admits a certain number of extensive conserved quantities\footnote{Integrals without domain specification are on the full space, which is $\R$ in continuous sytems or in macroscopic units, and $\Z$ in systems with discrete space and in microscopic units.}.
\beq
	Q_i = \int \dd \rx\,q_i(\rx,\micro t), \quad
	\p_{\micro t} Q_i = 0
\eeq
where $q_i(\rx,\micro t)$ is the associated local density. The index $i$ lies in some index set $\mathcal I$ which may be finite or not. The index may even be continuous like in integrable models where it represents the spectral space (such as velocities of quasiparticles); we still use the notation $\sum_i$. We assume that there are associated local currents $j_i(\rx,\micro t)$ giving rise to local conservation laws:
\beq\label{cons}
	\p_{\micro t} q_i(\rx,\micro t) + \p_\rx  j_i(\rx, \micro t) = 0.
\eeq
Total charges $Q_i$, on infinite volume, do not make sense as observables because they are infinite, but make sense within connected correlation functions, which is how we will use them.

When discussing hydrodynamics at the diffusive order, the concept of PT symmetry plays an important role, see e.g.~\cite{chaikin1995principles}; for our purposes, the discussion in \cite{de2019diffusion} is the most adapted. We assume that there is an involution ${\mathcal PT}$ of the algebra of observables that reverses space and time, and that {\em preserves all conserved densities and their currents}. Further, we will only consider local observables $o$ that are real or hemitian, and {\em that are preserved by the PT transformation}. That is,
\beq
	{\mathcal PT} (q_i (\rx,\micro t)) = q_i(-\rx,-\micro t),\quad
	{\mathcal PT}(j_i(\rx,\micro t)) = j_i(-\rx,-\micro t),\quad
	{\mathcal PT}(o(\rx,\micro t)) = o(-\rx,-\micro t).
\eeq
This simplifies the discussion in subtle ways. Note that in general, there may be observables $o$ with negative eigenvalue under the ${\mathcal PT}$ involution; however, our results do not apply to such observables.

The macrocanonical ensemble is that which maximises entropy under the constraints of the extensive conserved quantities. Formally, it has the Gibbs form
\beq\label{macrodef}
	\bra \cdots\ket_{\underline\beta}
	=\lim_{\mathcal V\to\R} \frc{\int \dd\mu \,\re^{-\sum_{i} \beta^i Q_i}
	\cdots}{\int \dd\mu \,\re^{-\sum_{i} \beta^i Q_i}}
\eeq
where $\underline\beta = (\beta^1,\beta^2,\ldots)$ are the ``Lagrange parameters'', the charges $Q_i = \int_{\mathcal V} \dd^d \rx\,q_i(\v\rx)$ lie on the volume $\mathcal V$, and the large-volume limit $\mathcal V\to\R$ is assumed to exist for averages of local observables. $\mu$ is a time-invariant, homogeneous prior measure, for instance in classical Hamiltonian systems with canonical coordinates $(\rx_a,\rp_a)$ one chooses $\dd\mu = \sum_{N=0}^\infty (N!)^{-1} \prod_{a=1}^N \dd \rx_a \dd \rp_a$ (direct sum). We assume that $\beta^i$'s are away from phase transitions, so that  connected correlation functions are short-range in these states, and averages of local observables are smooth in $\underline\beta$. Likewise, the microcanonical ensemble is denoted (by a slight abuse of notation)
\beq
	\bra \cdots\ket_{\underline{\mathsf q}}
	=\lim_{\mathcal V\to\R} \frc{\int_{Q_i/|\mathcal V|\,\in\,[\mathsf q_i-\ep,\mathsf q_i+\ep]\,\forall i} \dd\mu \,
	\cdots}{\int_{Q_i/|\mathcal V|\,\in\,[\mathsf q_i-\ep,\mathsf q_i+\ep]\,\forall i} \dd\mu}
\eeq
where $\ep\to0$ as $\mathcal V\to\R$ in an appropriate fashion \cite{aizenman1978conditional,brandao2015equivalence}. The covariance matrix,
\beq\label{Cmatrix}
	\mathsf C_{ij} = -\frc{\p\bra q_j\ket_{\underline\beta}}{\p \beta^i} = \bra Q_i,q_j\ket^{\rm c}_{\underline\beta} = \int \dd\rx \,\bra q_i(\rx),q_j( 0)\ket_{\underline\beta}^{\rm c} =
	\int \dd\rx \,\big(\bra q_i(\rx)q_j(0)\ket_{\underline\beta}
	-
	\bra q_i(\rx)\ket_{\underline\beta}\bra q_j( 0)\ket_{\underline\beta}\big)
\eeq
is always non-negative, and must be assumed to be positive for conserved quantities to have a hydrodynamic meaning\footnote{This guarantees convexity of the free energy, a fundamental concept of statistical mechanics. See e.g.~\cite{doyon2022hydrodynamic} for a general theorem concerning this in the context of the linearised Euler equation.}. That is,
\beq
	\mathsf C^{\rm T} = \mathsf C,\quad \mathsf C>0,
\eeq
and the map $\underline\beta\mapsto\bra\underline q\ket_{\underline\beta}$ is invertible (because then the Jacobian of this map is non-singular, by \eqref{Cmatrix}). This defines functions $\beta^i(\underline{\mathsf q}) = \beta^i(\mathsf q_1,\mathsf q_2,\ldots)$ for all $i$'s. There is equivalence between microcanonical and macrocanonical ensembles \cite{aizenman1978conditional,brandao2015equivalence,touchette2015equivalence}: for every local obsevable $o$, we have $\bra o\ket_{\underline{\mathsf q}} = \bra o\ket_{\underline\beta(\underline{\mathsf q})}$. It is crucial here that the volume be taken to infinity, as otherwise the equivalence does not hold, a point that will become important below. Throughout, we denote the resulting average of $o$, as a function of conserved densities $\underline{\mathsf q}$, by using the ``sans-serif'' font for the observable,
\beq\label{defo}
	\mathsf o(\underline{\mathsf q}) := \bra o\ket_{\underline{\mathsf q}} = \bra o\ket_{\underline\beta(\underline{\mathsf q})}.
\eeq
See Appendix \ref{apphydro} for hydrodynamic matrices, velocities and normal modes.

For our general discussion of hydrodynamics, we consider states which are not space-time stationary, but with variations on long wavelengths $\ell$. It is useful to think of states of the form
\beq\label{lw}
	\bra \cdots\ket_\ell
	= \frc{\int \dd\mu \,\re^{-\sum_{i} \int\dd \rx \,\beta^i(\rx/\ell) q_i(\v\rx)}
	\cdots}{\int \dd\mu \,\re^{-\sum_{i} \int\dd\rx \,\beta^i(\rx/\ell) q_i(\v\rx)}}
\eeq
for a large, macroscopic scale $\ell$. At initial time, this state is a local-equilibrium state,
\beq
	\lim_{\ell\to\infty} \bra o(\ell x,0)\ket_\ell = \bra o\ket_{\underline\beta(x)}.
\eeq
It also has short range correlations: for every fixed $x_1,\ry_1,x_2,\ry_2\in\R$, one has
\beq\label{initcorrlw}
	\bra o_1(\ell x_1 + \ry_1,0) ,o_2(\ell x_2 + \ry_2,0)\ket_\ell^{\rm c} =
	\Big(\bra o_1(\ry_1,0), o_2(\ry_2,0)\ket_{\beta(x_1)}^{\rm c}+\mathcal O(\ell^{-2})\Big)\delta_{x_1,x_2} + \mathcal O(\ell^{-\infty})
\eeq
where $\delta_{x_1,x_2} = 1$ if $x_1=x_2$ and $0$ otherwise; here $\mathcal O(\ell^{-\infty})$ means that the corrections decay faster than any power of $\ell$ (we expect them to be exponentially decaying). A more precise expression of these initial correlations, as distributions on the macroscopic coordinates, is given in \eqref{initcondq} below. At later times it develops long-range correlations \cite{doyon2023emergence}. More general states, with long-range correlations or with long-wavelength source insertions at various macroscopic times, can also be considered.

\section{Local relaxation with fluctuations}\label{sectbeyond}

At the basis of hydrodynamics is the idea of separation of scales: some observables vary and fluctuate slowly in space and time, while other do so much more quickly. By assuming that fast observables relax quickly, there remains a theory for slow observables, which are the local conserved densities admitted by the system. The Boltzmann Gibbs principle for hydrodynamics at the Euler scale says that linear fluctuations of observables project onto those of densities under such a fast relaxation \cite{kipnis2013scaling}. Its nonlinear version says that every local observable is a fixed, non-fluctuating function of conserved densities, which are themselves fluctuating. This gives access to ballistic-scaling large deviations \cite{doyon2023emergence,doyon2023ballistic,doyon2025nonlinear}. Beyond the Euler scale, one must add noise and diffusive effects. This is at the basis fluctuating hydrodynamics.

In this section we write the corresponding nonlinear fluctuating Boltzmann-Gibbs principle, and in order to justify its various parts, provide heuristic arguments for it. We consider the ballistic scaling of space and time \eqref{ballisticscale}, and its corrections in $1/\ell$. Most of this discussion is relevant to other scalings as well, such as the diffusive scaling, see \cite{gopalakrishnan2024distinct,krajnik2022exact,yoshimura2025anomalous,gopalakrishnan2024non,hydrodynamic2025Yoshimura}.

\subsection{Nonlinear fluctuating Boltzmann-Gibbs principle}\label{secNLFBG}

A fluid cell is an abstract construct that is useful in order to determine the large-scale properties of many-body system. In one dimension, it is an interval of size $L$ much larger than microscopic lengths $\ell_{\rm micro}$ but much smaller than variation lengths $\ell$,
\beq\label{limit}
	\ell\gg L\gg\ell_{\rm micro}.
\eeq
Both $\ell_{\rm micro}$ (determined by the microscopic model) and $\ell$ (determined by the initial state) are measurable. $L$ is not, but it is a convenient theoretical concept. We may consider $\ell_{\rm micro}$ to be finite (say 1), by appropriate choice of microscopic units. We write
 \beq
	V_L = [-L/2,L/2]
\eeq
and define the fluid-cell mean of an observable -- the coarse grained observable -- at macroscopic space-time coordinates $x,t$ as
\beq\label{cellmean}
	\overline o(x,t):= \frc1{L} \int_{V_L} \dd \ry\,
	o(\ell x+\ry ,\ell t).
\eeq
This precise form of the fluid cell is not mandatory; one may for instance consider an even weight function $w(\rx)= w(-\rx)$ such that $|w(\rx)|<a \re^{-b|\rx|}$ for some $a,b>0$ and $\int_\R\dd\rx \,w(\rx) = 1$, and define
\beq
	\overline o(x,t):= \frc1{L} \int_{\R} \dd \ry\,
	w(\ry/L) o(\ell x+\ry ,\ell t).
\eeq
In order to simplify the form of the initial correlations in the state \eqref{lw}, it is important that the fluid-cell mean be {\em balanced}: the fluid-cell averaging is invariant under $\ry\to-\ry$.

We argue in Sec.~\ref{ssectarg} that a {\em nonlinear, fluctuating Boltzmann-Gibbs principle} holds. That is, coarse-grained observables can be written solely in terms of fluctuating fields representing coarse-grained conserved densities, which by a slight abuse of notation we denote $q_i(x,t)$, and emergent Gaussian white noise fields $\xi_o(x,t)$'s representing the ``forgotten'' fast microscopic degrees of freedom for the observable $o$. Both of these families of fields fluctuate, and we represent averages in the corresponding emergent fluctuating theory as $\dbra\cdots\dket_\ell$. The emergent fluctuating theory implicitly depends on $L$ via our choice of fluid cell, but it only gives ``universal'' results, independent of how $L$ is chosen, in the limit \eqref{limit}.

Before we express this emergent fluctuating theory, we mention that it is expected to describe the large-$\ell$ asymptotic expansion of all connected correlation functions at distinct macroscopic space-time coordinates $(x_i,t_i)\neq (x_j,t_j)\;\forall i\neq j$, as
\beq\label{corrfctexp}
	\bra
	\overline{o_1}(x_1,t_1),\ldots, \overline{o_n}(x_n,t_n)\ket^{\rm c}_\ell = \dbra o_1(x_1,t_1),\ldots,o_n(x_n,t_n)
	\dket^{\rm c}_\ell  + \mathcal O(\ell^{-1-n})
\eeq
where $\dbra o_1(x_1,t_1),\ldots,o_n(x_n,t_n) \dket^{\rm c}_\ell$ are distributions, of the form
\beq\label{ldcorr}
	\dbra o_1(x_1,t_1),\ldots,o_n(x_n,t_n)
	\dket^{\rm c}_\ell = \ell^{1-n} \mathsf S_{o_1,\ldots,o_n}(x_1,t_1;\cdots;x_n,t_n)
	+ \ell^{-n} \delta \mathsf S_{o_1,\ldots,o_n}(x_1,t_1;\cdots;x_n,t_n)
	+ \ldots.
\eeq
The leading large-deviation scaling $\ell^{1-n}$ is explained in \cite{doyon2018exact,doyon2023ballistic,doyon2025nonlinear}.

The nonlinear Boltzmann-Gibbs principle says that the random variable $o(x,t)$ in the fluctuating theory $\dbra\cdots\dket_\ell$, representing the coarse-grained $\overline o(x,t)$, has three parts: microcanonical, diffusive and stochastic (Gaussian white noise)\footnote{In correlation functions with $n\geq 3$, if the Euler hydrodynamics admits shocks, we expect \eqref{NLFBG} to hold almost everywhere in space-time. The specification ``almost everywhere'' is because at shocks the local relaxation arguments at its basis do not hold, and this, even at the diffusive order; and for $n\geq 3$, the nonlinear expansion required may be affected by shocks. In no-shock systems, such as with linear degeneracy, it is expected to hold everywhere except at colliding space-time positions. See the discussion in \cite{doyon2025nonlinear}.}. For all\footnote{For $t<0$, the sign of the diffusive part is positive instead of negative, so in general we must replace $-\longrightarrow -\sgn t$ in front of the diffusive term.} $t>0$,
\beq\label{NLFBG}\boxed{
\begin{aligned}
	\lefteqn{\overline o( x, t)\ \mbox{in $\bra\cdots\ket_\ell$}} && \\
	&\longrightarrow&
	o(x,t):=\underbrace{\mathsf o^{\rm reg}(\underline q(x,t)) }_{\text{microcanonical}} - \underbrace{\frc{1}{2\ell}\sum_i\h{\mathfrak D}_o^{~i}(\underline q(x,t))
	\p_x q_i(x,t)}_{\text{diffusive}} + \underbrace{\frc1\ell\xi_o(x,t)}_{\text{noise}}
	\ \mbox{in $\dbra\cdots\dket_\ell$}
	\end{aligned}}
\eeq
and in particular, for conserved densities,
\beq\label{repconsdens}
	\overline{q_i}(x,t) \longrightarrow q_i(x,t).
\eeq

For the initial state \eqref{lw}, the noise $\xi_o(x,t)$ is exactly zero at $t=0$,
\beq\label{blabla2}
	\xi_o(x,0) = 0,
\eeq
but it is nonzero generically for all $t>0$, see below. The initial state gives rise to the following one- and two-point correlations for the emergent conserved density fields,
\beqa\label{initcondq}
	\dbra q_i(x,0)\dket_\ell &=&
	\bra q_i\ket_{\underline{\beta}(x)} +\mathcal O(\ell^{-2})\\ 
	\dbra q_i(x,0) q_j(x',0)\dket_\ell &=&
	\ell^{-1} \mathsf C_{ij}(x) \delta(x-x') + \mathcal O(\ell^{-3})\no
\eeqa
where $\mathsf C_{ij}(x)$ is the covariance matrix evaluated within the state $\underline{\beta}(x)$. Note that $n$-point functions receive $\mathcal O(\ell^{-1})$ relative corrections, PT symmetry and because the fluid-cell mean is balanced, as we show in Appendix \ref{appPT}.

The fluctuating theory $\dbra\cdots\dket_\ell$ is {\em completely determined by the initial correlations $S_{q_{i_1},\ldots,q_{i_n}}(x_1,0;\cdots,x_n,0)$ and $\delta S_{q_{i_1},\ldots,q_{i_n}}(x_1,0;\cdots,x_n,0)$, the replacement \eqref{NLFBG}, and the conservation laws}
\beq
	\p_t q_i(x,t) + \p_x j_i(x,t)=0.
\eeq
Indeed, these then provide the dynamics and hence the  correlations of conserved densities at later times: this is through the {\em nonlinear fluctuating hydrodynamic equation} (here for $t>0$)
\beq\label{hydroeq}\boxed{
	\p_t q_i(x,t) + \p_x \Big(
	\mathsf j_i^{\rm reg}(\underline q(x,t)) - \frc{1}{2\ell}\sum_k\h{\mathfrak D}_{i}^{~k}(\underline q(x,t))\,
	\p_x q_k(x,t) + \frc1\ell\xi_i(x,t) \Big)= 0 }
\eeq
where we use the simplified notation
\beq\label{notationD}
	\h{\mathfrak D}_{i}^{~k} := \h{\mathfrak D}_{j_i}^{~k},\quad
	\xi_i := \xi_{j_i}.
\eeq
The replacement \eqref{NLFBG} along with \eqref{hydroeq} gives the large-$\ell$ asymptotic expansion \eqref{ldcorr}.

Let us discuss the various elements in \eqref{NLFBG}. The quantities $\h{\mathfrak D}_o^{~i}(\underline q)$ are the ``bare'' diffusion functions for the observable $o$, which account for fluctuation-dissipation effects from the noise. They represent irreversible effects (thus the $\sgn(t)$ factor in general).

$\mathsf o^{\rm reg}(\underline q)$ is the finite-size-corrected microcanonical average \eqref{defo} in the fluid cell. Recall that the equivalence between microcanonical and macrocanonical ensembles breaks down at finite volumes. Here, in order to account for all diffusive-order effects, one must indeed consider this breaking of equivalences, and use the {\em microcanonical average}, and its leading finite-size corrections. The result, in our effective theory, is implemented as a ``regularisation,'' $\mathsf j_i^{\rm reg}(\mathsf q)$, of the random variable $\mathsf j_i(\mathsf q)$. This regularisation is {\em crucial in order to make sense of singularities that arise because of short-range correlations}, as introduced by the initial state, Eq.~\eqref{initcondq}.

Noise fields $\xi_o(x,t)$ are linear functions of $o$. As discussed in different contexts in \cite{klimontovich1990ito,sokolov2010ito,escudero2023versus,bhattacharyay2025brownian,sharma2025entropic}, physically meaningful noise is to be treated in the It\^o convention on time. That is, given $\mathcal F(t) = \{q_i(x',t'):x'\in\R,\,t'\in [0,t],i\in \mathcal I \}$, the noise fields $\xi_o(x,t)$, for all observables $o$ (including currents $j_i$) and for all $x\in\R$, are such that $\xi_o(x,t)\dd t= \dd w_o(x,t) := w_o(x,t+\dd t)-w_0(x,t)$ are centered Gaussian {\em forward} increment, with
\beq\label{itow}
	\dbra \dd w_o(x,t)\dd w_{o'}(x',t)|\mathcal F(t)\dket_\ell
	= \hat{\mathfrak L}_{o,o'}\big(\underline q(x,t)\big) \delta(x-x')\dd t
\eeq
for some symmetric noise covariance $\hat{\mathfrak L}_{o,o'}(\underline q)$, function of the densities.

Note that by this formula, the noise covariance $\hat{\mathfrak L}_{o,o'}(\underline q)$ is such that for any set of local observables $\mathcal S$, the matrix $\hat {\mathfrak L}(\underline{q})$ with elements $\hat{\mathfrak L}_{o,o'}(\underline q):(o,o')\in \mathcal S\times \mathcal S$ is non-negative. For $o=q_i$ a conserved density, according to \eqref{repconsdens} the noise must vanish, $\hat{\mathfrak L}_{q_i,q_i}(\underline q)=0$ (therefore $\hat{\mathfrak L}_{q_i,o}(\underline q)=0\,\forall\,o$ by the Cauchy-Schwartz inequality).

This gives the noise fields correlations at equal times. In order to express them at different times, we need to take away the explicit dependence of their covariance on the densities. Indeed, these are fluctuating, and correlate non-trivially with previous histories of noise fields because of the fluctuating hydrodynamic equation \eqref{hydroeq}. For this purpose, fix a set of observables $\mathcal S$ not linearly related to each other and with non-vanishing noise. Then $\mathcal S$ is a basis for its span and the matrix $\hat{\mathfrak L}(\underline{q})$ is strictly positive on this basis. Define
\beq
	\hat \xi_o = \sum_{o'\in\mathcal S}\Big(\sqrt{\hat {\mathfrak L}^{-1}}\Big)_{oo'}\xi_{o'}.
\eeq
Then the associated increments $\dd \hat w_o(x,t) := \hat\xi_o(x,t)\dd t$ satisfy
\beq
	\dbra \dd \hat w_o(x,t)\dd \hat w_{o'}(x',t)|\mathcal F(t)\dket_\ell
	= \delta_{o,o'}\delta(x-x')\dd t,\quad o,o'\in\mathcal S.
\eeq
It is natural to assume that these are now independent, delta-correlated noise fields over all times,
\beq\label{noisecorr}
	\dbra \hat \xi_o(x,t)\dket_\ell = 0,\quad \dbra \hat\xi_{o}(x,t) \hat\xi_{o'}(x',t')\dket_\ell
	= \delta_{o,o'}\delta(x-x')\delta(t-t'),\quad o,o'\in\mathcal S.
\eeq

In general, correlations $\dbra \xi_{o}(x,t) \xi_{o'}(x',t')\dket_\ell$ at different times will be very non-trivial, and $\xi_{o}(x,t)$ is non-Gaussian, with non-zero higher cumulants at different times. However, for the purpose of the fluctuating theory as a theory for the leading and first sub-leading order for correlation functions, Eqs.~\eqref{corrfctexp}, \eqref{ldcorr}, noise averages only occur in evaluating the diffusive $\mathcal O(\ell^{-1})$ relative correction. At this order, the cumulant expansion makes the noise covariance non-fluctuating, evaluated on the average densities. Hence, from this viewpoint $\xi_{o}(x,t)$ may be considered Gaussian, and we have
\beq\label{noisecorr}
	\dbra \xi_o(x,t)\dket_\ell = 0,\quad \dbra \xi_{o}(x,t) \xi_{o'}(x',t')\dket_\ell
	= \hat{\mathfrak L}_{o,o'}\big(\dbra\underline q(x,t)\dket_\ell\big) \,\delta(x-x')\delta(t-t').
\eeq
When used in conjunction with the fluctuating hydrodynamic equation \eqref{hydroeq}, this makes the latter not a standard stochastic PDE, but rather a {\em self-consistent stochastic PDE}, where the noise statistics is fixed by the {\em averages densities solving this evolution equation in a self-consistent manner}.

The meaning of the various terms in \eqref{NLFBG}, as well as the emergence of the It\^o convention, is discussed heuristically in Sec.~\ref{ssectarg}. These terms will be evaluated in Sec.~\ref{secconst} by evaluating, with this fluctuating theory, particular equilibrium quantities including the Green-Kubo formula.

\begin{rema}
Eq.~\eqref{NLFBG} is the general form expected in fluctuating hydrodynamics (see e.g.~\cite{spohn2014nonlinear}). However, the constitutive functions $\hat{\mathfrak L}_{o,o'}\big(\underline q\big)$ and $\h{\mathfrak D}_o^{~i}(\underline q)$ representing the corrections to the Euler scale must be evaluated, as well as the regularisation $\mathsf o^{\rm reg}(\underline q)$, which is typically not discussed. In systems that are purely diffusive, for current observables $o=j_k$ only the diffusive and stochastic part remain, as the microcanonical part is independent of $\underline{q}(x,t)$. This is the basis for the Macroscopic Fluctuation Theory (MFT) \cite{bertini2015macroscopic}. Without regularisation, diffusive and noise term, the result is the Ballistic Macroscopic Fluctuation Theory \cite{doyon2023ballistic}. These are large-deviation theories, which avoid singularities due to the noise and initial conditions.
\end{rema}

\begin{rema}
Usually one only considers noise terms associated to the local currents, and takes $\mathcal S = \{j_i:i\in\mathcal I\}$. However, in general, every observable $o$ has its associated, independent noise $\xi_o$, generically correlated to all other observables, including currents. It is simple to see from the evolution equation \eqref{hydroeq} that noise fields for observables that are not currents decouple from the theory, so may be omitted if one looks only at densities and currents.
\end{rema}

\subsection{Coarse graining and the origin of hydrodynamic noise}\label{ssectarg}

\begin{figure}
\bc
\includegraphics[width=14cm]{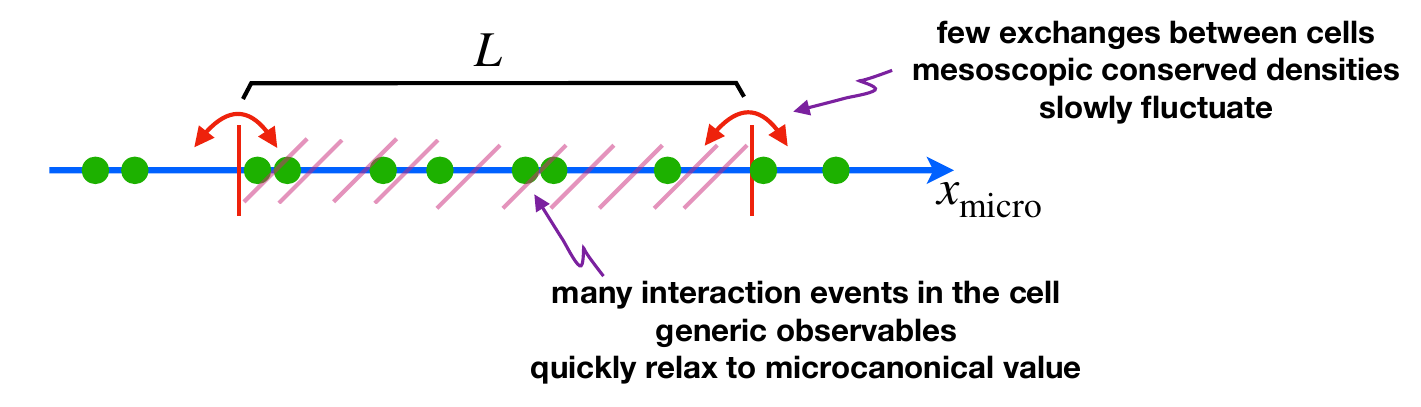}\ec
\caption{A fluid cell of length $L$ over time $T$. There are any more interactions within the cell than there are boundary effects, leading to a separation of scales between fluctuations of arbitrary local observables and fluctuations of conserved densities.}
\label{figfluidcell1}
\end{figure}

We now provide general heuristic arguments for why \eqref{NLFBG} holds. Although these general arguments suggest the form of each correction, they do not allow us to compute them. However, Eq.~\eqref{NLFBG}, as a principle for the large-$\ell$ asymptotic expansion of connected correlation functions, gives us a framework from which we will evaluate \`a la Kubo, in the next section, all constitutive elements in terms of equilibrium correlation functions, in linearly degenerate systems.

Recall the fluid-cell mean \eqref{cellmean}. The main argument is that, because of the conservation laws, fluid-cell means of conserved densities are affected only by boundary effects during the dynamics, while generic observables are affected by every interactions within the volume of the cell. This leads to a separation of scales between conserved densities and generic observables. See Fig.~\ref{figfluidcell1}.

Let us be more quantitative. By the conservation law, the time derivative of the fluid-cell mean of a conserved density, with respect to time $\micro t = \ell t$ at position $\rx = \ell x$ in microscopic units, is
\beq
	\frc{\p}{\p \micro t}\overline{q_i}(x,t) = L^{-1} 
	\big(j_i(\rx-L/2,\micro  t) - j_i(\rx+ L/2,\micro t)\big).
\eeq
Assuming finite densities and finite typical microscopic velocities, the current observables are of order 1 and vary on a time scale $T$ of order 1 as well. Therefore, $\overline {q_i}(x,t)$ has variations of order $\mathcal O(L^{-1})$, and this happen on a time scale of order 1. Physically, these are due to surface effects, such as particles entering or exiting the fluid cell, and interaction between particles through the interface of the cell (which are the points $\rx \pm L/2$); such surface effects are small. Then, on $T=\mathcal O(1)$, $\overline {q_i}(x,t)$ can be considered to be invariant, and in particular non-fluctuating.

Now consider the fluid-cell mean of an observable that is {\em not conserved}, $\overline{o}(x,t)$. Because it is composed of $\mathcal O(L)$ essentially independent observables, each affected by independent fluctuations, the quantity $\overline{o}(x,t)$ varies on a time scale of order $1/L$, each variation being of order $1/L$. Therefore, $\overline{o}(x,t)$ fluctuates quickly, with $\mathcal O(L)$ fluctuations on a time $T$ of order 1. Physically, these are due to volume effects: every ``interaction event'' between particles, such as collision of particles, within the volume of the cell, affect the fluid-cell mean of the observable, and on a time of order 1, there are $\mathcal O(L)$ such interaction events. They can be seen as independent fluctuations as they occur at random places within the volume of the cell and change the fluid-cell mean in incoherent ways. Averaging over $T$, which averages over these fluctuations, and considering that conserved densities are invariant and non-fluctuating, the fluid-cell mean of an arbitrary observable should therefore relax to its microcanonical value $\mathsf o(\overline{q_1}(x,t),\overline{q_2}(x, t),\ldots)$. See Fig.~\ref{figfluidcell2}.

\begin{figure}
\bc
\includegraphics[width=10cm]{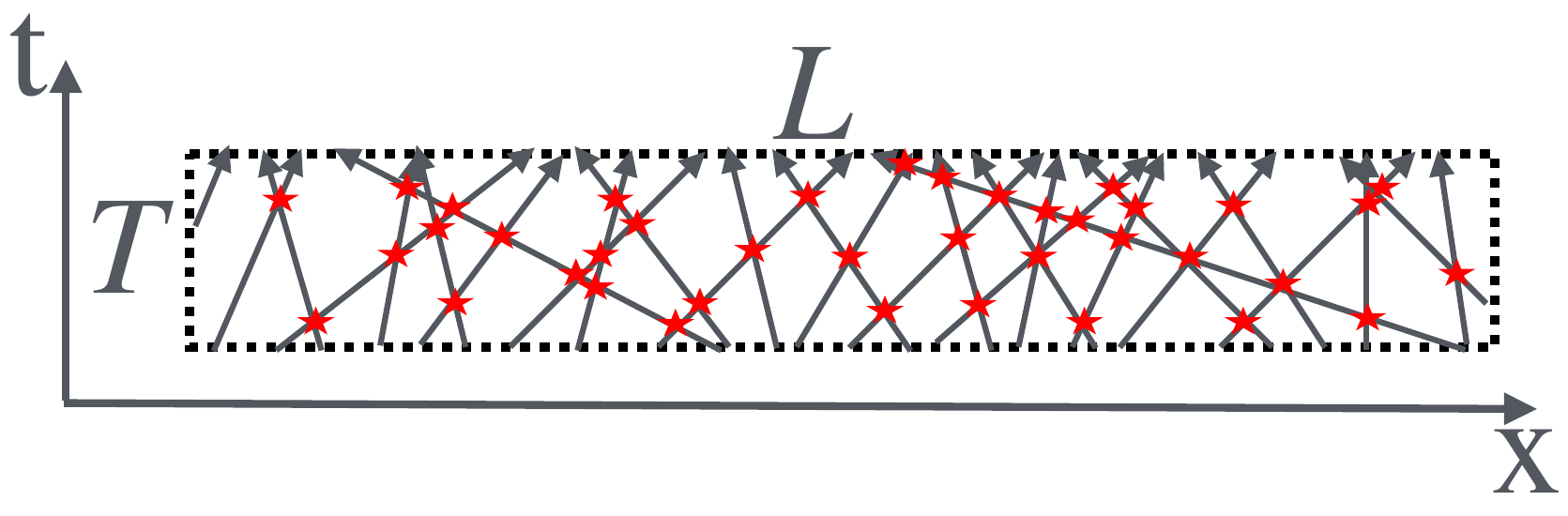}\ec
\caption{A fluid cell with the typical particle trajectories within it over a time $T$. There are $\mathcal O(LT)$ crossings, displayed as red stars, roughly representing independent interactions within this space-time region, while only $\mathcal O(T)$ boundary effects.}
\label{figfluidcell2}
\end{figure}

In fact, by typicality \cite{goldstein2006canonical,lanford2007entropy,allori2020statistical} the average over $T$ can be argued not to be necessary. This is because typicality ensures that an average on space implements an average over time in the past, as randomness in the spatial distribution encodes what happened in the past. That is, the $\mathcal O(L)$ fluctuations of $\overline{o}(x,t)$ over time $T=\mathcal O(1)$ re-organise the system's configuration in $V_L$, under the constraint given by the essentially fixed conserved densities. This re-organisation is enough to implement the small variation of the conserved densities and update $\overline{o}(x,t+T/\ell)$ so that it always take its microcanonical value determined by the $\overline{q_i}(x,t)'s$.

This suggests that
\beq\label{meanproj}
	\overline o(x,t) \stackrel{\ell\gg L\gg\ell_{\rm micro}}\to
	\mathsf o(\overline{\underline q}(x, t))
	= \mathsf o(\overline{q_1}(x,t),\overline{q_2}(x, t),\ldots).
\eeq
This is the hypothesis of {\em local relaxation of fluctuations} \cite{doyon2023emergence,doyon2023ballistic}. It is a nonlinear Boltzmann-Gibbs principle, where fluctuating local observables keep their full nonlinear dependence on the conserved densities.

The above can only be true for $\ell$ large enough. There are two concepts at play. The larger the mesoscopic length $L$ (here we follow the numbering below): (1) the closer is the average itself to the infinite-volume microcanonical average, which equals the macrocanonical average under the right identification of generalised temperatures (Sec.~\ref{secconsqties}); and (3) the more fluctuations are taken into account, hence the less random corrections there are. Yet the cell size cannot be larger than the macroscopic variation scale $\ell$, because the relaxation argument made above requires spatial homogeneity.

Similarly, the larger the time $T$, the more accurate is microcanonical relaxation as well, because more re-organisation has occurred, and also (2) the less there is a time delay between the effective time at which relaxation has occured, and the time at which the values of the conserved densities are taken to determine the state towards which relaxation occurs. But $T$ must not be too large so that conserved densities can be considered fixed and non-fluctuating.

Our conjecture is that the optimal choice is obtained for
\beq
	LT/\ell_{\rm micro} = \mathcal O(\ell).
\eeq
Note that neither $L$ nor $T$ are univeral quantity -- they are artefact of our fluid-cell construction. But $\ell$ is physical -- it is the physical variation length scale. Below we often simply set $\ell_{\rm micro}=1$, which we can do without loss of generality -- this sets the scale of our microscopic coordinates.

For instance, $L = \ell^{1-\ep}$ and $T=\ell^\ep$ for $\ep>0$ small. The time $T$ is not finite but very small on the scale $\ell$, so the asymptotic expansion as $\ell\to\infty$ introduces small mistakes of order $\ep$. The limit $\ep\to0$ is then taken after the asymptotic expansion as $\ell\to\infty$. Another example is $L=\ep\ell$ and $T=1/\ep$. Here it is the fluid cell that is too large, introducing mistakes of order $\ep$, so that, again, the limit $\ep\to0$ is taken after the asymptotic expansion in $\ell$. This is what is meant by the limit $\ell\gg L\gg\ell_{\rm micro}$.

What are the corrections at order $1/\ell$ or $1/L$ beyond the $\ell\to\infty$ limit? The argument above suggests that there are three potential corrections.

\medskip
{\em 1. Corrections to the microcanonical average, $L<\infty$.} Because the cell size is not infinite, the exact average at fixed conserved densities is not the infinite-volume microcanonical average. Hence on the right-hand side of \eqref{meanproj} there is a correction
\beq\label{corr1}
	+\,L^{-1} \delta \mathsf o(\overline{\underline q}(x, t)).
\eeq
We gather this into
\beq\label{regsub}
	\mathsf o^{\rm reg}(\overline{\underline q}(x, t)) := \mathsf o(\overline{\underline q}(x, t)) + L^{-1} \delta \mathsf o(\overline{\underline q}(x, t)).
\eeq
This is the finite-$L$ microcanonical average up to, including, $L^{-1}$ corrections. We will see in Sec.~\ref{ssectpointsplit} that these corrections are indeed of order $L^{-1}$. The result appears to be non-universal; as mentioned, contrary to $\ell$, the fluid cell size $L$ is not physical but simply an artefact of our construction. However, the emergent fluctuation theory $\dbra\cdots\dket_\ell$ implicitly depends on $L$ because it is a theory for fluctuations of fluid-cell means, and we will show in Sec.~\ref{ssectpointsplit} how, in linearly degenerate systems, these $L$-dependences effectively cancel out and the result is, in fact, universal.

\medskip
{\em 2. Time delay, $T>0$: diffusion term.} According to the argument above, there is a delay between the slow variations of the densities and relaxation, so that it is $\overline{o}(x,t+T/\ell)$ that takes the microcanonical value $\mathsf o^{\rm reg}(\overline{\underline q}(x, t))$. The $\ell^{-1}$ expansion gives an additional term. This is expected to be proportional to the spatial variations $\p_{\rx} \overline q_i(x,t) = \ell^{-1}\p_{x} \overline q_i(x,t)$, because $L$ must be taken smaller if the variations are larger, thus $T$ must be taken larger to keep $LT$ the same. Hence on the right-hand side of \eqref{meanproj} there is another correction, for $t>0$:
\beq\label{corr2}
	-\frc1{2}\ell^{-1}\sum_i \h{\mathfrak D}_o^{~i}\big(\overline{\underline q}(x,t)\big) \,\p_x\overline {q_i}(x,t)
\eeq
for some set of diffusion functions $\h{\mathfrak D}_o^{~i}(\overline{\underline q}(x,t))$ associated to the observable $o$, where the factor $-\frc12$ is conventional. For evolution backwards in time, $t<0$, we use PT symmetry, giving the factor $\sgn t$ as both $o$ and $q_i$'s are assumed to be PT invariant. Note how the direction of time is fundamental in this part of the argument: there is a {\em delay} in reaching the microcanonical average.

\medskip

{\em 3. Emergent noise, $LT<\infty$.} Microcanonical relaxation, at finite $L$, receives stochastic corrections according to the central limit theorem. Intuitively, this is the result of partially remembering the microscopic degrees of freedom that have been projected out when projecting onto the conserved densities in the nonlinear Boltzmann-Gibbs principle. Hence on the right-hand side of \eqref{meanproj} there is a noise correction,
\beq\label{corr3}
	\ell^{-1}\xi_o(x,t)
\eeq
with the set of $\xi_o(x,t)$'s, for all local observables $o$'s, being centered Gaussian white noises, delta-correlation in space-time. We need to explain three aspects: (a) why the scaling is $\ell^{-1}$; (b) why the emergent noise is delta-correlated in space-time; and (c) why it is the It\^o convention that must be used.

(a) The scaling in $\ell^{-1}$ is obtained as follows. The time-averaged fluid cell over $L\times T$ has $LT=\mathcal O(\ell)$ independent fluctuations, therefore a variance due to the noise of order $\mathcal O(\ell^{-1}$). Let us write a stochastic contribution $\xi_o^{\rm micro}(\micro x,\micro t)$ in microscopic coordinates, representing the roughness in both space and time. This can be chosen centered, however it is not Gaussian because it encodes small-scale effects, and it is not a white noise because there are correlations on lengths $\mathcal O(\ell_{\rm micro})= \mathcal O(1)$. We now argue that its two-point correlations can be taken as $\dbra \xi_{o}^{\rm micro}(\micro x,\micro t) \xi_{o'}^{\rm micro}(\micro x',\micro t')\dket_\ell
	= \hat{\mathfrak L}_{o,o'}\delta_{\ell_{\rm micro}}(\micro x-\micro x',\micro t-\micro t')$ for some $\hat{\mathfrak L}_{o,o'}$, with mollified space-time delta-function on scale the linear $\ell_{\rm micro}$. Indeed, with this, the variance of its mean over a fluid cell in space-time is then $\mathcal O(\ell^{-1})$, as it should:
\beq
	{\rm var}\Big(\frc1{LT} \int_{-L/2}^{L/2} \dd \rx \int_0^T \,\dd \micro t\, \xi_o^{\rm micro}(\micro x,\micro t)\Big) \sim
	\frc1{LT} \hat{\mathfrak L}_{o,o'} = 
	\mathcal O(\ell^{-1})\qquad (\ell\to\infty).
\eeq
A similar calculation holds for covariances. Transforming to macroscopic coordinates $\delta_{\ell_{\rm micro}}(\ell x-\ell x',\ell t-\ell t') \to \ell^{-2} \delta(x-x')\delta(t-t')$, we write $\xi_{o}^{\rm micro}(\ell x,\ell t) = \ell^{-1} \xi_{o}( x,t)$ and we obtain \eqref{corr3} with \eqref{noisecorr}, and on that scale the noise becomes Gaussian as higher-point correlations are expected to vanish faster with $\ell^{-1}$.

(b) The delta-correlation in space-time is argued for as follows. It is known that correlations of physical observables in space (out of equilibrium) and in time (in and out of equilibrium) can be strong, and that this is due to hydrodynamic mode propagation. Since hydrodynamic modes are waves of conserved densities, such strong correlations are destroyed when fixing the space-time configuration of conserved densities on every fluid cell. There remains only short-range correlations on scale $\ell_{\rm micro}$, thus giving $\dbra \xi_{o}^{\rm micro}(\micro x,\micro t) \xi_{o'}^{\rm micro}(\micro x',\micro t')\dket_\ell
	= \hat{\mathfrak L}_{o,o'}\delta_{\ell_{\rm micro}}(\micro x-\micro x',\micro t-\micro t')$.

(c) Finally, the It\^o convention is again an effect of time delay. Fluctuations occur on a time following that at which the essentially fixed fluid-cell means of conserved densities are taken. Thus \eqref{itow} holds, where Gaussian forward increments have a variance that is determined by the conserved densities at the initial time of the increment. Likewise \eqref{noisecorr} holds thanks to this.

\begin{rema}
In stochastic systems, the emergent noise can often be obtained in a mathematically accurate fashion. But such an understanding is still lacking in general in Hamiltonian systems.
\end{rema}

\section{Constitutive relations}\label{secconst}

We now evaluate the beyond-Euler components of the nonlinear fluctuating Boltzmann-Gibbs principle \eqref{NLFBG}. For this purpose, we evaluate various equilibrium correlation functions. It turns out that in order to evaluate correlation functions directly using \eqref{NLFBG}, it is sufficient to know the asymptotic form \eqref{ldcorr}, and to use the cumulant expansion principles from Malyshev's formula (see  \cite{doyon2025nonlinear}). In particular, the Euler-scale projection formulas from \cite{doyon2023ballistic,doyon2025nonlinear} for 2- and 3-point functions are sufficient for quantities in \eqref{NLFBG} with a factor of $\ell^{-1}$.

The results below hold in $d=1$ ``no-shock'' systems which are strictly hyperbolic. Strict hyperbolicity means that all hydrodynamic velocities are distinct. Further, a no-shock system is a many-body system whose Euler-scale equation does not develop shock in non-equilibrium situations. In such systems the solutions to Euler equations are unique at all times, and nonlinear Euler-scale projections are valid everywhere in space-time \cite{doyon2025nonlinear}. The most important examples (perhaps the only ones!) of such systems are {\em linearly degenerate systems} \cite{lax2005hyperbolic,ferapontov1991integration,el2011kinetic,pavlov2012generalized,bressan2013hyperbolic}, as indeed these do not develop shocks \cite{rozlinedeg,liu1979development,bressan2013hyperbolic}. See Appendix \ref{applindeg} for the definition of such systems. It is also known that no shocks are produced in integrable systems \cite{hubner2024new,hubner2024existence}, which are, in a formal sense, also linearly degenerate \cite{doyon2020lecture}.

We show in Appendix \ref{applindeg} that, in no-shock systems, the following results hold:

1. Correlation functions of conserved densities keep their local-equilibrium form at later time even in long-wavelength non-equilibrium states such as those emanating from \eqref{lw}, at the Euler scale:
\beq\label{equilcorr}
	\dbra q_{i_1}(x_1,t), q_{i_2}(x_2,t)\dket_\ell^{\rm c}
	= \ell^{-1}\Big( \mathsf C_{i_1,i_2}(x_1,t) \delta(x_1-x_2) + \mbox{regular}\Big) + \mathcal O(\ell^{-2}).
\eeq
Here the equilibrium covariance matrix is given in \eqref{Cmatrix}, and  it is to be evaluated at the state characterised by $\underline{\mathsf q}(x_1,t) = \dbra \underline q(x_1,t)\dket_\ell$. This is consistent with the initial condition \eqref{initcondq}. A partial proof for Eq.~\eqref{equilcorr} was given using the BMFT in \cite{doyon2023ballistic}, and a related calculation in \cite[Suppl Mat]{PhysRevLett.134.187101}. We provide a complete proof in Appendix \ref{applindeg}, which requires strict hyperbolicity.

2. For every local observable, the three-point couplings, Eq.~\eqref{3ptdef}, in space-time stationary, maximal entropy states vanish for normal modes of equal velocities:
\beq\label{3point}
	\bra Q_{I},Q_{I'},o_1^-\ket^{\rm c}_{\underline\beta} = 0\quad \mbox{if}\ \mathsf v_I=\mathsf v_{I'}.
\eeq
Thus, as we assume strict hyperbolicity, vanishing occurs whenever $I\neq I'$ (but the more general result shown here holds also without strict hyperbolicity). This implies that there is no superdiffusion \cite{spohn2014nonlinear,doyon2022diffusion}, and was shown from linear degeneracy in \cite{PhysRevLett.134.187101}, Supplementary Material Sec.~``Degenerate three-point coupling from linear degeneracy''. For readability, we reproduce this proof in Appendix \ref{applindeg}.

\medskip

Using these properties of no-shock systems, we will show:
\bi
\item[I.] {\bf Point-splitting regularisation.} The microcanonical regularisation $\mathsf o^{\rm reg}(\underline q(x,t))$ is given by the {\em point-splitting prescription} introduced in \cite{PhysRevLett.134.187101},
\beq\label{pointsplitting}
	\Big[q_{i_1}(x,t)\cdots
	q_{i_k}(x,t)\Big]^{\rm reg}
	=
	\frc1{k!}\sum_{\sigma\in S(k)}
	q_{i_1}(x+\sigma_i0^+,t)\cdots q_{i_k}(x+\sigma_k 0^+,t)
\eeq
where $S(k)$ is the set of permutations of $(1,2,\ldots,k)$, and $0^+$ is a fixed, positive infinitesimal number. Note how this is a sum over various orderings in space, with infinitesimal displacements.
\item[II.] {\bf Projected Kubo formula.} The noise covariance coefficient is given by the {\em projected Kubo formula}: with $\bra\cdots\ket = \bra\cdots\ket_{\underline{\mathsf q}}$ the space-time stationary maximal-entropy state with averages $\bra q_i\ket = \mathsf q_i$,
\beq\label{Lprojected}
	\h{\mathfrak L}_{o_1,o_2}(\underline{\mathsf q}) =
	\mathfrak L_{o_1,o_2}(\underline{\mathsf q})
	- \frc12 \sum_{I,I':\mathsf v_I\neq \mathsf v_I'} 
	\frc{\bra o_1^-,Q_{I},Q_{I'}\ket^{\rm c}
	\bra o_2^-,Q_{I},Q_{I'}\ket^{\rm c}}{|\mathsf v_I-\mathsf v_{I'}|}.
\eeq
Here $Q_I$ are the total conserved quantities in the normal mode basis (see \eqref{normalmodes}), $\mathsf v_I$ are the associated hydrodynamic velocities, and $\mathfrak L_{o_1,o_2}$ is given by the conventional Green-Kubo formula
\beq\label{Onsager}
	\mathfrak L_{o_1,o_2}(\underline{\mathsf q}) := \int_{-\infty}^\infty \dd \micro t\,\Big(
	\int \dd \rx\,\bra o_1(\rx,\micro t),o_2(0,0)\ket^{\rm c} - \mathsf D_{o_1,o_2}\Big)
\eeq
with ``Drude weight''
\beq\label{drudedef}
	\mathsf D_{o_1,o_2} := \lim_{\micro t\to\pm\infty} \int \dd \rx\,\bra o_1(\rx,\micro t),o_2(0,0)\ket^{\rm c}
	= \sum_{i_1,i_2} \frc{\p \mathsf o_1}{\p \mathsf q_{i_1}} \frc{\p \mathsf o_2}{\p \mathsf q_{i_2}}
	\mathsf C_{i_1,i_2}
\eeq
where the last equality follows from a general theorem \cite{doyon2022hydrodynamic}. Note that the quantity $\mathfrak L_{ik} := \mathfrak L_{j_i,j_k}$ is the usual Onsager matrix.
\item[III.] {\bf (Extended) Einstein relation.} The diffusion coefficients $\h{\mathfrak D}_{o}^{~k}(\underline{\mathsf q})$ are related to the noise covariance by an {\em Einstein relation}, here extended to diffusion for arbitrary observables $o$ instead of only currents,
\beq\label{eins}
	\hat{\mathfrak L}_{o,j_i} (\underline{\mathsf q}) = 
	\sum_k\h{\mathfrak D}_{o}^{~k}(\underline{\mathsf q})\mathsf C_{ki},\quad
	\h{\mathfrak D}_o^{~k} (\underline{\mathsf q}) = \sum_i\h{\mathfrak L}_{o,j_i} (\underline{\mathsf q})\mathsf C^{ik}.
\eeq
\ei
It is also instructive to evaluate the Drude weight \eqref{drudedef} within our formalism, in order to confirm that it is consistent. This is done in Appendix \ref{appdrude}.

Our derivation of Point I below gives a full justification of the point-splitting regularisation that was proposed in \cite{PhysRevLett.134.187101}.

Our main result is Point II, the projected Kubo formula. Our proof uses a calculation that follows somewhat that done in \cite{medenjak2020diffusion}: the ``diffusion from convection'' term is subtracted. Crucially, the subtraction corresponds to projecting out the quadratic charges forming the ``scattering subspace'' of the ``diffusive space'' constructed in \cite{doyon2022diffusion}, $\mathcal H_{\rm scat}\subset \mathcal H_{\rm dif}$. The diffusive space is a Hilbert space constructed with respect to the sesquilinear form given by the Green-Kubo formula
\beq\label{inner}
	(o_1,o_2)_{\rm dif} := \mathfrak L_{o_1,o_2}(\underline{\mathsf q}) = \int_{-\infty}^\infty \dd \micro t\,\Big(
	\int \dd \rx\,\bra o_1(\rx,\micro t),o_2(0,0)\ket^{\rm c} - \mathsf D_{o_1,o_2}\Big)
\eeq
(recall that our observables $o_1,o_2$ are assumed to be real or Hermitian, however this is easily extended by sesquilinearity). That is, the space $\mathcal H_{\rm dif}$ is obtained by a Gelfand-Naimark-Segal construction: it is easy to show that $(o_1,o_2)_{\rm dif}$ is positive semi-definite, so its null space is moded out and the resulting space of equivalence classes is completed with respect to the norm $||\cdot||_{\rm dif} = \sqrt{(\cdot,\cdot)_{\rm dif}}$. The scattering subspace is that spanned by particular quadratic elements parametrised by two normal mode indices $I,J$:
\beq
	\mathcal H_{\rm scat} := {\rm span}\Big\{w_{IJ} = \lim_{X\to\infty} \frc{|\mathsf v_I-\mathsf v_J|}{2X}
	\int_{-X}^X\dd \rx\,q_I(\rx)q_J\Big\}\subset\mathcal H_{\rm dif}.
\eeq
The fact that these lie within $\mathcal H_{\rm dif}$ is nontrivial and argued for in \cite{doyon2022diffusion}. With the projection $\mathbb P_{\rm scat}:\mathcal H_{\rm scat}\to \mathcal H_{\rm dif}$, it turns out that the second term in \eqref{Lprojected} is exactly \cite{doyon2022diffusion}
\beq
	(\mathbb P_{\rm scat} o_1,\mathbb P_{\rm scat}o_2)
	=
	 \frc12 \sum_{I,I':\mathsf v_I\neq \mathsf v_I'} 
	\frc{\bra o_1^-,Q_{I},Q_{I'}\ket^{\rm c}
	\bra o_2^-,Q_{I},Q_{I'}\ket^{\rm c}}{|\mathsf v_I-\mathsf v_{I'}|}.
\eeq
That is,
\beq
	\h{\mathfrak L}_{o_1,o_2}
	=
	\big((1-\mathbb P_{\rm scat})o_1,(1-\mathbb P_{\rm scat})o_2\big)_{\rm dif}.
\eeq
In \cite{doyon2022diffusion}, the subspace $\mathcal H_{\rm scat}$ is physically interpreted as describing the aggregated effect of nonlinear two-wave scattering events of hydrodynamic modes emanating from initial fluctuations and transported by the Euler equation. Nonlinear scattering of deterministically transported fluctuations is the source of long-range correlations, as explained in \cite{doyon2023emergence}. As these have $\mathcal O(\ell^{-1})$ strength (see \eqref{ldcorr}), they contribute to $\mathfrak L_{o_1,o_2}$ at the same order as the correlation between hydrodynamic noise $\xi_{o_1}$ and $\xi_{o_2}$. This is why this subspace must be subtracted to obtain the noise covariance. Thus, the present result finally puts together these various concepts in a coherent theory.

Point III is expected -- for current observables $o=j_k$, the diffusion functions $\h{\mathfrak D}_k^{~i}(\underline q(x,t))$ (using the notation \eqref{notationD}) are usually assumed to be related to the noise covariance by the Einstein relation. However, as far as we are aware, this has not been derived before in the context of \eqref{NLFBG}.

We believe the results of Points II and III still hold beyond linearly degenerate systems, with an appropriate projection including certain unbounded elements of the scattering Hilbert space that makes the resulting noise covariance finite. However we do not know of a derivation yet.

We note that in \cite{hydrodynamic2025Yoshimura}, the same projected Kubo formula \eqref{Lprojected} is obtained for linearly degenerate systems by using a fluctuation principle of the type \eqref{NLFBG} but under the diffusive scaling, instead of the large-deviation scaling. Importantly, the result is verified by explicit calculations in a stochastic model, thus confirming the projected Kubo formula. The calculation methods are quite different, because of the different scaling.

\subsection{Correction to the microcanonical average and point splitting}\label{ssectpointsplit}

Let us consider the microcanonical correction $\delta\mathsf o({\underline q}(x,t))$, Eq.~\eqref{corr1}, for one-point averages generically in non-equilibrium states. It can be evaluated by studying the average of $o = o(0,0)$ in a stationary, maximal entropy state $\bra\cdots\ket_{\underline{\beta}}$, and taking into account that it has short-range correlations. Recall that by the equivalence of ensembles,
\beq\label{equieval}
	\mathsf o(\underline{\mathsf q}) = \bra o\ket_{\underline \beta(\underline{\mathsf q})}
\eeq
where the left- (right-)hand side is the microcanonical (macrocanonical) ensemble. Here the Lagrange parameters $\beta^i = \beta^i(\underline{\mathsf q})$ are functions of the average conserved densities defined by $\mathsf q_i = \bra q_i\ket_{\underline\beta(\underline{\mathsf q})} = \dbra q_i\dket_{\underline{\beta}(\underline{\mathsf q})}$. Let us denote for lightness of notation $\bra\cdots\ket = \bra\cdots\ket_{\underline{\beta}(\underline{\mathsf q})}$ and likewise $\dbra\cdots\dket = \dbra\cdots\dket_{\underline{\beta}(\underline{\mathsf q})}$. Then
\beq
	\mathsf o(\dbra \underline{q}\dket) = \mathsf o( \underline{\mathsf q})  =\bra o\ket =
	\bra \overline o\ket
	= \dbra \mathsf o\big(\underline q\big)\dket
	+
	\frc1{L} \dbra\delta\mathsf o\big({\underline q}\big)
	\dket
\eeq
where we used homogeneity to write the average as an average of a fluid-cell mean, and then used \eqref{NLFBG} and the vanishing of noise and diffusive parts in a maximal entropy state. Recall that $\underline q = (q_i)_i$ and $q_i(x,t)$ are the fluctuating conserved densities in the effective emergent theory, whose averages are represented by $\dbra\cdots\dket$ (Eqs.~\eqref{corrfctexp}, \eqref{NLFBG}, here without the $\ell$ index because the state is space-time stationary); that $\mathsf q_i, \mathsf o$ are the averages of the microscopic observables $q_i(\micro x,\micro t)$ and $o(\micro x,\micro t)$ in a space-time stationary, maximal entropy state characterised by $\beta^i$'s, or seen as functions of $\underline{\mathsf q}$ (Eq.~\eqref{defo}), and that $\overline{o}(x,t)$ is the fluid-cell mean of the microscopic observable $o(\micro x,\micro t)$ (Eq.~\eqref{cellmean}).

Thus we find that the correction term should simply ``allow'' for the averaging to pass inside the function $\mathsf o(\underline{q})$. Recall that $\dbra\cdots\dket$ implicitly depends on $L$ because the fluctuating fields $q_i(x,t)$ represents fluctuating fluid-cell means. Let us evaluate equal-point connected correlation functions of these fluctuating fields using their fluid-cell mean definition. Assuming short-range correlations,
\beqa
	\dbra q_{i_1}(0,0),\cdots ,q_{i_k}(0,0)\dket^{\rm c} &=& \frc1{L^{k }}\int_{(V_L)^{\times k}} \prod_{i=1}^k\dd \rx_i\,\bra q_{i_1}(\rx_1,0)\cdots q_{i_k}( \rx_k,0)\ket^{\rm c} \n
	&=& 
	\frc{1}{L^{k-1}}\bra Q_{i_1},\ldots,Q_{i_{k-1}},q_{i_k}\ket^{\rm c}
	\label{tyu}
\eeqa
where corrections (not written) are exponentially small as $L\to\infty$.
Therefore we may use a cumulant expansion to get
\beqa
	\dbra \mathsf o\big( \underline{q}\big)\dket
	&=&
	\mathsf o\big(\underline {\mathsf q}\big)
	+ \frc12\sum_{ij}\frc{\p^2\mathsf o(\underline{\mathsf q})}{\p \mathsf q_i\p\mathsf q_j} \dbra q_i( 0,0)q_j(0,0)\dket_{\underline\beta}^{\rm c} + \mathcal O(L^{-2})\n
	&=&
	\mathsf o\big(\underline {\mathsf q}\big)+
	\frc1{2L}
	\sum_{ij}\frc{\p^2\mathsf o(\underline{\mathsf q})}{\p \mathsf q_i\p\mathsf q_j}\mathsf C_{ij}
	+ \mathcal O(L^{-2})\label{fhdh}
\eeqa
where we used \eqref{Cmatrix}. Again by the cumulant expansion we have
\beq
	\frc1{L} \dbra\delta\mathsf o\big({\underline q}\big)
	\dket
	=
	\frc1{L}  \delta\mathsf o\big({\underline {\mathsf q}}\big)
	+ \mathcal O(L^{-2})
\eeq
and hence from \eqref{equieval} we identify
\beq\label{resdeltao}
	\delta\mathsf o\big({\underline {\mathsf q}}\big)
	=
	-\frc1{2}
	\sum_{ij}\frc{\p^2\mathsf o(\underline{\mathsf q})}{\p \mathsf q_i\p\mathsf q_j}\mathsf C_{ij}.
\eeq
That is, the correction term simply gets rids of the effects of short-range correlations, the singularities arising at equal positions.

In general evaluating correlation functions using \eqref{regsub} with \eqref{resdeltao} is rather complicated: we need to keep track of all $1/L$ singularities of $\mathsf o(\underline{q}(x,t))$, and all corrections $L^{-1}\delta\mathsf o\big({\underline { q}}\big)$. These should cancel each other to give something universal (independent of $L$). However, there is a trick that simplifies this procedure and shows universality.

Consider a more general non-equilibrium state $\bra\cdots\ket_\ell$ such as \eqref{lw}, and recall the property \eqref{equilcorr}. Then a similar calculation as \eqref{tyu} and \eqref{fhdh} shows that, in every such state, $L^{-1}\delta\mathsf o\big({\underline { q}}\big)$ cancels the $1/L$ singularities of $\mathsf o(\underline{q}(x,t))$. In this calculation, Eq.~\eqref{equilcorr} is used only to evaluate the leading $1/L$ correction, hence we can neglect the $1/\ell$ correction in the delta-function coefficient in \eqref{equilcorr}.  But we can implement this cancellation simply as
\beq
	\dbra \mathsf o(\underline{q})\dket_\ell +
	\frc1L \dbra \delta \mathsf o(\underline q)\dket_\ell
	=
	\dbra \mathsf o^{\rm reg}(\underline{q})\dket_\ell
\eeq
by defining the regularisation as acting, on monomial of conserved densities at equal space-time points, by avoiding the ``fat diagonal'' region where coordinates are at microscopic distances from each other:
\beq\label{regfcm}
	\Big[\overline{ q}_{i_1}(x,t)\cdots
	\overline{ q}_{i_k}(x,t)\Big]^{\rm reg}
	=
	\frc1{L^{k}}\int_{(V_L)_{\neq}^{\times k}} \prod_{j=1}^k\Big(\dd \rx_j\,
	q_{i_j}(\ell x+\rx_i,\ell t)\Big)
\eeq
with
\beq
	(V_L)_{\neq}^{\times k}
	= \{( x_1,\ldots,x_k)\in (V_L)^{\times k}:
	|x_a- x_b|\gg \ell_{\rm micro}\, \forall a\neq b\}.
\eeq
Indeed, it is the integration on this fat diagonal that gives the $1/L$ singularity of \eqref{tyu}. But, out of equilibrium, inside any sector of a given ordering of spatial coordinates, $x_{\sigma_1}>\cdots>x_{\sigma_k}$ for some $\sigma\in P(k)$, correlations are continuous at the Euler scale, according to BMFT (see e.g.~\cite{hubner2024new,PhysRevLett.134.187101}). As these correlations give $\ell^{-1}$ subleading corrections to the representation of the observable $o$ within the fluctuating theory, they are sufficient to understand the diffusive scale. Therefore, we may replace the regularised fluid-cell mean \eqref{regfcm}, where the fat diagonal is omitted, simply by the point-splitting regularisation \eqref{pointsplitting}.

Finally, as we can reproduce correlation functions by functional differentiations with respect to external fields in such states (see \cite{doyon2018exact,doyon2025nonlinear}), this regularisation works for correlation functions as well. This is a universal procedure: the explicit $L$ dependence has disappeared, and all equal-point singularities are avoided.

Therefore, we have derived the point-splitting regularisation \eqref{pointsplitting} conjectured in \cite{PhysRevLett.134.187101}  from the finite-size correction to the microcanonical averaging in fluid cells.

\subsection{Noise correlations via projected Kubo formula}\label{ssectnoise}

We now evaluate the correlation matrix $\hat{\mathfrak L}_{o,o'}$ in \eqref{noisecorr}. We will show that it is not the usual Kubo formula, but instead a {\rm projected Kubo formula}, where the quadratic charge subspace \cite{doyon2022diffusion} has been projected out.

For this purpose, let us consider again a stationary, maximal entropy state $\bra\cdots\ket$, and evaluate, using \eqref{NLFBG}, the Kubo formula
\beqa\label{kuboint}
	\mathfrak L_{o_1,o_2} &=& \int_{-\infty}^\infty \dd \micro t\,\Big(
	\int \dd \rx\,\bra o_1(\rx,\micro t),o_2(0,0)\ket^{\rm c} - \mathsf D_{o_1,o_2}\Big)\n
	&=&
	\ell
	\int_{-\infty}^\infty \dd  t\,\Big(
	\ell \int \dd x\,\dbra o_1(x,t),o_2(0,0)\dket^{\rm c} - \mathsf D_{o_1,o_2}\Big).
\eeqa
In the second line we have written it in terms of the emergent fluctuating theory, with scaled space-time coordinates, where we will use \eqref{NLFBG}. We note that the Drude weight \eqref{drudedef} can be written as
\beq
	\mathsf D_{o_1,o_2} = \ell \sum_{i_1,i_2}\frc{\p \mathsf o_1}{\p \mathsf q_{i_1}} \frc{\p \mathsf o_2}{\p \mathsf q_{i_2}}\int \dd x\,
	\dbra q_{i_1}(x,t),q_{i_2}(0,0)\dket^{\rm c}
\eeq
for every $t\in\R$, therefore we may write
\beq\label{Lproj}
	\mathfrak L_{o_1,o_2}
	=
	\ell^2
	\int_{-\infty}^\infty \dd  t
	\int \dd x\,\dbra\, o_1^-( x, t),o_2^-(0,0)\,\dket^{\rm c}
\eeq
in terms the observable $o$ from which we project out its overlap with the conserved densities:
\beq\label{proj}
	o^- = o - \sum_{i,j}\bra o,Q_i\ket^{\rm c}\mathsf C^{ij}q_j
	= o - \sum_{i}\frc{\p{\mathsf o}}{\p \mathsf q_i}q_i.
\eeq

Using \eqref{NLFBG}, we analyse the terms that contribute to $\dbra o_1(x_1, t_1),o_2(x_2, t_2)\dket^{\rm c} $ at orders $\ell^{-1}$ and $\ell^{-2}$; and in particular, those that contribute to \eqref{kuboint} (equivalently \eqref{Lproj}), hence that are nonzero under space-time integration. For each of the observables $o_1$ and $o_2$, we consider the microcanonical, diffusive and noise terms, and evaluate their correlations.

First, the microcanonical-microcanonical correlation must be evaluated to leading and first subleading order in $\ell^{-1}$, by the cumulant expansion (or Malyshev's formula), see the principles explained in \cite{doyon2025nonlinear}. The result is
\beqa\label{o1o2}
	\lefteqn{\dbra \mathsf o_1^{\rm reg}(\underline q(x_1,t_1)) ,\mathsf o_2^{\rm reg}(\underline q(x_2,t_2))\dket^{\rm c}} && \\ &\sim&
	\sum_{i_1,i_2} \frc{\p \mathsf o_1}{\p \mathsf q_{i_1}} \frc{\p \mathsf o_2}{\p \mathsf q_{i_2}}
	\dbra q_{i_1}(x_1,t_1) ,q_{i_2}(x_2,t_2)\dket^{\rm c}
	\n &&
	+\,
	\frc12 
	\sum_{i_1,i_2,i_1',i_2'} \frc{\p^2 \mathsf o_1}{\p \mathsf q_{i_1}\p \mathsf q_{i_1'}} \frc{\p^2 \mathsf o_2}{\p \mathsf q_{i_2}\p \mathsf q_{i_2'}}
	\dbra q_{i_1}(x_1,t_1) ,q_{i_2}(x_2,t_2)\dket^{\rm c}
	\dbra q_{i_1'}(x_1,t_1), q_{i_2'}(x_2,t_2)\dket^{\rm c}
	\n &&
	+\,
	\frc12 
	\sum_{i_1,i_2,i_1'}\frc{\p^2 \mathsf o_1}{\p \mathsf q_{i_1}\p \mathsf q_{i_1'}} \frc{\p \mathsf o_2}{\p \mathsf q_{i_2}}
	\frc12\sum_{\pm} \dbra q_{i_1}(x_1\pm 0^+,t_1) ,q_{i_1'}(x_1\mp 0^+,t_1) , q_{i_2}(x_2,t_2)\dket^{\rm c}
	\n &&
	+\,
	\frc12 
	\sum_{i_1,i_2,i_2'}\frc{\p \mathsf o_1}{\p \mathsf q_{i_1}} \frc{\p^2 \mathsf o_2}{\p \mathsf q_{i_2}\p \mathsf q_{i_2'}}
	\frc12\sum_{\pm} \dbra q_{i_1}(x_1,t_1),q_{i_2}(x_2\pm 0^+,t_2),q_{i_2'}(x_2\mp 0^+,t_2)\dket^{\rm c}
	+ \mathcal O(\ell^{-3})\no
\eeqa
where the first line on the right-hand side is $\mathcal O(\ell^{-1})$ and the last three lines are $\mathcal O(\ell^{-2}$). In the second line, the point-splitting prescription has been dropped, as it will become clear now that it does not influence the result. Expression \eqref{o1o2} simplifies drastically when written for the projected-out observables \eqref{Lproj}, and this fully accounts for the Drude weight subtraction:
\beqa\label{o1o2proj}
	\lefteqn{\dbra (\mathsf o_1^-)^{\rm reg}(\underline q(x_1,t_1)) ,(\mathsf o_2^-)^{\rm reg}(\underline q(x_2,t_2))\dket^{\rm c}}
	&&\\
	&=&
	\frc12 
	\sum_{i_1,i_2,i_1',i_2'}\frc{\p^2 \mathsf o_1}{\p \mathsf q_{i_1}\p \mathsf q_{i_1'}} \frc{\p^2 \mathsf o_2}{\p \mathsf q_{i_2}\p \mathsf q_{i_2'}}
	\dbra q_{i_1}(x_1,t_1) ,q_{i_2}(x_2,t_2)\dket^{\rm c}
	\dbra q_{i_1'}(x_1,t_1), q_{i_2'}(x_2,t_2)\dket^{\rm c}
	+\mathcal O(\ell^{-3}).
	\no
\eeqa
On the right-hand side, we set $x_1=x,t_1=t$ and $x_2=t_2=0$, we evaluate
\beq
	\frc{\p^2 \mathsf o_1}{\p \mathsf q_{i_1}\p \mathsf q_{i_1'}}
	=
	\sum_{j_1,j_1'}\bra o_1^-,Q_{j_1},Q_{j_1'}\ket^{\rm c}
	\mathsf C^{j_1,i_1}\mathsf C^{j_1',i_1'}
\eeq
and pass to normal modes using \eqref{normalmodes}, to obtain
\beq
	\sum_{I_1,I_1',I_2,I_2'}\bra o_1^-,Q_{I_1},Q_{I_1'}\ket^{\rm c}
	\bra o_2^-,Q_{I_2},Q_{I_2'}\ket^{\rm c}
	\dbra q_{I_1}(x,t),q_{I_2}(0,0)\dket^{\rm c}
	\dbra q_{I_1'}(x,t),q_{I_2'}(0,0)\dket^{\rm c}.
\eeq
This can be evaluated exactly to its leading $\mathcal O(\ell^{-2})$ order by using the Euler-scale formula for the two-point functions \eqref{2pointeq},
\beq\label{blabla}
	\sum_{I,I'}\ell^{-2}\frc12\bra o_1^-,Q_{I},Q_{I'}\ket^{\rm c}
	\bra o_2^-,Q_{I},Q_{I'}\ket^{\rm c}
	\delta(x-\mathsf v_{I}t)
	\delta(x-\mathsf v_{I'}t).
\eeq
Therefore, taking into account \eqref{3point} from linear degeneracy, we obtain, in \eqref{Lproj},
\beqa
	\mathfrak L_{o_1,o_2} &\stackrel{\text{micro-micro}}=&
	\frc12\sum_{I,I'}\int \dd t\int \dd x\,
	\bra o_1^-,Q_{I},Q_{I'}\ket^{\rm c}
	\bra o_2^-,Q_{I},Q_{I'}\ket^{\rm c}
	\delta(x-\mathsf v_{I}t)
	\delta(x-\mathsf v_{I'}t)\n
	&=& 
	\frc12 \sum_{I, I':\mathsf v_I\neq \mathsf v_{I'}} \int \dd t\,
	\bra o_1^-,Q_{I},Q_{I'}\ket^{\rm c}
	\bra o_2^-,Q_{I},Q_{I'}\ket^{\rm c}
	\delta((\mathsf v_I-\mathsf v_{I'})t)
	\n &=&
	\frc12 \sum_{I, I':\mathsf v_I\neq \mathsf v_{I'}} 
	\frc{\bra o_1^-,Q_{I},Q_{I'}\ket^{\rm c}
	\bra o_2^-,Q_{I},Q_{I'}\ket^{\rm c}}{|\mathsf v_I-\mathsf v_{I'}|}.
	\label{quadraticterm}
\eeqa

Second, the noise-noise correlation is simply
\beq\label{noisenoise}
	\dbra \xi_{o_1}(x_1,t_1)\xi_{o_2}(x_2,t_2)\dket^{\rm c}
	 =\ \ell^{-2}\hat{\mathfrak L}_{o_1,o_2}\delta(x_1-x_2)\delta(t_1-t_2)
\eeq
so that
\beq\label{noiseterm}
	\mathfrak L_{o_1,o_2} \stackrel{\text{noise-noise}}=
	\hat{\mathfrak L}_{o_1,o_2}.
\eeq

Finally, we show that all remaining terms are of order $\mathcal O(\ell^{-3})$ or higher orders in $\ell^{-1}$. Let us use for lightness of notation
\beq
	o^{\rm d}(x,t) = -\frc{\sgn t}2\sum_i\mathfrak D_o^{~i}(\underline q(x,t))
	\p_x q_i(x,t).
\eeq
The diffusive-diffusive correlation is
\beq
	\dbra \ell^{-1}o_1^{\rm d}(x_1,t_1),\ell^{-1}o_2^{\rm d}(x_2,t_2)\dket^{\rm c} = \mathcal O(\ell^{-3})
\eeq
because the two-point function has leading behaviour $\mathcal O(\ell^{-1})$, Eq.~\eqref{ldcorr}. Similarly, the diffusive-noise correlation is $\dbra \ell^{-1} o_1^{\rm d}(x_1,t_1),\ell^{-1}\xi_{o_2}(x_2,t_2)\dket^{\rm c} = \mathcal O(\ell^{-3})$ because any function of conserved densities evolved in time has a noisy component, coming from integrating the current noise term in the conservation law \eqref{hydroeq}, that is of order $\ell^{-1}$, thus giving the additional factor $\mathcal O(\ell^{-1})$.

The diffusive-microcanonical correlation is more subtle. By the cumulant expansion, it is
\beq\label{regdif}
	\dbra o_1^{\rm reg}(x_1,t_1),o_2^{\rm d}(x_2,t_2)\dket^{\rm c}
	=
	-\frc{\sgn t_2}{2\ell}
	\sum_{i_1,i_2}
	\frc{\p \mathsf o_1}{\p \mathsf q_{i_1}}
	\mathfrak D_{o_2}^{~i_2}(\mathsf q)
	\p_{x_2} \dbra q_{i_1}(x_1,t_1),q_{i_2}(x_2,t_2)\dket^{\rm c}
	+ \mathcal O(\ell^{-3})
\eeq
where we have used the fact that $\dbra \p_{x_2} q_{i_2}(x_1,t_1)\dket=0$ by homogeneity of the state, so that other terms in the cumulant expansion vanish. This is in general of order $\ell^{-2}$. However, it is a total derivative. Because in \eqref{kuboint} we integrate over the spatial coordinate, this contribution vanishes in \eqref{kuboint}. Finally, the microcanonical-noise correlation is, by the cumulant expansion,
\beq
	\dbra o_1^{\rm reg}(x_1,t_1),\ell^{-1}\xi_{o_2}(x_2,t_2)\dket^{\rm c}
	= \ell^{-1}\sum_i \frc{\p \mathsf o_1}{\p \mathsf q_{i}}
	\dbra q_{i}(x_1,t_1),\xi_{o_2}(x_2,t_2)\dket^{\rm c}
	+ \mathcal O(\ell^{-3})
\eeq
which again is nonzero at order $\ell^{-2}$, because of the $\mathcal O(\ell^{-1})$ noisy component to $q_i(x_1,t_1)$ coming from integrating the hydrodynamic equation. However, under spatial integration the result vanishes:
\beq
	\dbra Q_i,\xi_{o_2}(x_2,t_2)\dket^{\rm c}=0
\eeq
because the total charge $Q_i = \int\dd x\,q_i(x)$ does not have noisy contribution, as the noise in the hydrodynamic equation appears inside a spatial derivative.

Hence, combining \eqref{noiseterm} with \eqref{quadraticterm}, we obtain
\beq
	\mathfrak L_{o_1,o_2}
	= \frc12 \sum_{I, I':\mathsf v_I\neq \mathsf v_{I'}} 
	\frc{\bra o_1^-,Q_{I},Q_{I'}\ket^{\rm c}
	\bra o_2^-,Q_{I},Q_{I'}\ket^{\rm c}}{|\mathsf v_I-\mathsf v_{I'}|}
	+ \h{\mathfrak L}_{o_1,o_2}
\eeq
from which we deduce the noise strength
\beq
	\h{\mathfrak L}_{o_1,o_2} =
	\mathfrak L_{o_1,o_2}
	- \frc12 \sum_{I, I':\mathsf v_I\neq \mathsf v_{I'}} 
	\frc{\bra o_1^-,Q_{I},Q_{I'}\ket^{\rm c}
	\bra o_2^-,Q_{I},Q_{I'}\ket^{\rm c}}{|\mathsf v_I-\mathsf v_{I'}|}.
\eeq
This is our main result: in linearly degenerate one-dimensional systems, the noise strength is given by the projected Kubo formula.

\subsection{Extended Einstein relation}\label{ssecteinst}

Finally, we establish the relation between the diffusion function $\h{\mathfrak D}_o^{~i}(\underline q)$ in \eqref{NLFBG} (equiv.~\eqref{NLFBG}), and the noise covariance $\h{\mathfrak L}_{o,j_i}(\underline q)$ for the observable $o$ and currents $j_i$. For this purpose, let us write, using space-time translation invariance of the state and the conservation laws,
\beqa
	\mathfrak L_{o,j_i}
	&=&
	\ell^2
	\int_{-\infty}^\infty \dd  t
	\int \dd x\,\dbra o^-( x, t),j_i(0,0)\dket^{\rm c}
	\n &=&
	\ell^2
	\int_{-\infty}^\infty \dd  t
	\int \dd x\,\dbra o^-( 0, 0),j_i(-x,-t)\dket^{\rm c}
	\n &=&
	- \ell^2
	\int_{-\infty}^\infty \dd  t
	\int \dd x\,x\p_x\dbra o^-( 0, 0),j_i(-x,-t)\dket^{\rm c}
	\n &=&
	\ell^2
	\int_{-\infty}^\infty \dd  t
	\int \dd x\,x\p_t\dbra o^-( 0, 0),q_i(-x,-t)\dket^{\rm c}
	\n &=&
	\ell^2
	\lim_{T\to\infty}
	\int \dd x\,x
	\dbra o^-( 0, 0),
	\big(q_i(-x,-T)
	- q_i(-x,T)\big) \dket^{\rm c}
	\n &=&
	\ell^2
	\lim_{T\to\infty}
	\int \dd x\,x
	\dbra \big(o^-( x, T) - o^-( x, -T)\big),
	q_i(0,0)\dket^{\rm c}.\label{manip}
\eeqa
We now evaluate this expression by considering in turn the microcanonical, diffusive and noise parts of $o^-(x,T)$ and $o^-(x,-T)$.

For the microcanonical part, the only term that remains from \eqref{o1o2} is the third one,
\beq
	\mathfrak L_{o,j_i} \stackrel{\text{micro}}=
	\sum_{\pm'}\pm' \ell^2
	\lim_{T\to\infty}
	\int \dd x\,x
	\frc12 
	\sum_{k,k'}
	\frc{\p^2 \mathsf o}{\p \mathsf q_{k}\p \mathsf q_{k'}} 
	\frc12\sum_{\pm} \dbra q_{k}(x\pm 0^+,\pm'T) ,q_{k'}(x\mp 0^+,\pm'T) , q_{i}(0,0)\dket^{\rm c}
\eeq
which we can write, by the inverse of the manipulations \eqref{manip}, as
\beq\label{rbag}
	\frc{\ell^2}4
	\int \dd t\int \dd x\,
	\sum_{k,k'}
	\frc{\p^2 \mathsf o}{\p \mathsf q_{k}\p \mathsf q_{k'}} 
	\sum_{\pm} \dbra q_{k}(x\pm 0^+,t) ,q_{k'}(x\mp 0^+,t) , j_{i}(0,0)\dket^{\rm c}.
\eeq
We now use the nonlinear projection formula for 3-point functions \cite{doyon2023ballistic,doyon2025nonlinear}
\beqa
	\lefteqn{\dbra q_{k}(x\pm 0^+,t) ,q_{k'}(x\mp 0^+,t) , j_{i}(0,0)\dket^{\rm c}} &&\n
	&=&
	\sum_l\frc{\p \mathsf j_i}{\p\mathsf q_l}\dbra q_{k}(x\pm 0^+,t) ,q_{k'}(x\mp 0^+,t) , q_{l}(0,0)\dket^{\rm c} \\
	&&
	+\, \sum_{l,l'}
	\frc{\p^2\mathsf j_i}{\p\mathsf q_l\p\mathsf q_{l'}}\dbra q_{k}(x\pm 0^+,t) ,q_l(0,0)\dket^{\rm c} \dbra q_{k'}(x\mp 0^+,t) , q_{l'}(0,0)\dket^{\rm c}.
\eeqa
The first term, when put in \eqref{rbag}, gives a total charge $Q_l$ and hence the factor
\beq
	\frc{\p}{\p\beta^l}
	\dbra q_{k}(\pm 0^+,t) ,q_{k'}(\mp 0^+,t)\dket^{\rm c} = 0
\eeq
which vanishes by the short-range correlations of the stationary state, and by the point-splitting. The second term is
\beq
	\frc{\ell^2}2
	\int \dd t\int \dd x\,
	\sum_{k,k'}
	\frc{\p^2 \mathsf o}{\p \mathsf q_{k}\p \mathsf q_{k'}}
	\sum_{l,l'}
	\frc{\p^2\mathsf j_i}{\p\mathsf q_l\p\mathsf q_{l'}}
	\dbra q_{k}(x,t) ,q_l(0,0)\dket^{\rm c} \dbra q_{k'}(x,t) , q_{l'}(0,0)\dket^{\rm c}
\eeq
which is exactly the term \eqref{o1o2proj} leading to the result \eqref{quadraticterm},
\beq
	\mathfrak L_{o,j_i} \stackrel{\text{micro}}=
	\frc12 \sum_{I, I':\mathsf v_I\neq \mathsf v_{I'}} 
	\frc{\bra o^-,Q_{I},Q_{I'}\ket^{\rm c}
	\bra j_i^-,Q_{I},Q_{I'}\ket^{\rm c}}{|\mathsf v_I-\mathsf v_{I'}|}.
\eeq

For the diffusive part, we use \eqref{regdif}
\beq
	\mathfrak L_{o,j_i} \stackrel{\text{diffusive}}=
	-\frc{\ell}2\sum_\pm \pm 
	\lim_{T\to\infty}
	\int \dd x\,x
	\sgn (\pm T)
	\sum_{k}
	\h{\mathfrak D}_{o}^{~k}(\mathsf q)
	\p_{x} \dbra q_{k}(x,\pm T),q_{i}(0,0)\dket^{\rm c}
\eeq
giving, by integration by part,
\beq
	\mathfrak L_{o,j_i} \stackrel{\text{diffusive}}=\ell
	\lim_{T\to\infty}
	\int \dd x\,
	\sum_{k}
	\h{\mathfrak D}_{o}^{~k}(\mathsf q)
	\dbra q_{k}(x,T),q_{i}(0,0)\dket^{\rm c}
	= \sum_k\h{\mathfrak D}_{o}^{~k}(\mathsf q)\mathsf C_{ki}.
\eeq
Finally, the noise part vanishes, $\dbra \xi_o( x, T), q_i(0,0)\dket^{\rm c}=0$, because $q_i(0,0)$ does not have any noisy component.

Hence we arrive at
\beq
	\mathfrak L_{o,j_i} = 
	\frc12 \sum_{I, I':\mathsf v_I\neq \mathsf v_{I'}} 
	\frc{\bra o^-,Q_{I},Q_{I'}\ket^{\rm c}
	\bra j_i^-,Q_{I},Q_{I'}\ket^{\rm c}}{|\mathsf v_I-\mathsf v_{I'}|}
	+\sum_k\h{\mathfrak D}_{o}^{~k}(\mathsf q)\mathsf C_{ki}
\eeq
and using \eqref{Lprojected} established above, we find \eqref{eins}.

\section{Absence of noise in integrable systems}\label{secabsence}

It was recently conjectured that the noise in integrable systems must vanish, see \cite{gopalakrishnan2024non,yoshimura2025anomalous,PhysRevLett.134.187101}. In particular, in \cite{PhysRevLett.134.187101} a nonlinear Boltzmann-Gibbs principle in the ballistic $1/\ell$ expansion, as in \eqref{NLFBG}, is used without noise, in order to evaluate diffusive-order corrections to the hydrodynamic equation, showing agreement with numerical simulation. That is, in integrable models, the contribution to the diffusive scale comes solely from the long-range correlations induced by the large-scale variations of the initial state \cite{doyon2023emergence}, which are of order $1/\ell$ and affect the (point-splitted) microcanonical average.

In this section we provide a proof that this is the case. We show that, under an appropriate choice of the current observables in integrable models -- a choice of ``gauge'' --, which we denote $[j_i]_\infty$, the nonlinear fluctuating Boltzmann-Gibbs principle simplifies to a non-fluctuating one at all orders of the hydrodynamic expansion:
\beq\label{NLFBGint}
	\overline{ [j_i]_\infty}( x, t)\ \mbox{in $\bra\cdots\ket_\ell$}
	\longrightarrow
	j_i(x,t):=\mathsf j_i^{\rm reg}(\underline q(x,t)) + \,\mathcal O(\ell^{-\infty})
		\ \mbox{in $\dbra\cdots\dket_\ell$}.
\eeq
At the diffusive order, with $\mathcal O(\ell^{-2})$ instead of $\mathcal O(\ell^{-\infty})$, the result holds for the PT-symmetric current $j_i$  instead of $[j_i]_\infty$ (that is, without the need to further fix the gauge).
\beq\label{NLFBGintdiff}
	\overline{ j_i}( x, t)\ \mbox{in $\bra\cdots\ket_\ell$}
	\longrightarrow
	j_i(x,t):=\mathsf j_i^{\rm reg}(\underline q(x,t)) + \,\mathcal O(\ell^{-2})
		\ \mbox{in $\dbra\cdots\dket_\ell$}.
\eeq

First, we show the result at the diffusive order \eqref{NLFBGintdiff}, with the PT-symmetric current $j_i$. This is done using previous, independent results on the Onsager matrix. We show that in integrable systems, the noise $\xi_{j_i}$  for the currents $j_i$, as appears at order $\ell^{-1}$ in \eqref{NLFBG}, vanishes, and as a consequence, the bare diffusion for all observables also vanishes. That is,
\beq
	\xi_{j_i} = 0
\eeq
and more specifically
\beq\label{Loji}
	\h{\mathfrak L}_{j_i,j_i}(\underline{\mathsf q}) = 0 \ \Rightarrow\ \h{\mathfrak L}_{o,j_i}(\underline{\mathsf q}) = \h{\mathfrak L}_{j_i,o}(\underline{\mathsf q})=0\ \Rightarrow\ 
	\h{\mathfrak D}_{o}^{~k}(\underline{\mathsf q}) = 0
\eeq
where the last equality follows from \eqref{eins}.

Second, we give a first-principle argument, based on the fluid-cell relaxation picture of Sec.~\ref{ssectarg}. This argument is important in two ways:

(1) It provides an {\em independent, first-principle derivation of the result for the Onsager matrix in integrable models}, Eq.~\eqref{Lintegrable}. Indeed, as we have shown that the noise strength is the projected Kubo formula, a corollary that this vanishes is that the un-projected, full Kubo formula is given by its projection onto quadratic charges. This reproduces the known expression for the Onsager matrix in integrable systems in terms of quasi-particles.

(2) It allows us to show that in fact, the noise for currents must vanish {\em at all orders in the hydrodynamic expansion}, Eq.~\eqref{NLFBGint}. This is true, beyond the diffusive scale, as long as current observables are chosen appropriately. Recall that charge densities and currents are defined only up to total derivatives of local observables -- this is the ``gauge'' freedom in determining densities and currents. At the diffusive level, this gauge freedom is lifted by choosing the PT gauge \cite{de2019diffusion}. But beyond it, there is no natural gauge, see however a discussion in \cite{de2023hydrodynamic}. Our proof provides a partially constructive way of choosing the gauge, that guarantees that the nonlinear Boltzmann-Gibbs principle for currents does not admit noise at all orders of the hydrodynamic expansion.

\medskip

{\em First proof.} We show that the absence of noise at the diffusive scale, Eq.~\eqref{NLFBGintdiff} or equivalently \eqref{Loji}, is a consequence of our main result \eqref{Lprojected}, along with the following observation.

In the hard rods model, the  Onsager matrix $\mathfrak L_{j_i,j_k}$, which is in fact an integral operator as it is taken in the basis of the conserved hard rods velocity $i = v\in\R$, was evaluated exactly, see e.g.~\cite{boldrighini1997one,spohn2012large,doyon2017dynamics}. Further, in \cite{de2018hydrodynamic,de2019diffusion} the Onsager matrix $\mathfrak L_{j_i,j_k}$ was evaluated in integrable models using quantum form factor techniques, generalising the hard-rod formula (see also \cite{gopalakrishnan2018hydrodynamics} where the diagonal part was evaluated using a quasi-particle picture). Then, it was shown in \cite{doyon2022diffusion} that this general formula in fact corresponds to a projection, within the diffusive Hilbert  space $\mathcal H_{\rm dif}$, onto the scattering space formed by quadratic charges (see the discussion in Sec.~\ref{secconst})
\beq\label{Lintegrable}
	\mathfrak L_{j_i,j_k}
	\stackrel{\text{integrable models}}= 
	\big(\mathbb P_{\rm scat}j_i,\mathbb P_{\rm scat}j_k\big)_{\rm dif}
	=
	\frc12 \sum_{I\neq I'} 
	\frc{\bra j_i^-,Q_{I},Q_{I'}\ket^{\rm c}
	\bra j_k^-,Q_{I},Q_{I'}\ket^{\rm c}}{|\mathsf v_I-\mathsf v_{I'}|}.
\eeq
Therefore, from these results and \eqref{Lprojected}, we conclude that the noise, at least at the leading order, has vanishing correlations
\beq
	\h{\mathfrak L}_{j_i,j_k} \stackrel{\text{integrable models}}= 0.
\eeq
By the Cauchy-Schwartz inequality (see \cite{doyon2022diffusion} for the inner-product structure of the Kubo formula), we have
\beq
	(\h{\mathfrak L}_{o,j_i})^2 \leq \h{\mathfrak L}_{o,o}
	\h{\mathfrak L}_{j_i,j_i}  \stackrel{\text{integrable models}}= 0
\eeq
which shows \eqref{Loji}.

\medskip

{\em Second proof.} We show \eqref{NLFBGintdiff}, and then more generally \eqref{NLFBGint}, using a general first-principle argument based on relaxation of fluid-cell means and fundamental properties of integrable systems.

In many-body integrable models, there exists an infinite number of commuting, independent ``higher'' time flows, besides the ``usual'' time evolution. If the integrable system is Hamitonian, then these are generated by infinitely many linearly and functionally independent commuting, extensive Hamiltonians. In translation-invariant systems, these time flows are also translation invariant. That is, we can evolve any local observable not just in time $\micro t=\micro t_1$, but also in times $\micro t_2,\,\micro t_3,\,\ldots$, which we denote by the commuting one-parameter groups of maps $\tau_2^{\micro t_2},\tau_3^{\micro t_3},\ldots$ acting on observables:
\beq
	o(\micro x,\micro t_1,\micro t_2,\micro t_3,\ldots) = \tau_2^{\micro t_2}\tau_3^{\micro t_3}\cdots o(\micro x,\micro t_1),\quad \micro t_1,\micro t_2,\micro t_3,\ldots\in\R.
\eeq
Crucially, because they are commuting flows, each conserved density $q_i$ is also a conserved density with respect to all higher times $t_k$, with associated currents $j_{ki}$:
\beq\label{tkcons}
	\p_{\micro t_k} q_i(\micro x,\micro t_1,\ldots,t_n) + \p_{\micro x} j_{ki}(\micro x,\micro t_1,\ldots,\micro t_n) = 0.
\eeq
Consistency requires $\p_{\micro t_k} j_{li} = \p_{\micro t_l}j_{ki}$. Further, using this, it was shown in \cite[App 4]{doyon2022diffusion} that the Onsager matrix is {\em invariant} under these flows,
\beq\label{Linv}
	\mathfrak L_{\tau_2^{\micro t_2}\tau_3^{\micro t_3}\cdots j_i,o} = 
	\mathfrak L_{j_i,o}.
\eeq
It is simple to see that the same holds for the Onsager matrix from which quadratic charges have been projected out, as these are invariant under all higher time flows,
\beq\label{Linv}
	\h{\mathfrak L}_{\tau_2^{\micro t_2}\tau_3^{\micro t_3}\cdots j_i,o} = 
	\h{\mathfrak L}_{j_i,o}.
\eeq

Now recall the argument in Sec.~\ref{ssectarg} for the origin of hydrodynamic noise. On a fluid cell of size $L$, after a time $T$, there has been $\mathcal O(LT) = \mathcal O(\ell)$ fluctuation events for the fluid-cell mean $\overline o$ of the observable $o$, while, thanks to the conservation laws, the conserved densities $\overline{q_i}$'s are essentially constant; see Fig.~\ref{figfluidcell2}. Hence, taking an average over a time $t\in[0,T]$ (or, equivalently, not taking the time average, but using typicality) relaxation to the microcanonical average occurs, plus a noise of variance $\propto 1/\ell$ by the central limit theorem.

Consider $o=j_i$ to be a current. Let us change this observable by averaging over $n$ of the higher time flows, with higher times in the range $[0,T']$ for some fixed $T'>1$ (not scaling with $\ell$),
\beq
	[j_i]= \frc1{{T'}^{n}} \int_0^{T'}\dd \micro t_2\cdots \int_0^{T'} \dd \micro t_{n+1} \prod_{k=2}^{n+1} \tau_k^{\micro t_k} j_i.
\eeq
Taking the time-average over $\micro t_1\in[0,T]$, the fluid-cell mean $\overline {[j_i]}$ is the average of the original fluid cell mean $\overline{j_i}$ over the time-hyperbox $[0,T]\times [0,T']^{n}$ of dimension $n+1$, with $\micro t_1\in[0,T]$ and $\micro t_k\in[0,T']$ for $k=2,\ldots,n+1$. On this hyperbox, fluid cell means of conserved densities $\overline{q_i}$'s are all essentially constant, because of \eqref{tkcons}. As each time flow is independent, the fluctuations induced, from typical configurations, by each flow are also independent. Thus, on the space-time hyperbox $[0,L]\times [0,T]\times[0,T']^{n}$ the configuration shows a number of independent fluctuation events (particle collisions, etc.). In the higher-dimensional equivalent of Fig.~\ref{figfluidcell2}, these fluctuation events lie on hypersurfaces of dimension $n+1$ criss-crossing the hyperbox. There are $\mathcal O(LT)$ such hypersurfaces, each of hypervolume $\propto {T'}^n$. It is natural to assume that on each such hypervolume, there is typically ${T'}^n/\ell_{\rm micro}^n$ independent changes, by an extension of the assumption of molecular chaos to this context with higher-dimensional time. Therefore, the fluid cell mean $\overline {[j_i]}$ has undergone $\propto {T'}^{n}$ times more fluctuations than $\overline{j_i}$.

Taking $n$ and/or $T'$ large after the $\ell\to\infty$ asymptotic, we thus find by the central limit theorem that the noise for the observable $[j_i]$ has variance
\beq
	\dbra \xi_{[j_i]}(x,t)\xi_{[j_i]}(x',t')\dket_\ell \sim c(T')\ell^{-1}\delta(x-x')\delta(t-t'),\quad c(T') = \mathcal O({T'}^{-n})
\eeq
hence
\beq
	\h{\mathfrak L}_{[j_i],[j_i]} = \mathcal O({T'}^{-n}).
\eeq
By \eqref{Linv}, this means
\beq
	\h{\mathfrak L}_{j_i,j_i} = \mathcal O({T'}^{-n}).
\eeq
As $n$ and $T'>0$ can be taken as large as we wish, this implies $\h{\mathfrak L}_{j_i,j_i} =0$, and by Cauchy Schwartz shows \eqref{Loji}.

We note that the presence of {\em a single} higher time flow, $n=1$, is sufficient for this argument, as we can take $T'>1$ as large as we wish (but according to integrability theory, a single higher-flow implies the presence of infinitely-many higher flows).

Now let us consider the observable $[j_i]$. Because the higher flows commute, this is a good current observable, for a conserved density that is likewise averaged over higher time flows:
\beq
	\p_\rx [q_i] + \p_{\micro t} [j_i] = 0.
\eeq
The above argument shows that as we make $n$ larger for any fixed $T'>1$, the emergent noise's variance becomes smaller. By the Bienaym\'e-Chebyshev theorem, this implies that all higher cumulants are likewise made smaller. Therefore, we find that the following particular choice of densities and currents,
\beq
	[q_i]_\infty = \lim_{n\to\infty} [q_i],\quad [j_i]_\infty = \lim_{n\to\infty} [j_i]
\eeq
are good choices of gauge that guarantee that {\em the associated current in the fluctuating hydrodynamic theory has vanishing noise at all orders of the hydrodynamic expansion}, and we obtain \eqref{NLFBGint}.

\begin{rema}
Note how this is related to the natural conjecture that semiclassical Bethe particles \cite{doyon2023ab,doyon2023generalised,doyon2024new} give a good representation of generalised hydrodynamics at all orders \cite{doyon2024new,urilyon2025simulating}, although still missing is the connection between the above choice of gauge and the choice that associates currents to these particles' flows.
\end{rema}

\section{Hydrodynamic equation and diffusive two-point functions}\label{sechydroeq}

Sec.~\ref{ssectnoise} and \ref{ssecteinst} illustrate some of the techniques involved in calculating using the nonlinear fluctuating Boltzmann-Gibbs principle \eqref{NLFBG}, in particular combining the point-splitting regularisation with the cumulant expansion.  In order to further illustrate this ``calculus'' and to show the general consistency of our framework, in this section we provide two simple applications. We obtain the hydrodynamic equation up to the diffusive correction, first for one-point function in non-equilibrium states emanating from \eqref{lw}, and second for two-point functions of conserved densities in stationary states.

For one-point function, the result is highly anomalous, taking the form of a coupled system of equations as in \cite{PhysRevLett.134.187101} for integrable systems, but generalised to the presence of bare diffusion. The calculation is a simple generalisation of that shown in the main text of \cite{PhysRevLett.134.187101}.

For two-point functions, the result is a normal diffusion equation, with the full diffusion matrix related by Einstein relation to the {\em full, un-projected Onsager matrix}. This is usually assumed to be valid and follows from sum-rule manipulations, see e.g.~\cite{spohn2012large,de2019diffusion}. Here we derive it from our fluctuating hydrodynamic theory, in which context it is somewhat non-trivial, as the effects of long-range correlations to the diffusive scale -- coming from the regularised microcanonical part of \eqref{NLFBG} -- and of bare diffusion must combine just in the right way to give the diffusion matrix related to the full Onsager matrix. This was also derived in \cite{PhysRevLett.134.187101} for integrable systems, although in this case there is no bare diffusion. It was derived by a different technique, using a special linear response theory on the hydrodynamic equation for one-point functions in non-equilibriun states. There, the full Onsager matrix -- without bare diffusion -- arises because the at infinitesimal macroscopic times, the long-range correlations exactly reproduce the projection onto quadratic charges, thus, for integrable systems the full Onsager matrix (as we showed in Sec.~\ref{secabsence}). We instead perform a direct calculation using the nonlinear fluctuating Boltzmann-Gibbs principle, which involves  3-point functions in stationary states recently evaluated \cite{doyon2025nonlinear}.

Thus we show that the present framework gives a consistent theoretical underpinning for both the special emergent hyrodynamic equation out of equilibrium, and normal diffusion at the linearised level. 

\subsection{The hydrodynamic equations at the diffusive scale}

Using our fluctuating hydrodynamic theory based on \eqref{NLFBG}, we first write down the diffusive-scale hydrodynamic equation, taking into account the results Points I-III. For definiteness the initial state is the long-wavelength state \eqref{lw}, but more general states would lead to the same equation.

The hydrodynamic equation is obtained from the conservation laws:
\beq
	\p_t \dbra q_i(x,t)\dket_\ell + \p_x \dbra j_i(x,t)\dket_\ell = 0.
\eeq
Inserting \eqref{NLFBG} for the case $o=j_i$ in this, using $\dbra\xi_{o}(x,t)\dket_\ell=0$ and the fact that for the diffusive order $\ell^{-1}$ we may use the leading cumulant-expansion form
\beq
	\sum_i\dbra \h{\mathfrak D}_o^{~i}(\underline q(x,t))
	\p_x q_i(x,t)\dket =
	\sum_i\h{\mathfrak D}_o^{~i}(\dbra \underline q(x,t)\dket_\ell)
	\p_x \dbra q_i(x,t)\dket_\ell + \mathcal O(\ell^{-1})
\eeq
and denoting for lightness $\mathsf q_i(x,t) = \dbra q_i(x,t)\dket_\ell$, this gives
\beq
	\p_t \mathsf q_i(x,t) + \p_x \dbra \mathsf j_i^{\rm reg}(\underline q(x,t))\dket_\ell = \frc1{2\ell} \sum_k\p_x\Big(\h{\mathfrak D}_i^{~k}(\underline {\mathsf q}(x,t))
	\p_x \mathsf q_k(x,t)\Big)
\eeq
where we used the notation \eqref{notationD}. The quantity $\dbra \mathsf j_i^{\rm reg}(\underline q(x,t))\dket_\ell$ was evaluated in \cite{PhysRevLett.134.187101}, and depends on the long-range correlations generated by the long-wavelength state \cite{doyon2023emergence}, which are at the diffusive order $\ell^{-1}$. These long-range correlations themselves satisfy a set of Euler-scale evolution equation. The result is a set of two combined equations:
\beqa
	\p_t \mathsf q_i(x,t) + \sum_k\mathsf A_i^{~k}(x,t)\p_x \mathsf q_k(x,t)
	+
	\frc1{4\ell} \sum_{I,J,\pm}\p_x\Big(
	\bra j_i^-,Q_I,Q_J\ket^{\rm c}_{\underline{\beta}(x,t)}
	E_{IJ}(x\pm 0^+,x\mp 0^+;t)\Big) \n &&
	\hspace{-5.5cm} = \frc1{2\ell} \sum_k\p_x \Big(\h{\mathfrak D}_i^{~k}(\underline {\mathsf q}(x,t))\,
	\p_x \mathsf q_k(x,t)\Big)\n
	\label{hydro1}
\eeqa
with
\beqa
	\lefteqn{\p_t E_{IJ}(x,y;t) + \p_x \big(\mathsf v_I(x,t) E_{IJ}(x,y;t)\big) + \p_y \big(\mathsf v_J(y,t) E_{IJ}(x,y;t) \big)} && \n
	&& \hspace{6cm} = \delta(x-y)\sum_{k}\bra j_k^-,Q_I,Q_J\ket^{\rm c}_{\underline{\beta}(x,t)}
	\p_x \beta^k(x,t).\n
	\label{hydro2}
\eeqa
Here $\beta^i(x,t) := \beta^i(\underline{\mathsf q}(x,t))$ (see Eq.~\eqref{defo}) and likewise for the hydrodynamic velocities $\mathsf v_I(x,t)$. In the state \eqref{lw}, the initial conditions are
\beq
	\beta^i(x,0) = \beta^i(x),\quad E_{rs}(x,y;0) = 0
\eeq
but these equations are more general, and valid also for initial states that admit non-trivial long-range correlations $E_{rs}(x,y;0)\neq 0$. Eqs.~\eqref{hydro1}, \eqref{hydro2} generalise \cite[Eqs 9, 10]{PhysRevLett.134.187101}, which was for integrable models, to the inclusion of the bare diffusion generically present away from integrability.

For integrable models, Eqs.~\eqref{hydro1} and \eqref{hydro2} are immediate to write in terms of quasi-particle rapidities, using that, in normal modes $I\in\R$ (which are of rapidity type), $\mathsf A_I^{~J} = \delta(I-J)v^{\rm eff}(I)$, and the three-point coupling in normal modes \eqref{Hessiannormal} has the expression displayed in \cite[Eq 103]{doyon2025nonlinear}. See \cite{urilyon2025simulating} where it is written in even more generality, including external force terms \cite{doyon2017note}.

\subsection{Normal diffusion for stationary-state two-point functions}

We restrict to stationary states
\beq
	\bra\cdots\ket = \bra\cdots\ket_{\underline{\beta}}
\eeq
and all constitutive functions are evaluated within this state (we keep their state dependence implicit).

We will show that the diffusive equation for two-point functions of conserved densities takes a diffusive form:
\beq\label{tpt}
	\p_t \bra \overline{q_i}(x,t),\overline{q_j}(0,0)\ket^{\rm c}
	+
	\sum_k\mathsf A_i^{~k} \p_x \bra \overline{q_k}(x,t),\overline{q_j}(0,0)\ket^{\rm c}
	=
	\frc1{2\ell} \sum_k\mathfrak D_i^{~k}\p_x^2
	\bra \overline{q_k}(x,t),\overline{q_j}(0,0)\ket^{\rm c},\quad
	t>0.
\eeq
This is in principle valid for fluid-cell means as written here, so that possible oscillatory behaviours are averaged out, but in practice often the fluid-cell mean is not required. Importantly, as mentioned, the diffusion matrix involved is related by the Einstein relation to the {\em full} Onsager matrix,
\beq\label{einstfull}
	\mathfrak L_{il} = \sum_k \mathfrak D_{i}^{~k}\mathsf C_{kl}.
\eeq
Here and below we use the simplified notation $\mathfrak L_{ik} = \mathfrak L_{j_i,j_k}$ paralleling \eqref{notationD}. Eq.~\eqref{einstfull} is the same relation as Eq.~\eqref{eins} for the bare diffusion matrix in terms of the projected Kubo formula.

That the full diffusion matrix be involved in the two-point function is non-trivial: the bare diffusion term easily gives a contribution to the diffusion term in \eqref{tpt}, however the long-range correlations, coming from the regularised microcanonical average, also give a contribution, and these sum up exactly to the full diffusion matrix. This is what we will show now. The proof is based on a result for the exact three-point function of conserved densities at the Euler scale that we obtained recently by projection techniques, see \cite[Sec 5.2]{doyon2025nonlinear}.

Using the conservation laws again, passing to the fluctuating theory, and using normal modes and the fact that two-point functions of conserved densities at the Euler scale are diagonal in this basis, Eq.~\eqref{2pointeq}, we must show that
\beq
	\dbra j_I(x,t),q_J(0,0)\dket^{\rm c}
	= \delta_{IJ}\mathsf v_I\dbra q_I(x,t),q_I(0,0)\dket^{\rm c}
	- \frc1{2\ell} \mathfrak L_{IJ}\p_x \dbra q_J(x,t),q_J(0,0)\dket^{\rm c},\quad t>0.
\eeq
As the second term on the right-hand side has a factor $1/\ell$, it can be evaluated at the Euler scale, so we must show
\beq\label{mustshowdelta}
	\dbra j_I(x,t),q_J(0,0)\dket^{\rm c}
	= \delta_{IJ}\mathsf v_I\dbra q_I(x,t),q_I(0,0)\dket^{\rm c}
	- \frc1{2\ell^2} \mathfrak L_{IJ}\delta'(x-\mathsf v_J t),\quad t>0.
\eeq

\proof
We use \eqref{NLFBG} for $o=j_I$. Using \eqref{noisecorr}, as the noise $\xi_I$ does not correlate with the fluctuation of $q_J(0,0)$, induced by the initial state, it does not contribute:
\beq
	\dbra j_I(x,t),q_J(0,0)\dket^{\rm c}\stackrel{\text{noise}}=0.
\eeq
The diffusive term can be evaluated by the cumulant expansion, at leading order giving simply
\beq\label{diffcontd}
	\dbra j_I(x,t),q_J(0,0)\dket^{\rm c}\stackrel{\text{diffusive}}=-\frc1{2\ell} \h{\mathfrak L}_{IJ} \p_x \dbra q_J(x,t),q_J(0,0)\dket^{\rm c}
	=
	-\frc1{2\ell^2} \h{\mathfrak L}_{IJ}\delta(x-\mathsf v_J t)
	,\quad t>0.
\eeq
This is the bare contribution to the diffusive term of the two-point function.

The most important part is that coming from the microcanonical average, with point-splitting regularisation. Using \eqref{o1o2}, this is
\beqa\label{jregj}
	\lefteqn{\dbra \mathsf j_I^{\rm reg}(\underline{q}(x,t)) q_J(0,0)\dket^{\rm c}}
	&&\\
	&=&
	\delta_{IJ} \mathsf v_I  
	\dbra q_{I}(x,t) ,q_{J}(0,0)\dket^{\rm c}
	+
	\frc12 
	\sum_{KL}
	\mathsf A_{I}^{~KL}
	\frc12\sum_{\ep=\pm 0^+} \dbra q_{K}(x+\ep,t) ,q_{L}(x- \ep,t) , q_{J}(0,0)\dket^{\rm c}.\no
\eeqa
We must now evaluate the three-point function. For this purpose, it is simpler to bring the space-time variables $x,t$ onto $q_J$ by translation invariance and PT symmetry:
\beq
	\dbra q_{K}(x+\ep,t) ,q_{L}(x- \ep,t) , q_{J}(0,0)\dket^{\rm c}
	=
	\dbra q_{K}(-\ep,0) ,q_{L}(+\ep,0) , q_{J}(x,t)\dket^{\rm c}.
\eeq
We then use the formula \cite[Eqs 98, 99]{doyon2025nonlinear}. This has two types of contribution:
\beq
	\dbra q_{K}(-\ep,0) ,q_{L}(\ep,0) , q_{J}(x,t)\dket^{\rm c}
	= \ell^{-2}(C_{\rm linear} - C_{\rm nonlinear}).
\eeq
The first is a direct, linear propagation. This contribution vanishes because of the point-splitting:
\beq
	C_{\rm linear} = \bra Q_K,Q_L,q_J\ket^{\rm c}
	\delta(x+\ep-\mathsf v_J t)\delta(x-\ep-\mathsf v_J t) = 0.
\eeq

The second is an indirect, nonlinear propagation, coming from the nonlinear part of the 3-point projection. Different cases in principle must be treated, depending if some of the velocities $\mathsf v_J,\,\mathsf v_K,\,\mathsf v_L$ are equal to each other. However, the expression in the case where only two velocities are equal to each other, can be obtained by taking the formal limit of equal velocities; while that where all velocities are equal to each other does not contribute because of the vanishing 3-point coupling \eqref{diagvanish}. Hence, it is sufficient to perform the full calculation from the expression for unequal velocities. Because of our choice of values of times, only the nonlinear propagation associated to $q_J$ is required, giving:
\beq
	C_{\rm nonlinear} = \mathsf A_{J}^{~KL}
	\delta'(v'u-vu')
	\big(|v'| (s(u'+v't) - s(u')) - |v| (s(u+vt)-s(u))\big)
	= \mathsf A_J^{~KL}c
\eeq
where $s(a) = \frc12\sgn(a)$ and
\beq
	u = x-\mathsf v_J t +\ep,\ u' = x-\mathsf v_J t -\ep,\ 
	v = \mathsf v_{JK},\ v' = \mathsf v_{JL}
\eeq
with the assumption $\mathsf v_J\neq \mathsf v_K,\,\mathsf v_J\neq \mathsf v_L,\,\mathsf v_K\neq \mathsf v_L$. Thus we must evaluate:
\beqa
	c &=& \delta'\big(\mathsf v_{KL}(x-\mathsf v_Jt) + \ep (2\mathsf v_J-\mathsf v_K-\mathsf v_L)\big)\times\\
	&&\times\,
	\Big(
	|\mathsf v_{JL}| (s(x-\mathsf v_Lt-\ep) - s(x-\mathsf v_J t-\ep))
	-
	|\mathsf v_{JK}|( s(x-\mathsf v_Kt+\ep) - s(x-\mathsf v_J t+\ep))
	\Big)\no
\eeqa
We evaluate the term with $s(x-\mathsf v_Jt-\ep)$ as
\beqa
	-\lefteqn{\frc{|\mathsf v_{JL}|}{\mathsf v_{KL}|\mathsf v_{KL}|}
	\delta'\Big(x-\mathsf v_Jt + \ep \frc{2\mathsf v_J-\mathsf v_K-\mathsf v_L}{\mathsf v_{KL}}\Big)s(x-\mathsf v_J t-\ep)} &&
	\n &\stackrel{|\ep|\to0}=&
	\frc{|\mathsf v_{JL}|}{\mathsf v_{KL}|\mathsf v_{KL}|}\Big(
	\delta'(x-\mathsf v_Jt)s\Big(\frc{\mathsf v_{JL}}{\mathsf v_{KL}}\Big)\sgn(\ep)
	+
	\delta(x-\mathsf v_Jt)
	\delta\Big(\frc{\mathsf v_{JL}}{\mathsf v_{KL}}\ep\Big)
	\Big)
	\n &=&
	\frc{|\mathsf v_{JL}|}{\mathsf v_{KL}|\mathsf v_{KL}|}
	\delta'(x-\mathsf v_Jt)s\Big(\frc{\mathsf v_{JL}}{\mathsf v_{KL}}\Big)\sgn(\ep)
\eeqa
and under the sum $\sum_{\ep=\pm0^+}$ in \eqref{jregj} the result vanishes. Likewise for $s(x-\mathsf v_J t+\ep)$,
\beq
	\frc{|\mathsf v_{JK}|}{\mathsf v_{KL}|\mathsf v_{KL}|}
	\delta'\Big(x-\mathsf v_Jt + \ep \frc{2\mathsf v_J-\mathsf v_K-\mathsf v_L}{\mathsf v_{KL}}\Big)s(x-\mathsf v_J t+\ep)
	\stackrel{|\ep|\to0}=
	-\frc{|\mathsf v_{JK}|}{\mathsf v_{KL}|\mathsf v_{KL}|}
	\delta'(x-\mathsf v_Jt)s\Big(\frc{\mathsf v_{JK}}{\mathsf v_{KL}}\Big)\sgn(\ep)
\eeq
which again vanishes under $\sum_{\ep=\pm0^+}$. For the term with $s(x-\mathsf v_L t-\ep)$ we get
\beqa
	\lefteqn{\frc{|\mathsf v_{JL}|}{\mathsf v_{KL}|\mathsf v_{KL}|}
	\delta'\Big(x-\mathsf v_Jt + \ep \frc{2\mathsf v_J-\mathsf v_K-\mathsf v_L}{\mathsf v_{KL}}\Big)s(x-\mathsf v_L t-\ep)} &&
	\n &\stackrel{|\ep|\to0}=&
	\frc{|\mathsf v_{JL}|}{\mathsf v_{KL}|\mathsf v_{KL}|}
	\big(\delta'(x-\mathsf v_Jt)s(\mathsf v_{JL} t)-
	\delta(x-\mathsf v_Jt)\delta(\mathsf v_{JL} t)
	\big)
	\n &\stackrel{t>0}=&
	\frc{\mathsf v_{JL}}{2\mathsf v_{KL}|\mathsf v_{KL}|}
	\delta'(x-\mathsf v_Jt)
\eeqa
and likewise for the term with $s(x-\mathsf v_K t+\ep)$,
\beq
	-\frc{|\mathsf v_{JK}|}{\mathsf v_{KL}|\mathsf v_{KL}|}
	\delta'\Big(x-\mathsf v_Jt + \ep \frc{2\mathsf v_J-\mathsf v_K-\mathsf v_L}{\mathsf v_{KL}}\Big)s(x-\mathsf v_K t+\ep)
	\stackrel{|\ep|\to0,\,t>0}=
	-\frc{\mathsf v_{JK}}{2\mathsf v_{KL}|\mathsf v_{KL}|}
	\delta'(x-\mathsf v_Jt).
\eeq
Combining, 
\beq
	\frc12\sum_{\ep=\pm 0^+}C_{\rm nonlinear} = 
	\mathsf A_J^{~KL}
	\frc1{2|\mathsf v_{KL}|}
	\delta'(x-\mathsf v_J t)
\eeq
and therefore
\beq\label{resAA}
	\frc12 
	\sum_{KL}
	\mathsf A_{I}^{~KL}
	\frc12\sum_{\ep=\pm 0^+} \dbra q_{K}(x+\ep,t) ,q_{L}(x- \ep,t) , q_{J}(0,0)\dket^{\rm c}
	\stackrel{t>0}=
	-\sum_{KL}\frc{\mathsf A_{I}^{~KL}\mathsf A_J^{~KL}}{4|\mathsf v_{KL}|\ell^2}
	\delta'(x-\mathsf v_J t).
\eeq
With \eqref{jregj}, \eqref{diffcontd} and \eqref{resAA}, we therefore find
\beq
	\dbra j_I(x,t),q_J(0,0)\dket^{\rm c}
	=
	\delta_{IJ} \mathsf v_I  
	\dbra q_{I}(x,t) ,q_{J}(0,0)\dket^{\rm c}
	-\frc1{2\ell^2}
	\Big(\h{\mathfrak L}_{IJ}
	+
	\sum_{KL}\frc{\mathsf A_{I}^{~KL}\mathsf A_J^{~KL}}{2|\mathsf v_{KL}|}
	\Big)
	\delta'(x-\mathsf v_J t).
\eeq
Thus from \eqref{Lprojected} and \eqref{Hessiannormal}, we have shown that \eqref{mustshowdelta} holds. \eproof

\section{Conclusion}\label{sectconclusion}

We have developed a hydrodynamic fluctuation theory for linearly degenerate systems in one spatial dimension, and more generally ``no-shock'' systems, up to, including, the diffusive order $1/\ell$ as $\ell\to\infty$ in the ballistic scaling of space and time, $\rx = \ell x,\,\micro t= \ell t$. The theory is presented as a stochastic PDE in a $1/\ell$ expansion, which is well defined up to the first subleading order. This is based on a nonlinear fluctuating Boltzmann-Gibbs principle, where generic observables are projected onto conserved densities, with the addition of diffusive and noise terms. This nonlinear fluctuating Boltzmann-Gibbs principle is justified by heuristic arguments that we have discussed carefully, based on the separation of relaxation time-scales between coarse-grained conserved densities and other coarse-grained observables. Relaxation of arbitrary observables to their microcanonical averages is modified by noise via the central limit theorem, by ``bare'' diffusion because of relaxation time delay (fluctuation-dissipation), and by a non-fluctuating, non-diffusive correction to the microcanonical average, representing the correction to the microcanonical average due to the finite size of the region (the fluid-cell) on which it is taken. The noise covariance is shown to be given by a projected Kubo formula, where diffusive-scale effects of long-range correlations \cite{PhysRevLett.134.187101} are subtracted (projected out, from a Hilbert space perspective \cite{doyon2022diffusion}). The diffusion is shown to be obtained by Einstein relation from this, and the microcanonical correction is shown to give a point-splitting regularisation \cite{PhysRevLett.134.187101} that makes the stochastic PDE, with delta-correlations both in noise and initial conditions, well-defined. We show, in two different ways, that the noise vanishes at the diffusive scale in integrable models, proving conjectures made previously \cite{gopalakrishnan2024non,yoshimura2025anomalous,PhysRevLett.134.187101}. We show that, under an appropriate choice of currents, their noise in fact vanishes at all orders in the hydrodynamic expansion. We write down the full hydrodynamic equation for linearly degenerate systems, including effects of long-range correlations \cite{PhysRevLett.134.187101} and bare diffusion. We also derived, within our fluctuating hydrodynamic formalism, the diffusive hydrodynamic equation for dynamical correlation functions in stationary states, where we showed that diffusion is associated to the full Onsager matrix instead of the projected one, as it should be.

% The use of PT-symmetry is known to guarantee that the Einstein relation holds between hydrodynamic diffusion and Onsager matrix \c; however here the diffusion is a microscopic one
 
% In a sense ld scaling more universal, as it does not require to know if  diffusive or superdiffusive effects
 
In the heuristic arguments we have provided to explain the form of the nonlinear Boltzmann-Gibbs principle in Sec.~\ref{ssectarg}, noise emerges under the usual assumption of molecular chaos, here expressed as the assumption that interaction events are essentially independent and non-correlated when separated enough in space and time (see Fig.~\ref{figfluidcell2}). Clearly, as we show that noise in fact vanishes for currents in integrable models, there may in fact be correlations that reduce the noise even for certain observables which are not conserved densities. In Hamiltonian systems, one of the leading open problems in the hydrodynamic theory remains to more accurately understand the emergence of Gaussian white noise, perhaps by expanding on the heuristic arguments we have expressed.

In order to make the absence of noise for currents in integrable models more explicit, and to put the heuristic arguments of Sec.~\ref{ssectarg} on a more solid basis, one should investigate numerically fluctuations of observables on fluid cells\footnote{As we mentioned, shortly after the first version of this paper was posted, paper \cite{hubner2025hydrodynamics} performed this in the hard rods model, confirming our results.}.

It has recently been shown with mathematical rigour that in (generalised) hard rods models, a certain fluctuating hydrodynamic equation emerges for linear fluctuating fields, where noise appears, albeit with strong correlations in space \cite{ferrari2025macroscopic,chahal2025stochastic}. This noise is {\em not} of the type that appears in \eqref{NLFBG}, which, as we have shown, vanishes in integrable models. Instead, we believe this space-correlated noise is a ``remnant'', at the linear level, of long-range correlations that are known to occur from the present nonlinear version of the fluctuating hydrodynamics. More investigation on this aspect would be appropriate.

It would be interesting to evaluate explicitly, using the present theory, the first subleading corrections to correlation functions and to large-deviation functions beyond the Euler scale \cite{doyon2023emergence,doyon2023ballistic,doyon2025nonlinear}. It would also be interesting to generalise this to higher dimensions (see \cite{doyon2025nonlinear} for the nonlinear projection principles for correlation functions at the leading Euler scale), where long-range correlations do not affect the diffusive scale but may affect higher scales of the hydrodynamic expansion.

As we have shown, the no-shock condition guarantees the absence of ``cubic'' singularities in the Euler current, $\mathsf A_k^{~II}=0$. An interesting question is as to the effects of quartic singularities: these are known to lead to logarithmic corrections at least in certain situations \cite{gopalakrishnan2024non}, yet our theory, as an asymptotic $1/\ell$ expansion, does no appear to suffer from any such corrections. Does the no-shock condition imply absence of quartic singularities as well?

Finally, going beyond linearly degenerate systems would be very interesting. It is possible that a regularisation similar to that which we introduced in order to make the fluctuating hydrodynamic theory well defined in its asymptotic expansion in $1/\ell$, be related to the regularisation necessary to make the KPZ equation well defined \cite{hairer2013solving}. We also propose that bare diffusion and hydrodynamic noise are given by similar projection principles even more generally, in one-dimensional systems with relevant nonlinearities, in such a way that, for the hydrodynamic noise, the infinite Onsager matrix (infinity being related to hydrodynamic superdiffusion) is projected to a subspace of the diffusive Hilbert space where it is finite.
 
\medskip

{\bf Acknowledgments.}
I am grateful to Friedrich H\"ubner, Tomohiro Sasamoto and Mayank Sharma for discussions, and especially Takato Yoshimura for exchanging ideas while working on the simultaneous paper \cite{hydrodynamic2025Yoshimura} where related results are obtained. I thank the Institute of Science Tokyo for hospitaly, and the workshop ``Hydrodynamics of low-dimensional interacting systems: Advances, challenges, and future directions", YITP-T-25-03, at the Yukawa Institute for Theoretical Physics, where part of this work was done. This work was supported by EPSRC under grants EP/W010194/1, and EP/Z534304/1 (UK Research and Innovation Horizon Europe Guarantee, Advanced Grant Scheme).

\appendix

\section{Hydrodynamic velocities and normal modes}\label{apphydro}

We recall that derivatives with respect to conserved densities are related to covariances in the {\em macrocanonical} ensemble. For instance, by the chain rule, we find
\beq\label{deroq}
	\frc{\p\mathsf o}{\p \mathsf q_i}
	= \sum_j \mathsf C^{ij} \bra Q_j,o\ket_{\underline \beta}^{\rm c} 
\eeq
where $\mathsf C^{ij}$ is the inverse of the covariance matrix $\mathsf C_{ij}$, that is $\sum_j\mathsf C_{ij}\mathsf C^{jk} = \delta_i^{~k}$,  and likewise
\beq
	\frc{\p^2\mathsf o}{\p \mathsf q_i \p \mathsf q_k}
	= \sum_{jl} \mathsf C^{ij}\mathsf C^{kl} \bra Q_j,Q_l,o^-\ket_{\underline \beta}^{\rm c},\quad
	o^- := o - \sum_{ij} q_i \mathsf C^{ij} \bra Q_j,o\ket_{\underline \beta}^{\rm c}
\eeq
where
\beq\label{3ptdef}
	\bra Q_j,Q_l,o^-\ket^{\rm c}_{\underline{\beta}}
	= \int \dd \rx\dd \rx'
	\bra q_j(\rx),q_l(\rx'),o^-(0)\ket^{\rm c}_{\underline{\beta}}
\eeq
is an integrated three-point connected correlation function in stationary states called the {\em 3-point coupling}.

The flux Jacobian ${\mathsf A}$ is the matrix obtained by differentiating the fluxes (average currents):
\beq\label{Amatrix}
	\frc{\p {\mathsf j}_i}{\p \mathsf q_l} = {\mathsf A}_i^{~l}
	=\sum_k \mathsf C^{lk}\bra Q_k, j_i\ket^{\rm c}_{\underline\beta}.
\eeq
For the Hessian of the current, we denote
\beq
	\mathsf A_i^{~jk} = \frc{\p^2 \mathsf j_i}{\p \mathsf q_j \p\mathsf q_k}.
\eeq

The flux Jacobian satisfies
\beq\label{AC}
	{\mathsf A}\mathsf C = \mathsf C {\mathsf A}^{\rm T}
\eeq
as shown in various contexts and levels of generality in \cite{toth2003onsager,grisi2011current,spohn2014nonlinear,castro2016emergent,de2019diffusion,karevski2019charge,doyon2020lecture,doyon2021free}. This implies that each element of the flux Jacobian vector has real eigenvalues, because the matrix $\sqrt{\mathsf C}^{-1}\mathsf A \sqrt{\mathsf C}$, obtained by a similarity transformation, is real and symmetric. In particular
\beq
	{\rm spec}\,\mathsf A = \{\mathsf v_I\}_I\subseteq \R
\eeq
where we use the capital letter $I$ to indicate that it runs over the {\em normal mode basis}. The quantities $\mathsf v_I$ have the interpretation as the hydrodynamic velocities, or generalised sound velocities, for wave propagation. Using the orthogonal matrix $M$ diagonalising the symmetric matrix $\sqrt{\mathsf C}^{-1}\mathsf A(\h{\v p}) \sqrt{\mathsf C}$, the transformation to the normal modes can be chosen as $\mathsf R = M\sqrt{\mathsf C}^{-1}$:
\beq\label{normalmodes}
	q_I =\sum_j \mathsf R_I^{~j}q_j,\quad
	\mathsf R \mathsf  A \mathsf  R^{-1} = {\rm diag}\,(\mathsf v_I)_I,\quad \mathsf R \mathsf C \mathsf R^{\rm T} = {\bf 1}.
\eeq
In particular, the Hessian in normal modes is
\beq\label{Hessiannormal}
	\mathsf A_{i}^{~JK} = \bra Q_J,Q_K,j_i^-\ket^{\rm c}_{\underline \beta}
	=
	\sum_{jk} (\mathsf R^{-1})_j^{~J}(\mathsf R^{-1})_k^{~K}\frc{\p^2 \mathsf j_i}{\p \mathsf q_j\p\mathsf q_k}.
\eeq
It is convenient to define the third cumulant, which is completely symmetric in its arguments,
\beq
	\mathsf C_{ijk} = \frc{\p \mathsf q_k}{\p\beta^i\p\beta^j}
	= \bra Q_i,Q_j,q_k\ket_{\underline\beta}^{\rm c}
\eeq
and in normal modes this is still a quantity that is generically non-diagonal and non-trivial,
\beq
	\mathsf C_{IJK} = 
	\sum_{ijk} ((\mathsf R^{-1})_i^{~I} (\mathsf R^{-1})_j^{~J}(\mathsf R^{-1})_k^{~K}\mathsf C_{ijk}.
\eeq
Clearly, by definition
\beq
	\mathsf A_i^{~JK} = \mathsf A_i^{~KJ}.
\eeq
It turns out, as shown in our separate work \cite{doyon2025nonlinear}, that when the lower index $k$ is transformed to normal modes, there is a partial symmetry relation involving this index as well:
\beq
	\mathsf A_I^{~JK} - \mathsf A_J^{~IK}
	= (\mathsf v_J-\mathsf v_I)\mathsf C_{IJK},
\eeq
which is a crucial relation for the consistency of the emergent fluctuating theory for 3-point correlation functions.

We refer to {\em velocity-diagonal 3-point couplings} and {\em diagonal 3-point couplings}, respectively, the quantities, here for the currents,
\beq
	\mathsf A_{k}^{~II} \quad \mbox{and}\quad
	\mathsf A_k^{~IJ} \mbox{ for $I,\,J$ such that $\mathsf v_I = \mathsf v_J$,\quad respectively},
\eeq
and more generally for arbitrary observables
\beq
	\bra Q_I,Q_I,o^-\ket^{\rm c}_{\underline \beta}
	\quad \mbox{and}\quad
	\bra Q_I,Q_J,o^-\ket^{\rm c}_{\underline \beta} \mbox{ for $I,\,J$ such that $\mathsf v_I = \mathsf v_J$,\quad respectively}.
\eeq

The Euler scale hydrodynamic equation at large scales $\rx = \ell x,\,\micro t = \ell t$, for local maximal entropy states parametrised by $\mathsf q_i(x,t)$, takes the form
\beq
	\p_t \mathsf q_i(x,t)+ \sum_j\mathsf A_i^{~j}(x,t)\p_x \mathsf q_j(x,t) = 0.
\eeq
In certain cases (but not all), it can be diagonalised by choosing appropriate, generically nonlinear functions $n_I = n_I (\underline{\mathsf q})$: if we can solve $\p n_I / \p \mathsf q_j = \mathsf R_I^{~j}(\underline{\mathsf q})$, we obtain
\beq\label{diageuler}
	\p_t n_I + \mathsf v_I \p_x n_I = 0.
\eeq
The $n_I$'s are {\em Riemann invariants}, or ``nonlinear normal modes''. Note how the linear normal modes $q_I$ differ from $n_I$: the notation $q_I$ means the linear transformation \eqref{normalmodes}, while $n_I$ is a nonlinear function of $\mathsf q_i$'s. At the linearised level the Euler equation can be diagonalised, and gives rise to the following simple Euler-scale correlation functions in the normal mode basis and for coarse-grained observables (see e.g.~\cite{doyon2020lecture}),
\beq\label{2pointeq}
	\bra \overline{q_I}( x, t),\overline{q_J}(0,0)\ket_{\underline{\beta}}^{\rm c} \sim \ell^{-1}
	\delta(x-\mathsf v_I t)\delta_{I,J}.
\eeq
or in the original basis
\beq
	\bra \overline{q_i}( x, t),\overline{q_j}(0,0)\ket_{\underline{\beta}}^{\rm c} \sim \ell^{-1}
	(\delta(x-\mathsf At)\mathsf C)_{ij}.
\eeq

\section{Initial conditions in the long-wavelength state}\label{appPT}

In this section all time coordinates are implicitly set to $t=0$.

Consider the one-point function $\dbra q_i(x)\dket_\ell = \bra\overline{q_i}(x)\ket_\ell$. We first perform the calculation by omitting the fluid-cell mean for clarity. Expanding \eqref{lw} for the one-point function $\bra q_i(\ell x)\ket_\ell$, we get
\beq\label{ghdja}
	\bra q_i(\ell x)\ket_\ell
	= \bra q_i\ket_{\underline{\beta}(x)}
	-\sum_j\frc{\p_x \beta^j(x)}{\ell} \int \dd\ry \,\ry\bra q_i(0),q_j(\ry)\ket^{\rm c}_{\underline{\beta}(x)} + \mathcal O(\ell^{-2}).
\eeq
By PT symmetry and translation invariance, we have
\beqa
	\int \dd\ry \,\ry\bra q_i(0),q_j(\ry)\ket^{\rm c}_{\underline{\beta}(x)}
	&\stackrel{\text{PT symmetry}}=& 
	\int \dd\ry \,\ry\bra q_i(0),q_j(- \ry)\ket^{\rm c}_{\underline{\beta}(x)} \n
	&\stackrel{\text{change }\ry\to-\ry}=& 
	-\int \dd\ry \,\ry\bra q_i(0),q_j(\ry)\ket^{\rm c}_{\underline{\beta}(x)}
	\n &=& 0\label{vanishPT}
\eeqa
showing the first equation of \eqref{initcondq}. The fluid cell mean does not modify the first term in \eqref{ghdja} by translation invariance; while in the second term the integral is modified to
\beq
	\frc1L \int_{-L/2}^{L/2}\dd\rx\,\int \dd\ry \,(\ry+\rx)\bra q_i(0),q_j(\ry)\ket^{\rm c}_{\underline{\beta}(x)}
\eeq
which, by PT symmetry is again zero. Note that it is essential that the fluid cell be {\em balanced}: the argument of the fluid-cell mean is the center of the segment over which we average (otherwise, there are nonzero contributions which are easily calculable).

For the two-point function in \eqref{initcondq}, by a scaling analysis we have in general
\beq
	\dbra q_i(x) q_j(x')\dket_\ell =
	\ell^{-1} \Big(\mathsf C_{ij}(x) + \ell^{-1}\delta^{(1)} \mathsf C_{ij}(x) \Big)\delta(x-x') +
	\ell^{-2}\delta^{(2)} \mathsf C_{ij}(x)  \delta'(x-x') + \mathcal O(\ell^{-3})\no
\eeq
for some regular functions $\delta^{(1,2)} \mathsf C_{ij}(x)$. We extract the $\delta(x-x')$ and $\delta'(x-x')$ by integrating against a slowly-varying smooth function of the microscopic coordinates. We do this on the correlation function in the original state $\bra\cdots\ket_\ell$. Agian we first perform the calculation by omitting the fluid-cell mean for clarify. For smooth $g(y)$, we get
\beqa
	\int \dd \ry \,g(\ry/\ell)\,\bra q_i(\ell x)q_j(\ell x+\ry)\ket^{\rm c}_\ell
	&=&
	\int \dd \ry \,g(\ry/\ell)\,\bra q_i(0)q_j(\ry)\ket^{\rm c}_{\underline{\beta}(x)} \n
	&& -\,\sum_k\frc{\p_x\beta^k(x)}\ell
	\int \dd \ry\dd\micro z \,\micro zg(\ry/\ell)\,\bra q_i(0)q_j(\ry)q_k(\micro z)\ket^{\rm c}_{\underline{\beta}(x)}
	+\mathcal O(\ell^{-2})\n
	&=&
	g(0)\mathsf C_{ij}(x) +\mathcal O(\ell^{-2})	
\eeqa
where for the first term in the parenthesis in  the last step, we have used \eqref{vanishPT}, and for the second term we have used a similar argument: PT-symmetry and the change $\ry,\micro z\longrightarrow -\ry,-\micro z$,
\beq
	\mathsf C_{ijk}^{(1)}(x) =
	\int \dd \ry\dd\micro z \,\micro z\,\bra q_i(0)q_j(\ry)q_k(\micro z)\ket^{\rm c}_{\underline{\beta}(x)}
	=
	-\int \dd \ry\dd\micro z \,\micro z\,\bra q_i(0)q_j(\ry)q_k(\micro z)\ket^{\rm c}_{\underline{\beta}(x)} = 0.
\eeq
Therefore, $\delta^{(1)} \mathsf C_{ij}(x) = \delta^{,2)} \mathsf C_{ij}(x) = 0$, and we recover the second equation of \eqref{initcondq}. The fluid cell mean must be applied to both observables. But again, because it is balanced, the PT-symmetry argument is not broken and the result remains true.

\section{No-shock systems: linear degeneracy}\label{applindeg}

By definition, a hydrodynamic system in $d=1$ is linearly degenerate if each hydrodynamic velocity satisfies \cite{lax2005hyperbolic,ferapontov1991integration,el2011kinetic,pavlov2012generalized,bressan2013hyperbolic}
\beq\label{lindegdef}
	\sum_i (\mathsf R^{-1})_i^{~I}\frc{\p \mathsf v_I}{\p \mathsf q_i} = 0 \quad \forall\;I.
\eeq
If in addition the Riemann invariants $n_I$ exist, then this is
\beq
	\frc{\p \mathsf v_I}{\p n_I} = 0 \quad \forall\;I.
\eeq
In fact, each $\mathsf v_I$ may be said to be linearly degenerate or not \cite{bressan2013hyperbolic}; here we use this terminology to mean that they all are.

For our purpose, the most crucial aspect of linearly degenerate systems is that, in the Euler-scale evolution from inhomogeneous state, no shock are produced \cite{rozlinedeg,liu1979development}, see also \cite[Sec 3.2, 3.3]{bressan2013hyperbolic}. Integrable systems, described by Generalised Hydrodynamics (GHD), possess a continuum of modes $I\to \theta\in\R$, but a certain finite-mode reduction are linearly degenerate \cite{el2011kinetic,pavlov2012generalized}, and in a formal sense even without reduction, they are linearly degenerate, see the lecture notes \cite{doyon2020lecture}. An independent proof of absence of shocks for a family of GHD equations is done in \cite{hubner2024new,hubner2024existence}.

Below, we will not use \eqref{lindegdef}. Instead, we use, in a subtle but crucial way, the absence of shock production. Thus, our results hold for no-shock systems, of which linearly degenerate systems are examples (perhaps the only ones).

\subsection{Proof that the velocity-diagonal 3-point couplings vanish}\label{app3ptcoupling}

Here we reproduce the main lines of the calculation in \cite[Sup Mat]{PhysRevLett.134.187101}, Section ``Degenerate three-point coupling from linear degeneracy''.  As far as we are aware, this calculation is the first proof of the vanishing of the diagonal 3-point coupling in no-shock systems. We provide here more details, rendering slightly more explicit tacit assumptions that were made, and pointing out steps that would require a more detailed analysis of correlation functions decay.

We show that, in no-shock systems,
\beq\label{diagvanish}
	\mathsf A_{k}^{~IJ} = 0\quad \mbox{whenever $\mathsf v_I=\mathsf v_J$}
\eeq
and more generally
\beq\label{3ptoo}
	\bra Q_I,Q_J,o^-\ket^{\rm c}_{\underline \beta}
	=0\quad \mbox{whenever $\mathsf v_I=\mathsf v_J$}
\eeq
for every local observable $o$.

This does not appear to follow from the Euler-scale calculus of linear degeneracy -- that is, it does not follow by direct calcualtion, by taking derivatives and going to the normal mode basis, from the basic definition \eqref{lindegdef}. Instead, it requires us to consider the statistical mechanical model underlying the hydrodynamic equation. Thus, it is an extra constraint on the hydrodynamic equation for it to emerge from a many-body system, in the case where it is linearly degenerate.

\proof
Consider a stationary state
\beq
	\bra\cdots\ket = \bra\cdots\ket_{\underline\beta}
\eeq
and the quantity
\beq
	U(o_1,o_2) = \lim_{T\to\infty}\frc1{T} \int_0^{T}\dd \micro t\int\dd \rx\,
	\bra o_1(\rx,\micro t),o_2^-(0,0)\ket^{\rm c}.
\eeq
We recall the notation
\beq
	o^-(\rx,\micro t) = o(\rx,\micro t) - \sum_{ij} q_i(\rx,\micro t)\mathsf C^{ij} \bra Q_j,o\ket.
\eeq
By the general projection result \cite[Thm 6.1]{doyon2022hydrodynamic}, we have
\beq\label{coo}
	U(o_1,o_2)=0
\eeq
for every local observable $o_1,o_2$.

Consider
\beq
	\dbra q_I(x,0),q_J(y,0),o^-(z,t)\dket^{\rm c}.
\eeq
By the nonlinear projection result for three-point function \cite{doyon2023ballistic} (see also \cite{doyon2025nonlinear}), we have
\beq
	\dbra q_I(x,0),q_J(y,0),o^-(z,t)\dket^{\rm c}
	=
	\sum_{ij}\frc{\p^2 \mathsf o}{\p \mathsf q_i\p \mathsf q_j}
	\dbra q_I(x,0)q_i(z,t)\dket^{\rm c}\,
	\dbra q_J(x,0)q_j(z,t)\dket^{\rm c}.
\eeq
As explained in \cite{doyon2025nonlinear}, with the absence of shock production, projection results hold for all values of space-time. Hence, in no-shock systems, the above holds for all space-time, and we can use the explcit solution for two-point functions. We pass to normal modes using the first equation in \eqref{3ptoo}, obtaining the result first established in \cite[App E]{doyon2022diffusion}:
\beq
	\dbra q_I(x,0),q_J(y,0),o^-(z,t)\dket^{\rm c}
	=
	\ell^{-2}\bra Q_I,Q_J,o^-\ket^{\rm c}
	\delta(z-x-\mathsf v_It)
	\delta(z-y-\mathsf v_Jt).
\eeq
With $q_I^0= q_I-\dbra q_I\dket{\bf 1}$, at $z=0$, and after replacing $t\to-t$, we have a result for the following second cumulant:
\beq
	\dbra q_I^0(x,t) q_J^0(y,t), o^-(0,0)\dket^{\rm c}
	=
	\ell^{-2}\bra Q_I,Q_J,o^-\ket^{\rm c}
	\delta(x-\mathsf v_It)
	\delta(y-\mathsf v_Jt).
\eeq

Re-interpreting this in terms of microscopic coordinates in the full stationary state, this is
\beq\label{qqomicro}
	\bra q_I^0(\micro x,\micro t) q_J^0(\micro y,\micro t), o^-(0,0)\ket^{\rm c}
	=
	\bra Q_I,Q_J,o^-\ket^{\rm c}
	f(\micro x,\micro y,\micro t).
\eeq
Here $f$ is finite and supported on $(\micro x,\micro y)\in [\mathsf v_I \micro t-L,\mathsf v_I \micro t+L]\times [\mathsf v_J \micro t-L,\mathsf v_J \micro t+L]$ for some $L$, plus corrections, outside this region, that vanish ``weakly on large scales'', that is, under integration with slowly-varying smooth functions $g(\rx/\ell,\ry/\ell)$ after taking $\micro t = \ell t$ with $\ell_{\rm micro}\ll L\ll \ell\to\infty$. We denote the total $\ry$ integral, which is finite, as
\beq
	F(\rx,\micro t) := \int \dd\ry\,f(\rx,\ry,\micro t)
\eeq
and note that it integrates to $\int \dd \rx\,F(\rx,\micro t) = 1$.

Now take a function $w(z)$ that is even, bounded, smooth and supported on $[-1,1]$, with $w(0)=1$, and define
\beq
	o'(\rx,\micro t) = \int \dd \ry\,
	w((\ry-\rx)/\ell)q_I^0(\rx,\micro t) q_J^0(\ry,\micro t).
\eeq
For all finite $\ell>0$, this is a local observable, as it has a finite support. Using \eqref{qqomicro} and the information about the support of $f$, we find, for every smooth functions $g$,
\beq
	\int \dd\rx\,g(\rx/\ell) \bra o'(\rx,\ell t),o^-(0,0)\ket^{\rm c} =	\bra Q_I,Q_J,o^-\ket^{\rm c} w((\mathsf v_I-\mathsf v_J)t)\,\int \dd\rx\,g(\rx/\ell) F(\rx,\ell t)
	+ \mathcal o(\ell_{\rm micro}/L,L/\ell).
\eeq
Taking $g=1$ and averaging over $t\in[0,T]$, we therefore obtain
\beq
	\frc1{\ell T} \int_0^{\ell T} \dd \micro t\,
	\int \dd\rx\,\bra o'(\rx,\micro t),o^-(0,0)\ket^{\rm c}
	=
	\bra Q_I,Q_J,o^-\ket^{\rm c} 
	R(T)
	+ \mathcal o(\ell_{\rm micro}/L,L/\ell)
\eeq
where
\beq
	R(T) = \lt\{\ba{ll}
	1
	& (\mathsf v_I=\mathsf v_J)\z
	\frc1{|\mathsf v_I-\mathsf v_J|T}\int_0^{\min(1,|\mathsf v_I-\mathsf v_J|T)} \dd z\,w(z)
	& (\mbox{otherwise}).
	\ea\rt.
\eeq
Taking the limit $\lim_{T\to\infty}$, we get
\beq
	U(o',o )
	=
	\lt\{\ba{ll}
	\bra Q_I,Q_J,o^-\ket^{\rm c} & (\mathsf v_I = \mathsf v_J) \z
	0 & (\mbox{otherwise})
	\ea\rt.
	+\, \mathcal o(\ell_{\rm micro}/L,L/\ell).
\eeq
From the general result \eqref{coo}, and as we can take $L,\,\ell$ as large as desired with $L\ll \ell$,  we have shown the vanishing of the 3-point coupling. \eproof

%\subsection{Evolution equation for the normal-mode transformation matrix}
%
%In this subsection we show a special property of the change of basis matrix $\mathsf R_I^{~j}$. This property does not seem to follow directly from the Euler equation itself, as a hyperbolic equation. Instead, it is an additional requirement on the structure of this equation, that follows from statistical mechanics, i.e.~from the fact that it must emerge from a short-range many-body system. Indeed, the property follows from the vanishing of the diagonal 3-point coupling, which, as we have seen above, is a consequence of short-range correlations in maximal entropy states.
%
%The property is that
%\beq
%	\frc{\p}{\p n_I} \mathsf R_J^{~k} = 0\quad\mbox{whenever}\ 
%	\mathsf v_I\neq \mathsf v_J
%\eeq
%This implies in particular that
%\beq
%	\p_t \mathsf R_I^{~k} + \mathsf v_I \p_x \mathsf R_I^{~k} = 0.
%\eeq

\subsection{Proof that the short-range two-point correlation in a large-wavelength non-equilibrium state is given by the equilibrium covariance matrix}

This proof is based on the calculation found in \cite[Sup Mat]{PhysRevLett.134.187101}, Section ``Dynamics of two-point correlations''. This calculation as it is, is not enough to derive the result we want here, so we complete it. Note that the result was claimed in \cite{doyon2023ballistic} using BMFT for integrable systems, however the poof there was not complete, as the commutant of the flux Jacobian was not considered.

For simplicity we assume that hydrodynamic velocities are non-degenerate $\mathsf v_I\neq \mathsf v_{I'}\,\forall\,I\neq I'$ (for generic states). That is, the hydrodynamic equation is strictly hyperbolic.

Consider the equal-time two-point Euler amplitude $S_{ij}(x,x';t) := \lim_{\ell\to\infty}\ell \dbra q_i(x,t),q_j(x',t)\dket_\ell^{\rm c}$ of conserved densities in a long-wavelength inhomogeneous state, at the Euler scale. The initial condition \eqref{initcondq} is
\beq\label{initcondS}
	S_{ij}(x,x';0) = \mathsf C_{ij}(x,0)\delta(x-x')
\eeq
where, as above, we use the notation $\mathsf C_{ij}(x,t) = \mathsf C_{ij}(\underline{\mathsf q}(x,t))$ for the equilibirum covariance matrix in the state determined by the $\mathsf q_i(x,t)$'s, Eq.~\eqref{Cmatrix}.

We will show that for all times,
\beq
	S_{ij}(x,x';t) = \mathsf C_{ij}(x,t)\delta(x-x') + \mbox{regular}
\eeq

\proof $S_{ij}(x,x';t)$ satisfies the Euler-scale propagation equation
\beq\label{twopointeq}
	\p_t S_{ij}(x,x',t)
	+ \sum_k\p_x \Big(\mathsf A_i^{~k}(x,t) S_{kj}(x,x',t)\Big)
	+ \sum_k\p_{x'} \Big(S_{ik}(x,x',t) \mathsf A_j^{~k}(x',t) \Big) = 0
\eeq
where $\mathsf A(x,t) = \mathsf A(\underline{\mathsf q}(x,t))$ is the flux Jacobian evaluated at the stationary state charactersied by $\mathsf q_i(x,t)$'s. This follows from the BMFT \cite{doyon2023emergence,doyon2023ballistic}. Equivalently, it follows from Euler-scale projections, that is the nonlinear Boltzmann Gibbs principle Eq.~\eqref{NLFBG} without bare diffusion and noise, along with the fact that no shocks are produced, so that the projection holds everywhere in space-time \cite{doyon2025nonlinear}.

The general solution structure is
\beq\label{assumedform}
	S_{ij}(x,x',t) = \t{C}_{ij}(x,t)\delta(x-x') + E_{ij}(x,x',t)
\eeq
where $\t{ C}_{ij}(x,t)$ and $E_{ij}(x,x',t)$ are ordinary functions. In particular $\t{ C}_{ij}(x,t)$ is not assumed to be a function of the local state determined by $\mathsf q_i(x,t)$'s. By integrating $\dbra q_i(x,t),q_j(x',t)\dket_\ell^{\rm c}$ on $x,x'\in[-\ep,\ep]$ and dividing by $\ep$, we obtain a non-negative quantity, the local covariance of conserved quantities. As conserved quantities must be non-degenerate (Sec.~\ref{secconsqties}), the result must in fact be strictly positive, hence $\t C>0$.

In \eqref{twopointeq}, the form \eqref{assumedform} leads to a term proportional to $\delta'(x-x')$, which turns out to have the form \cite[Sup Mat]{PhysRevLett.134.187101} (see Section ``Dynamics of two-point correlations'')
\beq
	\big(\mathsf A(x,t)\t{ C}(x,t) -\t{ C}(x,t)\mathsf A(x,t)^{\rm T}\big)_{ij}\,\delta'(x-x').
\eeq
As this must vanish, we must have $\mathsf A(x,t)\t{ C}(x,t) =\t{ C}(x,t)\mathsf A(x,t)^{\rm T}$. We may always write
\beq\label{tCM}
	\t{ C}(x,t) = \mathsf R(x,t)^{-1} m(x,t) \mathsf R(x,t)^{-\rm T}
\eeq
for some positive matrix $m(x,t)>0$, where $\mathsf R(x,t) = \mathsf R(\underline{\mathsf q}(x,t))$ is the local normal-mode transformation matrix, Eq.~\eqref{normalmodes}. Thus we obtain the condition
\beq
	[{\rm diag}(\mathsf v_I(x,t))_I,m(x,t)]=0.
\eeq
Under the assumption that the hydrodynamic velocities are non-degenerate $\mathsf v_I(x,t)\neq \mathsf v_{I'}(x,t)\,\forall\,I\neq I'$, we must have $m(x,t) = {\rm diag}(m(x,t)_I)_I$ with $m(x,t)_I> 0$. Because of \eqref{initcondS}, we also have
\beq\label{initcondm}
	m(x,0)_I = 1 \quad \forall \ I.
\eeq

%where $m(x,t) = {\rm diag}(m(x,t)_j)_j$ is diagonal with $m(x,t)_j\neq 0$. Because of \eqref{initcondS}, we have
%\beq\label{initcondm}
%	m(x,0)_j = 1 \quad \forall \ j.
%\eeq
%\beq\label{tCM}
%	\t{ C}(x,t) = M(x,t)\mathsf C(x,t)M(x,t)^{\rm T}
%\eeq
%and
%\beq
%	M(x,t) = \mathsf R(x,t)^{-1} m(x,t)\, \mathsf R(x,t)
%\eeq
%
%for $M(x,t)$ in the commutant of $\mathsf A(x,t)$, where $\mathsf C(x,t) = \mathsf C(\underline{\mathsf q}(x,t))$ is the local covariance matrix. Note that \eqref{tCM} guarantees $\t C(x,t)>0$ as long as $M(x,t)$ does not have any left eigenvector with eigenvalue 0. Under the assumption that the hydrodynamic velocities are non-degenerate $\mathsf v_I(x,t)\neq \mathsf v_{I'}(x,t)\,\forall\,I\neq I'$, the matrix $M(x,t)$ must have the form
%\beq
%	M(x,t) = \mathsf R(x,t)^{-1} m(x,t)\, \mathsf R(x,t)
%\eeq
%where $m(x,t) = {\rm diag}(m(x,t)_j)_j$ is diagonal with $m(x,t)_j\neq 0$. Because of \eqref{initcondS}, we have
%\beq\label{initcondm}
%	m(x,0)_j = 1 \quad \forall \ j.
%\eeq

For lightness of notation, from now on in this proof we do not write the explicit $x,t$ dependence of matrices. Let us construct the matrix $M(x,t)$ as
\beq
	M(x,t) = \mathsf R(x,t)^{-1} m(x,t)\, \mathsf R(x,t).
\eeq
hence we have
\beq
	\t C = \mathsf R^{-1}m\,\mathsf R^{-\rm T}=
	M \mathsf C = \mathsf C M^{\rm T}.
\eeq
%
%\beq
%	\t C = \mathsf R^{-1}m^2\mathsf c\,\mathsf R^{-\rm T}.
%\eeq
%\beq
%	\mathsf A\t C = \mathsf R^{-1}\mathsf v m^2 \mathsf c\,\mathsf R^{-\rm T}
%\eeq
%
The term proportional to $\delta(x-x')$ coming from $\t{ C}_{ij}\delta(x-x')$ is
\beq\label{termdelta}
	\big(\p_t \t{ C}+ \p_x (\mathsf A\t{ C})\big)_{ij}
	\delta(x-x').
\eeq
It is shown in \cite[Sup Mat]{PhysRevLett.134.187101} that, with the standard definition $\mathsf B:=\mathsf A\mathsf C$,
\beq
	\big(\p_t \mathsf C + \p_x \mathsf B\big)_{ij}
	= \bra Q_i,Q_j,j_k^-\ket^{\rm c} \mathsf C^{kl}\p_x\mathsf q_l
\eeq
thus the term \eqref{termdelta} is
\beq
	\Big(\mathsf C\p_tM^{\rm T} + \mathsf B\p_x M^{\rm T}  + \sum_{kl}\bra Q_\cdot,Q_\cdot,j_k^-\ket^{\rm c} \mathsf C^{kl}\p_x\mathsf q_l\Big)_{ij}\,
	\delta(x-x').
\eeq
So we must have
\beqa
	&& \p_t E_{ij}
	+ \sum_k\p_x \Big(\mathsf A_i^{~k} E_{kj}\Big)
	+ \sum_k\p_{x'} \Big(E_{ik} \mathsf A_j^{~k} \Big)\n
	&& = -\,
	\Big(\mathsf C\p_tM^{\rm T} + \mathsf B\p_x M^{\rm T}   + \sum_{kl}\bra Q_\cdot,Q_\cdot,j_k^-\ket^{\rm c} \mathsf C^{kl}\p_x\mathsf q_l\Big)_{ij}\,
	\delta(x-x').
\eeqa

On the left-hand side, only derivatives in $x,x'$ may produce a delta-function $\delta(x-x')$, and this, because of discontinuities. These can only come from $E_{ij}(x,x',t)$, and using the basis of derivatives $\p_\pm = \p_x\pm\p_{x'}$, can only come from $\p_- E_{ij}$. Passing to normal modes, and taking into consideration that neither $\mathsf A$ nor  the transformation matrix $\mathsf R$ can produce delta-functions $\delta(x-x')$, we get
\beq
	\mathsf v_I\p_x E_{IJ}
	+ \mathsf v_J\p_{x'} E_{IJ} \sim -
	\Big((\mathsf R \p_t M\mathsf R^{-1} + \mathsf v_I\mathsf R \p_x M\mathsf R^{-1})_{J}^{~I} + \sum_{kl}\bra Q_I,Q_J,j_k^-\ket^{\rm c}\mathsf C^{kl}\p_x \mathsf q_l\Big)\,
	\delta(x-x')
\eeq
For $I=J$, and using the fact that $\bra Q_I,Q_J,j_K^-\ket^{\rm c}=0$ whenever $\mathsf v_I = \mathsf v_J$ (Sec.~\eqref{app3ptcoupling}), we obtain
\beq
	\mathsf v_I (\p_x+\p_{x'}) E_{II}
	\sim
	-
	 (\mathsf R \p_t M\mathsf R^{-1} + \mathsf v_I 
	\mathsf R \p_x M\mathsf R^{-1})_{I}^{~I}\,
	\delta(x-x').
\eeq
As the left-hand side cannot produce $\delta(x-x')$, the coefficient of the delta function on the right-hand side must vanish. This is
\beq
	\Big(\p_t m +\mathsf v_I \p_x m
	+
	[m ,(\p_t \mathsf R + \mathsf v_I\p_x \mathsf R) \mathsf R^{-1}]
	\Big)_{I}^{~I}=0.
\eeq
Because $m$ is diagonal, the second term, the commutator, evaluated on the diagonal $(\cdot)_I^{~I}$, vanishes. Thus we have
\beq\label{eqm}
	\p_t m_I + \mathsf v_I\p_x m_I=0\ \forall\, I.
\eeq
Eq.~\eqref{eqm} says that $m_I$ is transported along the characteristics of the fluid. Because there are no shocks, all characteristics come from the initial time. Because of the initial condition \eqref{initcondm}, we then have
\beq
	m_I(x,t)=1.
\eeq
\eproof

\section{Drude weight from nonlinear fluctuating Boltzmann-Gibbs principle}\label{appdrude}

Recall that as a general theorem \cite{doyon2022hydrodynamic}, the Drude weight is given by the projection onto the conserved quantities:
\beq\label{drudeproj}
	\mathsf D_{o_1,o_2}=\frc{\p \mathsf o_1}{\p \mathsf q_{i_1}} \frc{\p \mathsf o_2}{\p \mathsf q_{i_2}}
	\mathsf C_{i_1,i_2}.
\eeq
It is instructive to verify that our formalism is in agreement with the general result \eqref{drudeproj}, by evaluating the Drude weight using the nonlinear fluctuating Boltzmann-Gibbs principle \eqref{NLFBG}.

Consider the definition \eqref{drudedef} written in terms of macrosocpic coordinates for fluid-cell means in the fluctuation theory,
\beq
	\mathsf D_{o_1,o_2} = \lim_{t\to\pm\infty} \ell \int \dd x\,\dbra o_1(x,t),o_2(0,0)\dket^{\rm c}.
\eeq
Consider $\dbra o_1( x_1, t_1),o_2( x_2, t_2)\dket^{\rm c} $, which depends on $x_1-x_2$ and $t_1-t_2$ by space-time translation invariance. We may integrate over $x_1$, or integrate over $x_2$, and we look at the limit $t_1-t_2\to\pm\infty$. Clearly, the noise-boise term \eqref{noisenoise} gives zero under the limit $t_1-t_2\to\pm\infty$, and by the results of Sec.~\ref{ssectnoise} all other correlations vanish except for that from the microcanonical terms in \eqref{NLFBG}, that is \eqref{o1o2}.

For the first line on the right-hand side of \eqref{o1o2}, we obtain \eqref{drudeproj}. Hence, we need to verify that the other lines give zero.

For the last line on the right-hand side of \eqref{o1o2}, we integrate over $x_1$ and notice that the result is independent of $t_1,t_2$:
\beqa
	\lefteqn{\int \dd x_1\,
	\dbra q_{i_1}(x_1,t_1),q_{i_2}(x_2\pm 0^+,t_2),q_{i_2'}(x_2\mp 0^+,t_2)\dket^{\rm c}} &&\\
	&=& \dbra Q_1,q_{i_2}(x_2\pm 0^+,t_2),q_{i_2'}(x_2\mp 0^+,t_2)\dket^{\rm c} = -\frc{\p}{\p\beta^1}\dbra q_{i_2}(\pm 0^+,0),q_{i_2'}(\mp 0^+,0)\dket^{\rm c} = 0\no
\eeqa
where we use the fact that the state has short-range correlations. Similarly, we obtain 0 for the third line on the right-hand side of \eqref{o1o2} by integrating over $x_2$.

For the second line on the right-hand side of \eqref{o1o2}, we set $x_1=x,t_1=t$ and $x_2=t_2=0$, and, as in Sec.~\ref{ssectnoise}, we pass to normal modes using
\beq
	\frc{\p^2 \mathsf o_1}{\p \mathsf q_{i_1}\p \mathsf q_{i_1'}}
	=
	\bra o_1^-,Q_{j_1},Q_{j_1'}\ket^{\rm c}
	\mathsf C^{j_1,i_1}\mathsf C^{j_1',i_1'}
\eeq
and the normal-mode transformation \eqref{normalmodes}, to obtain
\beq
	\bra o_1^-,Q_{I_1},Q_{I_1'}\ket^{\rm c}
	\bra o_2^-,Q_{I_2},Q_{I_2'}\ket^{\rm c}
	\dbra q_{I_1}(x,t),q_{I_2}(0,0)\dket^{\rm c}
	\dbra q_{I_1'}(x,t),q_{I_2'}(0,0)\dket^{\rm c}.
\eeq
This can be evaluated exactly to its leading $\mathcal O(\ell^{-2})$ order by using the Euler-scale formula for the two-point functions
\beq\label{blabla}
	\ell^{-2}\frc12\bra o_1^-,Q_{I},Q_{I'}\ket^{\rm c}
	\bra o_2^-,Q_{I},Q_{I'}\ket^{\rm c}
	\delta(x-\mathsf v_{I}t)
	\delta(x-\mathsf v_{I'}t).
\eeq
where we used \eqref{proj} for the projected observable $o^-$. Clearly, the integral over $x$ is 0 if for every values of the indices $I,I'$ we have either $\mathsf v_I \neq \mathsf v_{I'}$, or  $\bra o_1^-,Q_{I},Q_{I'}\ket^{\rm c}=0$, or $\bra o_2^-,Q_{I},Q_{I'}\ket^{\rm c}$. As we assume linear degeneracy, \eqref{3point} holds, hence the result vanishes.

%\bibliographystyle{unsrt} 
%\bibliographystyle{unsrturl} 
%\bibliography{bib}

%correlations for transport of initial fluctuations give non-gaussian parts of fluctuations
%
%same general formula for free cumulants?

%\bibliography{bib.bib}

\begin{thebibliography}{10}

\bibitem{mori1965transport}
H.~Mori.
\newblock Transport, collective motion, and brownian motion.
\newblock {\em Progress of theoretical physics}, 33(3):423--455, 1965.
\newblock \href {http://dx.doi.org/10.1143/PTP.33.423}
  {\path{doi:10.1143/PTP.33.423}}.

\bibitem{zwanzig1966statistical}
R.~W. Zwanzig.
\newblock Statistical mechanics of irreversibility.
\newblock In W.~E. Britten, B.~W. Downs, and J.~Downs, editors, {\em Lectures
  in Theoretical Physics III}, page 106. Interscience, Boulder, 1961.

\bibitem{brox1984equilibrium}
Th. Brox and H.~Rost.
\newblock Equilibrium fluctuations of stochastic particle systems: the role of
  conserved quantities.
\newblock {\em The Annals of Probability}, 12(3):742--759, 1984.
\newblock \href {http://dx.doi.org/10.1214/aop/1176993225}
  {\path{doi:10.1214/aop/1176993225}}.

\bibitem{demasi2006mathematical}
A.~DeMasi and E.~Presutti.
\newblock {\em Mathematical methods for hydrodynamic limits}.
\newblock Springer, Berlin, Heidelberg, 1991.
\newblock \href {http://dx.doi.org/10.1007/BFb0086457}
  {\path{doi:10.1007/BFb0086457}}.

\bibitem{spohn2012large}
H.~Spohn.
\newblock {\em Large scale dynamics of interacting particles}.
\newblock Springer Science \& Business Media, Berlin, Heidelberg, 1991.
\newblock \href {http://dx.doi.org/10.1007/978-3-642-84371-6}
  {\path{doi:10.1007/978-3-642-84371-6}}.

\bibitem{kipnis2013scaling}
C.~Kipnis and C.~Landim.
\newblock {\em Scaling limits of interacting particle systems}.
\newblock Springer, Berlin, Heidelberg, 1999.
\newblock \href {http://dx.doi.org/10.1007/978-3-662-03752-2}
  {\path{doi:10.1007/978-3-662-03752-2}}.

\bibitem{doyon2023emergence}
Benjamin Doyon, Gabriele Perfetto, Tomohiro Sasamoto, and Takato Yoshimura.
\newblock Emergence of hydrodynamic spatial long-range correlations in
  nonequilibrium many-body systems.
\newblock {\em Physical review letters}, 131(2):027101, 2023.

\bibitem{doyon2023ballistic}
B.~Doyon, G.~Perfetto, T.~Sasamoto, and T.~Yoshimura.
\newblock Ballistic macroscopic fluctuation theory.
\newblock {\em SciPost Physics}, 15(4):136, 2023.
\newblock \href {http://dx.doi.org/10.21468/SciPostPhys.15.4.136}
  {\path{doi:10.21468/SciPostPhys.15.4.136}}.

\bibitem{de2022correlation}
J.~De~Nardis, B.~Doyon, M.~Medenjak, and M.~Panfil.
\newblock Correlation functions and transport coefficients in generalised
  hydrodynamics.
\newblock {\em Journal of Statistical Mechanics: Theory and Experiment},
  2022(1):014002, 2022.
\newblock \href {http://dx.doi.org/10.1088/1742-5468/ac3658}
  {\path{doi:10.1088/1742-5468/ac3658}}.

\bibitem{wienand2024emergence}
J.~F. Wienand, S.~Karch, A.~Impertro, C.~Schweizer, E.~McCulloch, R.~Vasseur,
  S.~Gopalakrishnan, M.~Aidelsburger, and I.~Bloch.
\newblock Emergence of fluctuating hydrodynamics in chaotic quantum systems.
\newblock {\em Nature Physics}, 20(11):1732--1737, 2024.
\newblock \href {http://dx.doi.org/10.1038/s41567-024-02611-z}
  {\path{doi:10.1038/s41567-024-02611-z}}.

\bibitem{bertini2015macroscopic}
L.~Bertini, A.~De~Sole, D.~Gabrielli, G.~Jona-Lasinio, and C.~Landim.
\newblock Macroscopic fluctuation theory.
\newblock {\em Reviews of Modern Physics}, 87(2):593--636, 2015.
\newblock \href {http://dx.doi.org/10.1103/RevModPhys.87.593}
  {\path{doi:10.1103/RevModPhys.87.593}}.

\bibitem{zwanzig1961memory}
R.~Zwanzig.
\newblock Memory effects in irreversible thermodynamics.
\newblock {\em Physical Review}, 124(4):983, 1961.
\newblock \href {http://dx.doi.org/10.1103/PhysRev.124.983}
  {\path{doi:10.1103/PhysRev.124.983}}.

\bibitem{mori1973nonlinear}
H.~Mori and H.~Fujisaka.
\newblock On nonlinear dynamics of fluctuations.
\newblock {\em Progress of Theoretical Physics}, 49(3):764--775, 1973.
\newblock \href {http://dx.doi.org/10.1143/PTP.49.764}
  {\path{doi:10.1143/PTP.49.764}}.

\bibitem{kawasaki1973simple}
K.~Kawasaki.
\newblock Simple derivations of generalized linear and nonlinear langevin
  equations.
\newblock {\em Journal of Physics A: Mathematical, Nuclear and General},
  6(9):1289, 1973.
\newblock \href {http://dx.doi.org/10.1088/0305-4470/6/9/004}
  {\path{doi:10.1088/0305-4470/6/9/004}}.

\bibitem{fujisaka1976fluctuation}
H.~Fujisaka.
\newblock Fluctuation renormalization in nonlinear dynamics.
\newblock {\em Progress of Theoretical Physics}, 55(2):430--437, 1976.
\newblock \href {http://dx.doi.org/10.1143/PTP.55.430}
  {\path{doi:10.1143/PTP.55.430}}.

\bibitem{zubarev1983statistical}
D.~N. Zubarev and V.~G. Morozov.
\newblock Statistical mechanics of nonlinear hydrodynamic fluctuations.
\newblock {\em Physica A: Statistical Mechanics and its Applications},
  120(3):411--467, 1983.
\newblock \href {http://dx.doi.org/10.1016/0378-4371(83)90062-6}
  {\path{doi:10.1016/0378-4371(83)90062-6}}.

\bibitem{donev2011diffusive}
A.~Donev, J.~B. Bell, A.~de~La~Fuente, and A.~L. Garcia.
\newblock Diffusive transport by thermal velocity fluctuations.
\newblock {\em Physical review letters}, 106(20):204501, 2011.
\newblock \href {http://dx.doi.org/10.1103/PhysRevLett.106.204501}
  {\path{doi:10.1103/PhysRevLett.106.204501}}.

\bibitem{donev2011enhancement}
A.~Donev, J.~B. Bell, A.~De~la Fuente, and A.~L. Garcia.
\newblock Enhancement of diffusive transport by non-equilibrium thermal
  fluctuations.
\newblock {\em Journal of Statistical Mechanics: Theory and Experiment},
  2011(06):P06014, 2011.
\newblock \href {http://dx.doi.org/10.1088/1742-5468/2011/06/P06014}
  {\path{doi:10.1088/1742-5468/2011/06/P06014}}.

\bibitem{nakano2025looking}
H.~Nakano, Y.~Minami, and K.~Saito.
\newblock Looking at bare transport coefficients in fluctuating hydrodynamics.
\newblock {\em arXiv preprint arXiv:2502.15241}, 2025.
\newblock \href {http://dx.doi.org/10.48550/arXiv.2502.15241}
  {\path{doi:10.48550/arXiv.2502.15241}}.

\bibitem{saito2021microscopic}
K.~Saito, M.~Hongo, A.~Dhar, and S.~Sasa.
\newblock Microscopic theory of fluctuating hydrodynamics in nonlinear
  lattices.
\newblock {\em Physical Review Letters}, 127(1):010601, 2021.
\newblock \href {http://dx.doi.org/10.1103/PhysRevLett.127.010601}
  {\path{doi:10.1103/PhysRevLett.127.010601}}.

\bibitem{bandak2024spontaneous}
D.~Bandak, A.~A. Mailybaev, G.~L. Eyink, and N.~Goldenfeld.
\newblock Spontaneous stochasticity amplifies even thermal noise to the largest
  scales of turbulence in a few eddy turnover times.
\newblock {\em Physical Review Letters}, 132(10):104002, 2024.
\newblock \href {http://dx.doi.org/10.1103/PhysRevLett.132.104002}
  {\path{doi:10.1103/PhysRevLett.132.104002}}.

\bibitem{cheskidov2023dissipation}
A.~Cheskidov.
\newblock Dissipation anomaly and anomalous dissipation in incompressible fluid
  flows.
\newblock {\em arXiv preprint arXiv:2311.04182}, 2023.
\newblock \href {http://dx.doi.org/10.48550/arXiv.2311.04182}
  {\path{doi:10.48550/arXiv.2311.04182}}.

\bibitem{spohn2014nonlinear}
H.~Spohn.
\newblock Nonlinear fluctuating hydrodynamics for anharmonic chains.
\newblock {\em Journal of Statistical Physics}, 154:1191--1227, 2014.
\newblock \href {http://dx.doi.org/10.1007/s10955-014-0933-y}
  {\path{doi:10.1007/s10955-014-0933-y}}.

\bibitem{doyon2022diffusion}
B.~Doyon.
\newblock Diffusion and superdiffusion from hydrodynamic projections.
\newblock {\em Journal of Statistical Physics}, 186(2):25, 2022.
\newblock \href {http://dx.doi.org/10.1007/s10955-021-02863-6}
  {\path{doi:10.1007/s10955-021-02863-6}}.

\bibitem{PhysRevLett.134.187101}
F.~H\"ubner, L.~Biagetti, J.~De~Nardis, and B.~Doyon.
\newblock Diffusive hydrodynamics from long-range correlations.
\newblock {\em Physical Review Letters}, 134:187101, 2025.
\newblock \href {http://dx.doi.org/10.1103/PhysRevLett.134.187101}
  {\path{doi:10.1103/PhysRevLett.134.187101}}.

\bibitem{lax2005hyperbolic}
P.~D. Lax.
\newblock Hyperbolic systems of conservation laws ii.
\newblock {\em Communications in Pure and Applied Mathematics}, 10:537--566,
  1957.
\newblock \href {http://dx.doi.org/10.1002/cpa.3160100406}
  {\path{doi:10.1002/cpa.3160100406}}.

\bibitem{ferapontov1991integration}
E.~V. Ferapontov.
\newblock Integration of weakly nonlinear hydrodynamic systems in riemann
  invariats.
\newblock {\em Physics Letters A}, 158(3-4):112--118, 1991.
\newblock \href {http://dx.doi.org/10.1016/0375-9601(91)90910-Z}
  {\path{doi:10.1016/0375-9601(91)90910-Z}}.

\bibitem{el2011kinetic}
G.~A. El, A.~M. K., M.~V. Pavlov, and S.~A. Zykov.
\newblock Kinetic equation for a soliton gas and its hydrodynamic reductions.
\newblock {\em Journal of Nonlinear Science}, 21:151--191, 2011.
\newblock \href {http://dx.doi.org/10.1007/s00332-010-9080-z}
  {\path{doi:10.1007/s00332-010-9080-z}}.

\bibitem{pavlov2012generalized}
M.~V. Pavlov, V.~B. Taranov, and G.~A. El.
\newblock Generalized hydrodynamic reductions of the kinetic equation for a
  soliton gas.
\newblock {\em Theoretical and Mathematical Physics}, 171:675--682, 2012.
\newblock \href {http://dx.doi.org/10.1007/s11232-012-0064-z}
  {\path{doi:10.1007/s11232-012-0064-z}}.

\bibitem{bressan2013hyperbolic}
A.~Bressan.
\newblock Hyperbolic conservation laws: an illustrated tutorial.
\newblock In B.~Piccoli and M.~Rascle, editors, {\em Modelling and Optimisation
  of Flows on Networks, Lecture Notes in Mathematics: Cetraro, Italy 2009},
  pages 157--245. Springer, Heidelberg New York Dordrecht London, 2013.
\newblock \href {http://dx.doi.org/10.1007/978-3-642-32160-3_2}
  {\path{doi:10.1007/978-3-642-32160-3_2}}.

\bibitem{rozlinedeg}
B.~L. Rozdestvenskii and A.~D. Sidorenko.
\newblock On the impossibility of ‘gradient catastrophe’ for slightly
  nonlinear systems.
\newblock {\em USSR Computational Mathematics and Mathematical Physics}, 77,
  1967.
\newblock \href {http://dx.doi.org/10.1016/0041-5553(67)90105-X}
  {\path{doi:10.1016/0041-5553(67)90105-X}}.

\bibitem{liu1979development}
T.-P. Liu.
\newblock Development of singularities in the nonlinear waves for quasi-linear
  hyperbolic partial differential equations.
\newblock {\em Journal of Differential Equations}, 33(1):92--111, 1979.
\newblock \href {http://dx.doi.org/10.1016/0022-0396(79)90082-2}
  {\path{doi:10.1016/0022-0396(79)90082-2}}.

\bibitem{doyon2020lecture}
B.~Doyon.
\newblock Lecture notes on generalised hydrodynamics.
\newblock {\em SciPost Physics Lecture Notes}, page 018, 2020.
\newblock \href {http://dx.doi.org/10.21468/SciPostPhysLectNotes.18}
  {\path{doi:10.21468/SciPostPhysLectNotes.18}}.

\bibitem{hubner2024new}
F.~H{\"u}bner and B.~Doyon.
\newblock Quadrature for the generalized hydrodynamics equation and absence of
  shocks in the lieb-liniger model.
\newblock {\em Physical Review B}, 112:L241101, 2024.
\newblock \href {http://dx.doi.org/10.1103/5w4p-bxnt}
  {\path{doi:10.1103/5w4p-bxnt}}.

\bibitem{hubner2024existence}
F.~H{\"u}bner and B.~Doyon.
\newblock Existence and uniqueness of solutions to the generalized
  hydrodynamics equation.
\newblock {\em arXiv preprint arXiv:2411.04922}, 2024.
\newblock \href {http://dx.doi.org/10.48550/arXiv.2411.04922}
  {\path{doi:10.48550/arXiv.2411.04922}}.

\bibitem{gopalakrishnan2024non}
S.~Gopalakrishnan, E.~McCulloch, and R.~Vasseur.
\newblock Non-gaussian diffusive fluctuations in dirac fluids.
\newblock {\em Proceedings of the National Academy of Sciences},
  121(50):e2403327121, 2024.
\newblock \href {http://dx.doi.org/10.1073/pnas.2403327121}
  {\path{doi:10.1073/pnas.2403327121}}.

\bibitem{yoshimura2025anomalous}
T.~Yoshimura and Z.~Krajnik.
\newblock Anomalous current fluctuations from euler hydrodynamics.
\newblock {\em Physical Review E}, 111(2):024141, 2025.
\newblock \href {http://dx.doi.org/10.1103/PhysRevE.111.024141}
  {\path{doi:10.1103/PhysRevE.111.024141}}.

\bibitem{mcculloch2025ballistic}
E.~McCulloch, R.~Vasseur, and S.~Gopalakrishnan.
\newblock Ballistic modes as a source of anomalous charge noise.
\newblock {\em Physical Review E}, 111(1):015410, 2025.
\newblock \href {http://dx.doi.org/10.1103/PhysRevE.111.015410}
  {\path{doi:10.1103/PhysRevE.111.015410}}.

\bibitem{hydrodynamic2025Yoshimura}
T.~Yoshimura and Z.~Krajnik.
\newblock Hydrodynamic fluctuations of stochastic charged cellular automata.
\newblock {\em Journal of Statistical Mechanics: Theory and Experiment},
  2025:053209, 2025.
\newblock \href {http://dx.doi.org/10.1088/1742-5468/add513}
  {\path{doi:10.1088/1742-5468/add513}}.

\bibitem{de2018hydrodynamic}
J.~De~Nardis, D.~Bernard, and B.~Doyon.
\newblock Hydrodynamic diffusion in integrable systems.
\newblock {\em Physical Review Letters}, 121(16):160603, 2018.
\newblock \href {http://dx.doi.org/10.1103/PhysRevLett.121.160603}
  {\path{doi:10.1103/PhysRevLett.121.160603}}.

\bibitem{de2019diffusion}
J.~De~Nardis, D.~Bernard, and B.~Doyon.
\newblock Diffusion in generalized hydrodynamics and quasiparticle scattering.
\newblock {\em SciPost Physics}, 6(4):049, 2019.
\newblock \href {http://dx.doi.org/10.21468/SciPostPhys.6.4.049}
  {\path{doi:10.21468/SciPostPhys.6.4.049}}.

\bibitem{gopalakrishnan2018hydrodynamics}
S.~Gopalakrishnan, D.~A. Huse, V.~Khemani, and R.~Vasseur.
\newblock Hydrodynamics of operator spreading and quasiparticle diffusion in
  interacting integrable systems.
\newblock {\em Physical Review B}, 98(22):220303, 2018.
\newblock \href {http://dx.doi.org/10.1103/PhysRevB.98.220303}
  {\path{doi:10.1103/PhysRevB.98.220303}}.

\bibitem{doyon2025nonlinear}
B.~Doyon.
\newblock Nonlinear projection for ballistic correlation functions: a formula
  in terms of minimal connected covers.
\newblock {\em arXiv preprint arXiv:2506.05266}, 2025.
\newblock \href {http://dx.doi.org/10.48550/arXiv.2506.05266}
  {\path{doi:10.48550/arXiv.2506.05266}}.

\bibitem{medenjak2020diffusion}
M.~Medenjak, J.~De~Nardis, and T.~Yoshimura.
\newblock Diffusion from convection.
\newblock {\em SciPost Physics}, 9(5):075, 2020.
\newblock \href {http://dx.doi.org/10.21468/SciPostPhys.9.5.075}
  {\path{doi:10.21468/SciPostPhys.9.5.075}}.

\bibitem{boldrighini1997one}
C.~Boldrighini and Y.~M. Suhov.
\newblock One-dimensional hard-rod caricature of
  hydrodynamics:“navier--stokes correction” for local equilibrium initial
  states.
\newblock {\em Communications in Mathematical Physics}, 189(2):577--590, 1997.
\newblock \href {http://dx.doi.org/10.1007/s002200050218}
  {\path{doi:10.1007/s002200050218}}.

\bibitem{hubner2025hydrodynamics}
F.~H{\"u}bner.
\newblock Hydrodynamics without averaging--a hard rods study.
\newblock {\em SciPost Physics}, 20:081, 2026.
\newblock \href {http://dx.doi.org/10.21468/SciPostPhys.20.3.081}
  {\path{doi:10.21468/SciPostPhys.20.3.081}}.

\bibitem{bratteli1987operator}
O.~Bratteli and D.~W. Robinson.
\newblock {\em Operator algebras and quantum statistical mechanics I: C*-and
  W*-Algebras. Symmetry Groups. Decomposition of States}.
\newblock Springer, Berlin, Heidelberg, 1987.
\newblock \href {http://dx.doi.org/10.1007/978-3-662-02520-8}
  {\path{doi:10.1007/978-3-662-02520-8}}.

\bibitem{bratteli1997operator}
O.~Bratteli and D.~W. Robinson.
\newblock {\em Operator algebras and quantum statistical mechanics II:
  Equilibrium States Models in Quantum Statistical Mechanics}.
\newblock Springer, Berlin, Heidelberg, 1997.
\newblock \href {http://dx.doi.org/10.1007/978-3-662-03444-6}
  {\path{doi:10.1007/978-3-662-03444-6}}.

\bibitem{chaikin1995principles}
P.~M. Chaikin, T.~C. Lubensky, and T.~A. Witten.
\newblock {\em Principles of Condensed Matter Physics}.
\newblock Cambridge University Press, Cambridge, 1995.
\newblock \href {http://dx.doi.org/10.1017/CBO9780511813467}
  {\path{doi:10.1017/CBO9780511813467}}.

\bibitem{aizenman1978conditional}
M.~Aizenman, S.~Goldstein, and J.~L. Lebowitz.
\newblock Conditional equilibrium and the equivalence of microcanonical and
  grandcanonical ensembles in the thermodynamic limit.
\newblock {\em Communications in Mathematical Physics}, 62(3):279--302, 1978.
\newblock \href {http://dx.doi.org/10.1007/BF01202528}
  {\path{doi:10.1007/BF01202528}}.

\bibitem{brandao2015equivalence}
F.~G. S.~L. Brandao and M.~Cramer.
\newblock Equivalence of statistical mechanical ensembles for non-critical
  quantum systems.
\newblock {\em arXiv preprint arXiv:1502.03263}, 2015.
\newblock \href {http://dx.doi.org/10.48550/arXiv.1502.03263}
  {\path{doi:10.48550/arXiv.1502.03263}}.

\bibitem{doyon2022hydrodynamic}
B.~Doyon.
\newblock Hydrodynamic projections and the emergence of linearised euler
  equations in one-dimensional isolated systems.
\newblock {\em Communications in Mathematical Physics}, 391(1):293--356, 2022.
\newblock \href {http://dx.doi.org/10.1007/s00220-022-04310-3}
  {\path{doi:10.1007/s00220-022-04310-3}}.

\bibitem{touchette2015equivalence}
H.~Touchette.
\newblock Equivalence and nonequivalence of ensembles: Thermodynamic,
  macrostate, and measure levels.
\newblock {\em Journal of Statistical Physics}, 159(5):987--1016, 2015.
\newblock \href {http://dx.doi.org/10.1007/s10955-015-1212-2}
  {\path{doi:10.1007/s10955-015-1212-2}}.

\bibitem{gopalakrishnan2024distinct}
S.~Gopalakrishnan, A.~Morningstar, R.~Vasseur, and V.~Khemani.
\newblock Distinct universality classes of diffusive transport from full
  counting statistics.
\newblock {\em Physical Review B}, 109(2):024417, 2024.
\newblock \href {http://dx.doi.org/10.1103/PhysRevB.109.024417}
  {\path{doi:10.1103/PhysRevB.109.024417}}.

\bibitem{krajnik2022exact}
Z.~Krajnik, J.~Schmidt, V.~Pasquier, E.~Ilievski, and T.~Prosen.
\newblock Exact anomalous current fluctuations in a deterministic interacting
  model.
\newblock {\em Physical Review Letters}, 128(16):160601, 2022.
\newblock \href {http://dx.doi.org/10.1103/PhysRevLett.128.160601}
  {\path{doi:10.1103/PhysRevLett.128.160601}}.

\bibitem{doyon2018exact}
B.~Doyon.
\newblock Exact large-scale correlations in integrable systems out of
  equilibrium.
\newblock {\em SciPost Physics}, 5(5):054, 2018.
\newblock \href {http://dx.doi.org/10.21468/SciPostPhys.5.5.054}
  {\path{doi:10.21468/SciPostPhys.5.5.054}}.

\bibitem{klimontovich1990ito}
Y.~L. Klimontovich.
\newblock Ito, stratonovich and kinetic forms of stochastic equations.
\newblock {\em Physica A: Statistical Mechanics and its Applications},
  163(2):515--532, 1990.
\newblock \href {http://dx.doi.org/10.1016/0378-4371(90)90142-F}
  {\path{doi:10.1016/0378-4371(90)90142-F}}.

\bibitem{sokolov2010ito}
I.~M. Sokolov.
\newblock Ito, stratonovich, h{\"a}nggi and all the rest: The thermodynamics of
  interpretation.
\newblock {\em Chemical Physics}, 375(2-3):359--363, 2010.
\newblock \href {http://dx.doi.org/10.1016/j.chemphys.2010.07.024}
  {\path{doi:10.1016/j.chemphys.2010.07.024}}.

\bibitem{escudero2023versus}
C.~Escudero and H.~Rojas.
\newblock {It\^o versus H\"anggi-Klimontovich}.
\newblock {\em arXiv preprint arXiv:2309.03654}, 2023.
\newblock \href {http://dx.doi.org/10.48550/arXiv.2309.03654}
  {\path{doi:10.48550/arXiv.2309.03654}}.

\bibitem{bhattacharyay2025brownian}
A.~Bhattacharyay.
\newblock Brownian motion near a wall: the dilemma of it{\^o} or stratonovich.
\newblock {\em Journal of Physics A: Mathematical and Theoretical},
  58(21):213001, 2025.
\newblock \href {http://dx.doi.org/10.1088/1751-8121/add8cf}
  {\path{doi:10.1088/1751-8121/add8cf}}.

\bibitem{sharma2025entropic}
M.~Sharma and A.~Bhattacharyay.
\newblock Entropic pulling and diffusion diode in an it{\^o} process.
\newblock {\em Physics Letters A}, page 130709, 2025.
\newblock \href {http://dx.doi.org/10.1016/j.physleta.2025.130709}
  {\path{doi:10.1016/j.physleta.2025.130709}}.

\bibitem{goldstein2006canonical}
S.~Goldstein, J.~L. Lebowitz, R.~Tumulka, and N.~Zangh{\`\i}.
\newblock Canonical typicality.
\newblock {\em Physical Review Letters}, 96(5):050403, 2006.
\newblock \href {http://dx.doi.org/10.1103/PhysRevLett.96.050403}
  {\path{doi:10.1103/PhysRevLett.96.050403}}.

\bibitem{lanford2007entropy}
O.~E. Lanford.
\newblock Entropy and equilibrium states in classical statistical mechanics.
  lecture notes in physics.
\newblock In {\em Statistical mechanics and mathematical problems}, volume~20,
  pages 1--113. Springer, Berlin, Heidelberg, 1973.
\newblock \href {http://dx.doi.org/10.1007/BFb0112756}
  {\path{doi:10.1007/BFb0112756}}.

\bibitem{allori2020statistical}
V.~Allori.
\newblock {\em Statistical mechanics and scientific explanation: Determinism,
  indeterminism and laws of nature}.
\newblock World Scientific, 2020.
\newblock \href {http://dx.doi.org/10.1142/11591} {\path{doi:10.1142/11591}}.

\bibitem{de2023hydrodynamic}
J.~De~Nardis and B.~Doyon.
\newblock Hydrodynamic gauge fixing and higher order hydrodynamic expansion.
\newblock {\em Journal of Physics A: Mathematical and Theoretical},
  56(24):245001, 2023.
\newblock \href {http://dx.doi.org/10.1088/1751-8121/acd153}
  {\path{doi:10.1088/1751-8121/acd153}}.

\bibitem{doyon2017dynamics}
B.~Doyon and H.~Spohn.
\newblock Dynamics of hard rods with initial domain wall state.
\newblock {\em Journal of Statistical Mechanics: Theory and Experiment},
  2017(7):073210, 2017.
\newblock \href {http://dx.doi.org/10.1088/1742-5468/aa7abf}
  {\path{doi:10.1088/1742-5468/aa7abf}}.

\bibitem{doyon2023ab}
B.~Doyon and F.~H{\"u}bner.
\newblock Towards an ab initio derivation of generalised hydrodynamics from a
  gas of interacting wave packets.
\newblock {\em arXiv preprint arXiv:2307.09307}, 2023.
\newblock \href {http://dx.doi.org/10.48550/arXiv.2307.09307}
  {\path{doi:10.48550/arXiv.2307.09307}}.

\bibitem{doyon2023generalised}
B~Doyon, F~H{\"u}bner, and T~Yoshimura.
\newblock Generalised {$T\bar T$}-deformations of classical free particles.
\newblock {\em arXiv preprint arXiv:2312.14855}, 2023.
\newblock \href {http://dx.doi.org/10.48550/arXiv.2312.14855}
  {\path{doi:10.48550/arXiv.2312.14855}}.

\bibitem{doyon2024new}
Benjamin Doyon, Friedrich H{\"u}bner, and Takato Yoshimura.
\newblock New classical integrable systems from generalized {$T\bar
  T$}-deformations.
\newblock {\em Physical Review Letters}, 132(25):251602, 2024.
\newblock \href {http://dx.doi.org/10.1103/PhysRevLett.132.251602}
  {\path{doi:10.1103/PhysRevLett.132.251602}}.

\bibitem{urilyon2025simulating}
A.~Urilyon, L.~Biagetti, J.~Kethepalli, and J.~De~Nardis.
\newblock Simulating generalised fluids via interacting wave packets evolution.
\newblock {\em Physical Review B}, 113:014314, 2026.
\newblock \href {http://dx.doi.org/10.1103/b587-8yyt}
  {\path{doi:10.1103/b587-8yyt}}.

\bibitem{doyon2017note}
B.~Doyon and T.~Yoshimura.
\newblock A note on generalized hydrodynamics: inhomogeneous fields and other
  concepts.
\newblock {\em SciPost Physics}, 2(2):014, 2017.
\newblock \href {http://dx.doi.org/10.21468/SciPostPhys.2.2.014}
  {\path{doi:10.21468/SciPostPhys.2.2.014}}.

\bibitem{ferrari2025macroscopic}
P.~A. Ferrari and S.~Olla.
\newblock Macroscopic diffusive fluctuations for generalized hard rods
  dynamics.
\newblock {\em The Annals of Applied Probability}, 35(2):1125--1142, 2025.
\newblock \href {http://dx.doi.org/10.1214/24-AAP2137}
  {\path{doi:10.1214/24-AAP2137}}.

\bibitem{chahal2025stochastic}
S.~Chahal, I.~Mukherjee, A.~Dhar, H.~Spohn, and A.~Kundu.
\newblock Stochastic dynamics of quasiparticles in the hard rod gas.
\newblock {\em arXiv preprint arXiv:2510.19693}, 2025.
\newblock \href {http://dx.doi.org/10.48550/arXiv.2510.19693}
  {\path{doi:10.48550/arXiv.2510.19693}}.

\bibitem{hairer2013solving}
M.~Hairer.
\newblock Solving the kpz equation.
\newblock {\em Annals of mathematics}, pages 559--664, 2013.
\newblock \href {http://dx.doi.org/10.4007/annals.2013.178.2.4}
  {\path{doi:10.4007/annals.2013.178.2.4}}.

\bibitem{toth2003onsager}
B.~T{\'o}th and B.~Valk{\'o}.
\newblock Onsager relations and eulerian hydrodynamic limit for systems with
  several conservation laws.
\newblock {\em Journal of Statistical Physics}, 112:497--521, 2003.
\newblock \href {http://dx.doi.org/10.1023/A:1023867723546}
  {\path{doi:10.1023/A:1023867723546}}.

\bibitem{grisi2011current}
R.~M. Grisi and G.~M. Sch{\"u}tz.
\newblock Current symmetries for particle systems with several conservation
  laws.
\newblock {\em Journal of Statistical Physics}, 145:1499--1512, 2011.
\newblock \href {http://dx.doi.org/10.1007/s10955-011-0341-5}
  {\path{doi:10.1007/s10955-011-0341-5}}.

\bibitem{castro2016emergent}
O.~A. Castro-Alvaredo, B.~Doyon, and T.~Yoshimura.
\newblock Emergent hydrodynamics in integrable quantum systems out of
  equilibrium.
\newblock {\em Physical Review X}, 6(4):041065, 2016.
\newblock \href {http://dx.doi.org/10.1103/PhysRevX.6.041065}
  {\path{doi:10.1103/PhysRevX.6.041065}}.

\bibitem{karevski2019charge}
D.~Karevski and G.~Sch{\"u}tz.
\newblock Charge-current correlation equalities for quantum systems far from
  equilibrium.
\newblock {\em SciPost Physics}, 6(6):068, 2019.
\newblock \href {http://dx.doi.org/10.21468/SciPostPhys.6.6.068}
  {\path{doi:10.21468/SciPostPhys.6.6.068}}.

\bibitem{doyon2021free}
B.~Doyon and J.~Durnin.
\newblock Free energy fluxes and the {Kubo--Martin--Schwinger} relation.
\newblock {\em Journal of Statistical Mechanics: Theory and Experiment},
  2021(4):043206, 2021.
\newblock \href {http://dx.doi.org/10.1088/1742-5468/abefe3}
  {\path{doi:10.1088/1742-5468/abefe3}}.

\end{thebibliography}
%\bibliographystyle{hieeetr.bst}

\end{document}